\documentclass{aa}
\usepackage{natbib}
\bibpunct{(}{)}{;}{a}{}{,} 
\usepackage{graphicx}
\usepackage{subcaption}
\usepackage{url}
\usepackage{hyperref}
\usepackage{color}
\usepackage{float}

\usepackage{txfonts}
\newcommand{\jetset}{\texttt{JetSeT}}   

\begin{document} 

   \title{Radio-$\gamma$-ray response in blazars as a signature of adiabatic blob expansion}

   \subtitle{}

   \author{  
A. Tramacere  \and V. Sliusar \and R. Walter  \and J. Jurysek \and M. Balbo}

\institute{Department of Astronomy, University of Geneva, Chemin Pegasi 51, CH-1290 Versoix, Switzerland}

\date{Received August 11, 2021; accepted December 6, 2021}

 
  \abstract
   {Multi-wavelength  light curves in long-term campaigns show that, for
   several blazars, the radio emission occurs with a significant delay with respect
   to the $\gamma$-ray band, with timescales  ranging from weeks to years. Such
   observational evidence has long been a matter of debate, and is usually interpreted as a signature of the $\gamma$-ray emission originating upstream in the jet, with  the emitting region becoming radio transparent at larger scales.}
   {In this paper, we show, by means of self-consistent numerical modelling,
   that the adiabatic expansion of a relativistic blob can explain these delays,
   reproducing lags compatible with the observed timescales.}
   {We use the \jetset~ framework to reproduce the numerical modelling of the
   radiative and accelerative processes, reproducing the temporal evolution of a
   single blob, from the initial flaring activity and the subsequent expansion.
   We follow the spectral evolution and the corresponding light curves,
   investigating the relations among the observed parameters, rise time, delay,
   and decay time, and we identify the link with physical parameters.}
   {We find that, when adiabatic expansion is active, lags due to the shift of
   the synchrotron frequency occur. The corresponding  time lags have an offset
   equal to the distance in time between the flaring onset and the beginning of
   the expansion, whilst the rising and decaying timescales depend on the
   velocity of the expansion and on the time required for the source to exhibit a
   synchrotron self-absorption frequency below the relevant radio spectral
   window. We derive an inter-band response function, embedding the
   aforementioned parameters, and we investigate the effects of the competitions between radiative and  adiabatic cooling timescales on the response. We  apply the response function to long-term radio  and $\gamma-$ray light curves of Mrk 421, Mrk 501, and 3C 273, finding satisfactory agreement on the log-term behaviour, and we use  a Monte Carlo Markov Chain approach to estimate some relevant physical parameters. We discuss applications of the presented analysis to polarization measurements and to jet collimation profile kinematics. The collimation profiles observed in radio images are in agreement with the prediction from our model.
   }
  {}

\maketitle
%

\section{Introduction}
In the unified scenario of the active galactic nuclei (AGN), blazars are understood as sources in which a
relativistic jet is produced by the central engine and points close to our line of sight \citep{BR1978}. 
Blazars are comprised of two kinds of AGN: BL Lacs and flat-spectrum radio quasars (FSRQs). BL Lac objects are marked by featureless optical spectra, while FSRQs show broad emission lines from hot high-velocity gas presumably located deep in the gravitational well of the central black hole. In a few cases, FSRQs show the presence of a blue `bump' \citep{Brown1989,Pian1999ApJ}. Both BL Lacs and FSRQs show high flux from the radio to the $\gamma$-ray band, strong variability \citep{Ulrich1997}, and high polarisation observed in the optical band. Their spectral energy distributions (SEDs) show two main components: a low-energy component with power peaking in the range from the IR to the X-ray band, and a high-energy component  peaking in the $\gamma$-rays. In the most widely accepted scenario, the low-energy component is understood as synchrotron (S) radiation emitted by ultrarelativistic electrons, and the high-energy bump is interpreted as inverse Compton (IC) emission for the case of a purely leptonic scenario \citep{Blandford_1979ApJ...232...34B} or dominated by high-energy emission of ultrarelativistic protons in the case of hadronic models \citep{Bottcher2013}. Despite the general consensus on this scenario, several open questions remain as to their puzzling behaviour that challenge our understanding of their observed phenomenology.
In particular, long-term multiwavelength  campaigns for several blazars have shown radio emission occurring  with a significant delay with respect to the $\gamma$-ray band, with timescales  ranging from weeks to years  \citep{Pushkarev_2010ApJ,max-moerbeck_2014MNRAS.445..428M}.  This observational evidence is not compatible with different cooling timescales, and has been a matter of debate for several years. A possible interpretation was proposed by taking into account  the different distances of the $\gamma-$ray and radio transparent region,  with the moving region becoming  transparent to the $\gamma$-rays \citep{Blandford_1995ApJ...441...79B} and later to the radio frequencies \citep{Blandford_1979ApJ...232...34B}. 
Delays between the GeV and radio variability of blazars have already been reported several times. \cite{max-moerbeck_2014MNRAS.445..428M} found hints for GeV leading the radio in four blazars. In the case of Mrk\,421, a delay of 40 days was reported in agreement with \cite{hovatta_2015MNRAS.448.3121H}. A much longer delay of $140\pm 20$ days was found by \cite{esposito_2015A&A...576A.122E} in 3C 273 with a response decaying in $\sim190$ days. Finally, \cite{Rani_2014A&A} reported a delay of 82 days between the GeV and radio light curves of S5\,0716+714 with a correlated change of the VLBI jet position angle and suggested that a shock in the jet first affected the gamma-ray emission and later triggered morphology shifts in the radio core. A key aspect to investigate in this scenario is to understand the  role of the adiabatic expansion of the emitting region, and the consequences in terms of variation of the synchrotron self-absorption frequency (SSA), as presented in the seminal works by \cite{McCray1969} and \cite{VanDerLaan1969} and more recently \cite{Boula2018}, \cite{Potter2018}, and \cite{Boula2021}. A recent analysis by \cite{Zacharias2021} investigates the adiabatic expansion scenario, but focuses  on the impact of the expansion on the usual `double-humped' SED shape rather than on the possible relation with the radio$-\gamma$ delay.

In this paper, we derive phenomenological trends linking the relevant timescales of the delay to the physical parameters of the emitting region,  and we verify them by means of a self-consistent numerical modelling. We propose a response function based  on the relevant phenomenological timescales that is  able to reproduce the radio-delayed light curve as a response to the $\gamma-$ray, and we validate the phenomenological trends against the numerical simulations, investigating biases due to the competition between radiative and adiabatic cooling timescales. We  apply this response to  Mrk 421, Mrk 501, and 3C237, and obtain good agreement with the long-term radio trends. Finally, we employ a Monte Carlo Markov Chain (MCMC) approach to estimate physical parameters from the comparison between the response function convolution parameters and the prediction from the phenomenological trends.  
The paper is organised as follows. In section \ref{sect:phenom} we derive the phenomenological trends 
expected under the hypothesis of a moving blob expanding with uniform velocity, and we characterise the delay 
in terms of the velocity of expansion and of the consequent evolution of the SSA, finding a physical link between observed rise and decay timescales and the physical parameters of the blob  and jet.
In section \ref{sect:sim_setup}, we describe our  setup of  numerical simulations done with the \jetset~ code \citep{jetset2020,Tramacere2011,Tramacere2009}, taking into account  radiative, accelerative processes, and adiabatic expansion. The simulations reproduce the long-term temporal evolution of a single blob, from the initial flaring activity, and the subsequent expansion. In section \ref{sect:exp_vs_no_exp} we compare the results for the cases of  an expanding versus  a non-expanding  blob.
In section \ref{sect:response}  we follow the spectral evolution and the corresponding light curves for 
different values of the expansion velocity and  for different radio frequencies. We propose a response function ---embedding the relevant observed timescales---  able to reproduce the radio light curve as a convolution with the $\gamma$-ray one, and we  validate the phenomenological trends against the numerical simulations, studying the biases on the timescales embedded in the response functions 
resulting from competition between radiative and adiabatic cooling timescales. In section \ref{sect:data_comparison} we apply our model to observed  data for Mrk 421,  Mrk 501, and  3C 273, and we reproduce long-term radio light curves as convolution of the $\gamma$-ray light curve with the proposed  response function. In section \ref{sect:discussion}  we employ a MCMC approach to estimate physical parameters  of the jet from a comparison between the response function convolution parameters and the prediction from the phenomenological trends.
More specifically, we investigate estimates of the source size, the magnetic field index, the initial SSA frequency, the expansion velocity,  and the spectral index of the electron distribution. We also compare our results with similar works in the literature, and discuss some  implications of our model  regarding the impact on the Compton dominance, hadronic models, and polarisation, and we also speculate on other possible causes of the delays, such as jet bending and the connection to the jet profile observed in the VLBI radio analysis.
In section \ref{sect:conclusions} we summarise our findings and discuss our upcoming extension of the presented model. In section \ref{sect:code} we provide instructions to reproduce the analysis and the numerical modelling presented in this paper.   
\begin{figure*}
\centering
\includegraphics[width=1.0\textwidth,angle=0]{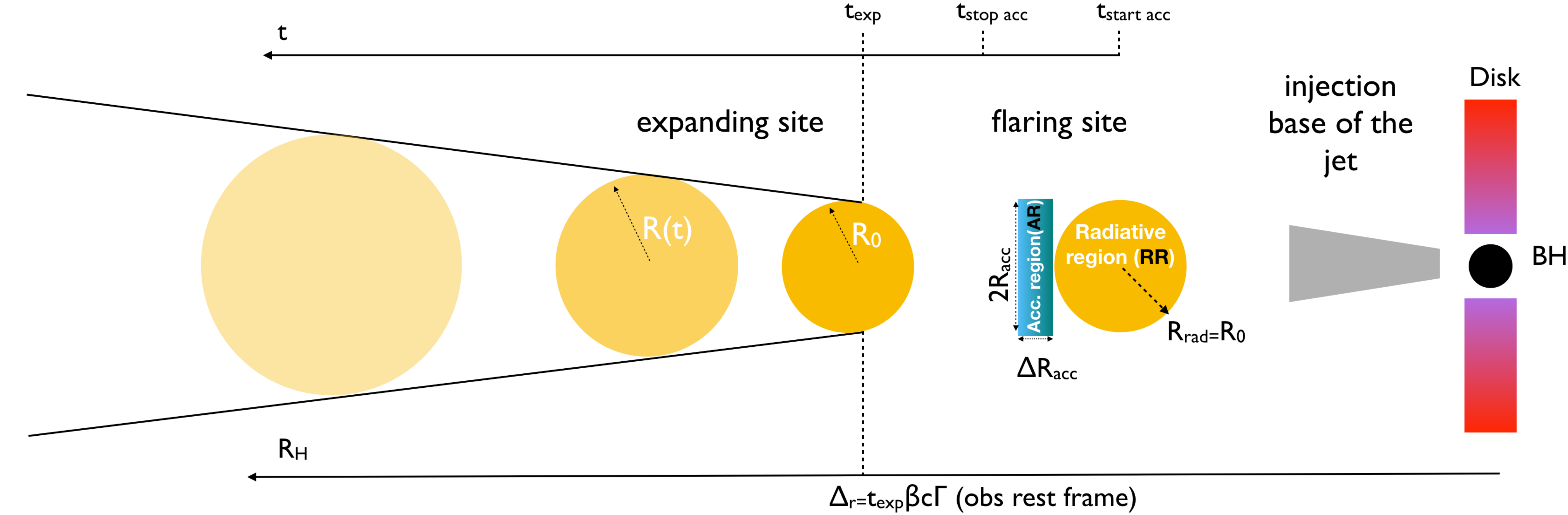}
\caption{Schematic representation of the model implemented in \jetset\  to simulate the flaring stage and 
the adiabatic expansion. At time $t_{\rm start ~acc }$, particles are injected and accelerated in the acceleration region  where they undergo both cooling and acceleration processes and diffuse towards the radiative region
, where only losses take place. The acceleration process ends at time $t_{\rm stop~ acc }$. After a time $t_{\rm exp}$, the expansion process takes place in the RR region.
}
\label{fig:jet_scheme}
\end{figure*}

\section{Phenomenological setup of an expanding  blob and synchrotron self-absorption }
\label{sect:phenom}
We assume that a spherical blob, characterised by an initial radius $R_0$ and
magnetic field $B_0$ expands with a constant velocity $\beta_{\rm exp}=v_{\rm exp}/c$, 
and that the expansion begins at a time $t_{\rm exp}$. All the quantities are measured in the frame of the emitting blob.
Quantities expressed in the observer frame are labelled by the $obs$ flag.
 The size of the
blob can be expressed as

\begin{equation}
\label{eq:R(t)}
R(t)=R_{0}+ \beta_{\rm exp}c(t-t_{\rm exp})H(t-t_{\rm exp}),
\end{equation}
where $H$ is the Heaviside step function.

The time-dependent law of the magnetic field, dictated by flux freezing  \citep{Begelman1984}  and
energy conservation, reads
\begin{equation}
\label{eq:B(t)}
B(t)=B_{0}\Big(\frac{R_{0}}{R(t)}\Big)^{m_B},
\end{equation}

where the index $m_B\in[1,2]$ depends on the geometric configuration of the
magnetic field, with $m_B=2$ for a fully poloidal configuration, and $m_B=1$ for
a fully toroidal configuration   \citep{Begelman1984}. The adiabatic cooling will read
\citep{SLongair2011}
\begin{equation}
\label{eq:ad_cool}
\dot \gamma_{ad}(t)=\frac{1}{3}\frac{\dot{V}}{V}\gamma = \frac{\dot R(t)}{R(t)}\gamma = \frac{\beta_{\rm exp}c}{R(t)}\gamma,  
\end{equation}
where $\gamma$ is the Lorentz factor of the electrons, and $V$ is the volume of the region that we assume to be spherical. The corresponding cooling time can be expressed as
\begin{equation}
\label{eq:t_ad_cool}
t^{ad}_{\rm cooling}(t)=\frac{\gamma}{\dot \gamma}= \frac{R(t)}{\beta_{\rm exp}c} = \frac{R_{0}+ \beta_{\rm exp}c(t-t_{\rm exp})H(t-t_{\rm exp})}{\beta_{\rm exp}c}
.\end{equation}
The evolution of the synchrotron self-absorption frequency can be expressed as
\citep{Ryb1986}

\begin{equation}
\label{eq:nu_ssa_generic}
\nu_{\rm SSA}(t) =\nu_L(t) \Big[\frac{\pi\sqrt{\pi}}{4} \frac{e R(t) N(t)}{B(t)} f_{k}(p)  \Big]^{\frac{2}{p+4}},
\end{equation}

where $e$ is the electron charge, $N(t)$ is the particle number density at time $t$, $p$ is the power-law index of the 
electron distribution at the Lorentz factor  most contributing to
$\nu_{\rm SSA}(t)$, and $\nu_L(t)=\frac{eB(t)}{2\pi m_e c}$ is the Larmor frequency.
The functions $f_{k}(p)$ are approximated to  
percent accuracy   as reported in \cite{Ghisellini2013b}. Assuming that
particles are confined ($R^3N(t)=N^{\rm tot}$), and plugging Equation \ref{eq:B(t)} and
\ref{eq:R(t)} into Equation \ref{eq:nu_ssa_generic} we obtain \begin{eqnarray}
\label{eq:nu_ssa_t_N_const}
\nu_{\rm SSA}(t)&\propto &\Big[B(t)^{\frac{p+2}{2}}\frac{N^{\rm tot}}{R(t)^2}\Big]^{\frac{2}{p+4}}
.\end{eqnarray}
Setting the initial self-absorption frequency $\nu^0_{\rm SSA}\equiv \nu_{\rm SSA}(t=0)$,
an increase in flux of the synchrotron emission at a given  frequency
$\nu^* >\nu^0_{\rm SSA}$ is expected at time $t^*$ such that
$\nu_{\rm SSA}(t^*) \equiv \nu^*_{\rm SSA} \simeq\nu^*$, when the source is characterised
by a size $R^* = R(t^*)$ and $B^*=B(t^*)$. Hence, at the time $t^*$ the values of $R^*$ and $B^*$ 
are such that the source is optically thin at frequencies $\nu\geq \nu^*$.
We use Equation \ref{eq:nu_ssa_t_N_const} to relate the two frequencies $\nu^0_{\rm SSA}$
and $\nu^*_{\rm SSA}$ to the corresponding blob radius $R^*$ :

\begin{eqnarray}
\label{eq:nu_ssa_R}
\frac{\nu^*_{\rm SSA}}{\nu^0_{\rm SSA}}& = &\Big[ \Big( \frac{B^*}{B_0}
\Big)^{\frac{p+2}{2}} \Big(\frac{R_0}{R^*}\Big)^2\Big]^{\frac{2}{p+4}}=\
\Big[\frac{R_0}{R^*}\Big]^{\frac{m_B(p+2)+4}{p+4}}
.\end{eqnarray}
This equation provides a link between the temporal evolution of the SSA
frequency and source radius, for a homogeneous blob expanding with a constant
velocity. We can easily invert this relation, and solve it in terms of $R^*$:
\begin{eqnarray}
\label{eq:R_transp}
R^*&=& R_0\Big( \frac{\nu^0_{\rm SSA}}{\nu^*_{\rm SSA}}\Big)^{\psi}\\ 
\psi&=&\frac{p+4}{{m_B(p+2)+4}}.\nonumber
\end{eqnarray}
This equation allows us to determine the time needed, starting from $t_{\rm exp}$, to move the initial
$\nu^0_{\rm SSA}$ to $\nu^*_{\rm SSA}$, which is also the time needed to expand the source from an initial 
radius $R_0$ to the  radius $R^*$. The corresponding time to reach the peak of the synchrotron light curve  
at the frequency $\nu^*_{\rm SSA}$ in the blob rest frame is\begin{eqnarray}
\label{eq:t_rise}
t_{\rm peak} =\Delta t_{R_0\rightarrow R^*}=    \frac{R^*-R_0}{\beta_{\rm exp}c}= 
\frac{R_0}{\beta_{\rm exp}c}\Big[ \Big( \frac{\nu^0_{\rm SSA}}{\nu^*_{\rm SSA}}\Big)^{\psi}  - 1 \Big]
.\end{eqnarray}
We stress that this equation holds as long as the synchrotron cooling is not the dominant cooling timescale, and we discuss this topic  more in detail in Section \ref{sect:response}.
The  total delay will be given by the sum of $t_{\rm exp}$ and $t_{\rm peak}$, that is,
\begin{eqnarray}
\label{eq:delta_T}
\Delta t_{\nu^0_{\rm SSA}\rightarrow \nu^*_{\rm SSA}}&=&
t_{\rm exp}+ t_{\rm peak}=t_{\rm exp}+\frac{R_0}{\beta_{\rm exp}c}\Big[ \Big( \frac{\nu^0_{\rm SSA}}{\nu^*_{\rm SSA}}\Big)^{\psi}  - 1 \Big]
.\end{eqnarray}
Finally, the adiabatic decay time will be proportional to the adiabatic cooling time at $R^*$:  
\begin{eqnarray}
\label{eq:t_decay_ad}
t_{\rm decay}^{ad}(t^*) &\propto& \frac{R^*}{\beta_{\rm exp}c}=
\frac{R_0}{\beta_{\rm exp}c} \Big( \frac{\nu^0_{\rm SSA}}{\nu^*_{\rm SSA}}\Big)^{\psi} 
.\end{eqnarray}
Of course, this is the relevant timescale at the time $t^*$ such that $R(t^*)=R^*$, and will increase 
according to Equation \ref{eq:t_ad_cool}.
It is relevant to note that the decaying time will also be affected by the purely geometric factor, depending 
on $B(t)$ and $R(t)$,  that is, the flux variation due to a change in $B(t)$ and $R(t)$, ignoring the cooling terms. This can be easily derived  from the expected synchrotron trend for an optical depth $\tau>>1$, $F_{\nu^{\rm SSA}}(t) \propto N(t) V(t) B(t)^{-0.5}$   \citep{Ghisellini2013b} , and taking into account that, for confined emitters, $N(t)V(t)$ is constant:
$F_{\nu^{\rm SSA}}(t) \propto B(t)^{-0.5}$, where $V(t)$ is the time-dependent value of the blob volume.
Hence, the time-dependent geometric decay time at $t=t^*$ will read
\begin{equation}
\label{eq:t_geom}
t_{\rm decay}^{geom}(t^*) \propto \frac{F_{\nu^{\rm SSA}}(t^*)}{\dot{F}_{\nu^{\rm SSA}}(t^*)} \propto 
\frac{ R^*}{ m_B\beta_{\rm exp}c}=\frac{ t_{\rm decay}^{ad}(t^*)}{ m_B}.
\end{equation}

\begin{figure*}
   \centering
   \begin{tabular}{lr}
   \includegraphics[width=0.4\textwidth,angle=0]{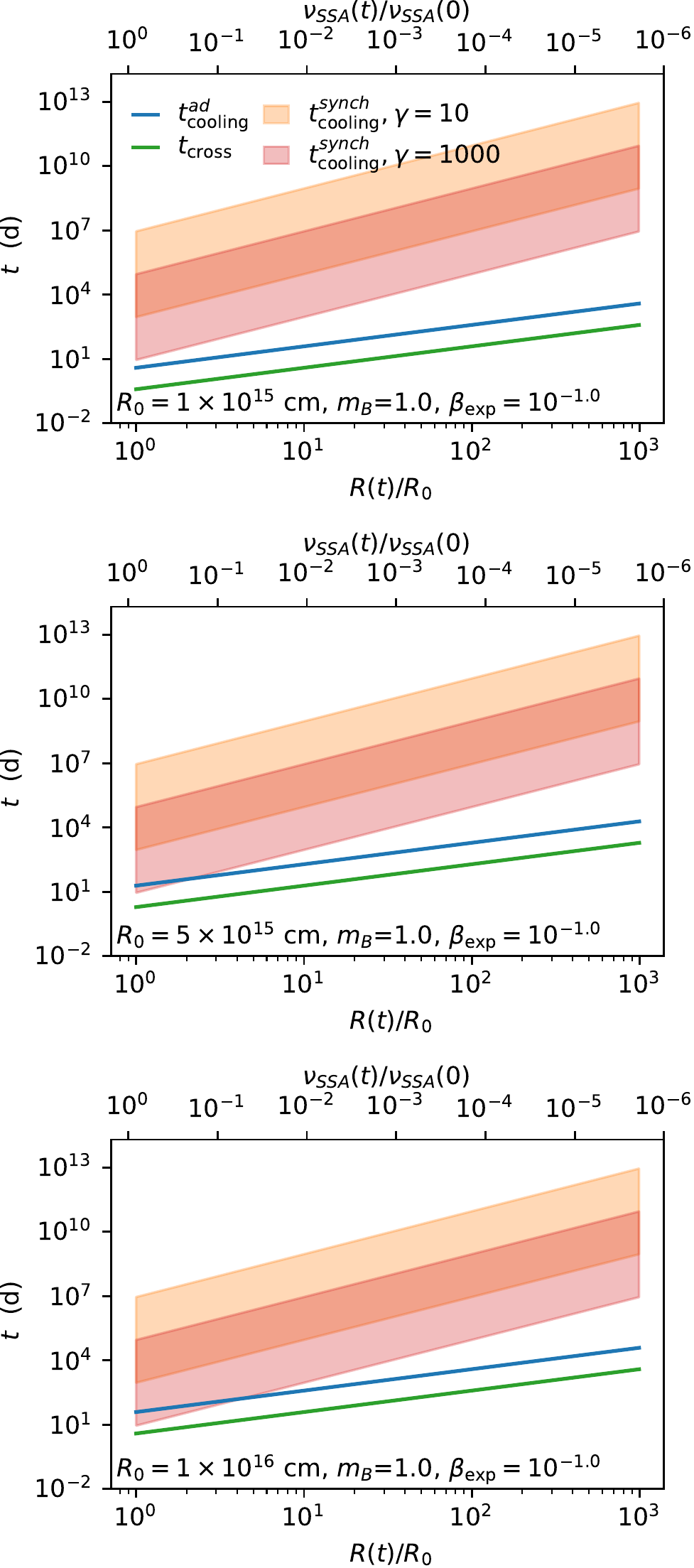}&
   \includegraphics[width=0.4\textwidth,angle=0]{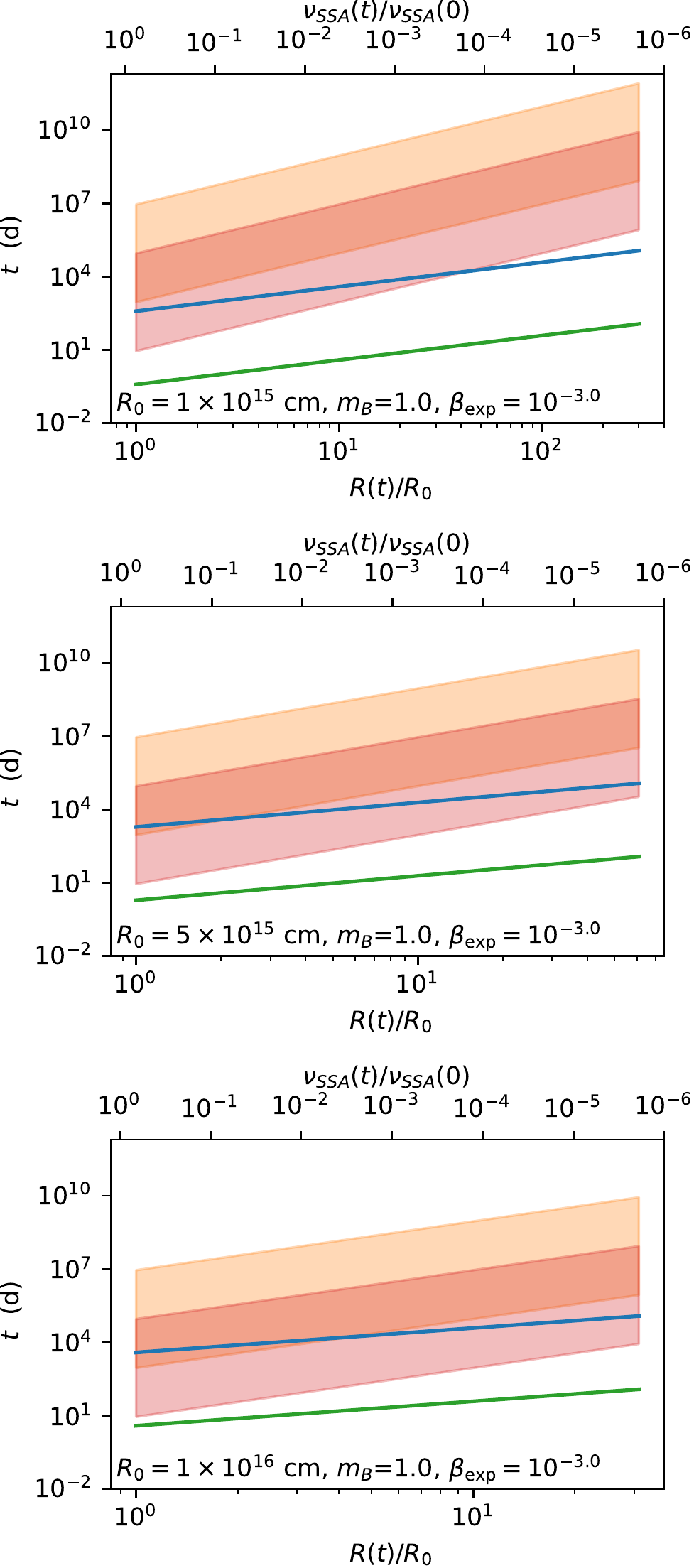}
   \end{tabular}
   \caption{
   \label{fig:time_scales}
   Competition among the different timescales. \textit{Left panels:} Case 
   of $\beta_{\rm exp}=0.1$.  \textit{Right panels:} Case of $\beta_{\rm exp}=0.001$. The top row refers to the case of $R_0=1\times 10^{15}$ cm, the middle row to the case of  $R_0=1\times 10^{15}$, and the bottom row to the case of $R_0=1\times 10^{16}$ cm. The times are in the  blob frame, and the total duration is such that $R(t_{stop} )=1000 R_0$. The $x$ axis reports on the bottom side the value of $R(t)/R_0$, whilst on the opposite side the corresponding value of $\nu_{\rm SSA}(t)/\nu_{\rm SSA}(0)$ is reported, for the case of no particle escape. The orange shaded area represents the synchrotron cooling timescales, with the lower bound corresponding to the case of $\gamma=10$ and $B_0=1.0$ G, and the upper bound corresponding to the case of $\gamma=10$ and    $B_0=0.01$ G. The red shaded area, represents the same  trend  for the case of  $\gamma=1000$. The blue line represents the adiabatic cooling time (Equation \ref{eq:t_ad_cool}), the black red line represents the geometrical decay time (Equation \ref{eq:t_geom}), and the green line the light crossing time ($R(t)/c$).
   }
\end{figure*}

Therefore, within the assumptions described above, and as  long as the adiabatic cooling timescale dominates over the synchrotron one, we can estimate the final decay timescale as 
\begin{eqnarray}
\label{eq:t_decay}
t_{\rm decay}(t^*)  \propto \frac{R_0}{m_B \beta_{\rm exp}c} \Big( \frac{\nu^0_{\rm SSA}}{\nu^*_{\rm SSA}}\Big)^{\psi} 
.\end{eqnarray}
The balance between the relevant timescales is quite complex, and in
 Figure \ref{fig:time_scales} we show the trends for different configurations. The left panels refer to  the case of $\beta_{\rm exp}=0.1$,  and the right panels to the case of $\beta_{\rm exp}=0.001$. 
 We select the boundaries $\gamma=10$ and $\gamma=1000$ to sample typical values of the Lorentz factor corresponding to electrons radiating by synchrotron emission in the radio band, for $B$ ranging in $[0.001,1]$ G, and a broad range of values of $\delta$ and $z$. 
 We notice that for $\beta_{\rm exp}=0.1$, the adiabatic and geometrical decay timescales dominate over the synchrotron timescale, except for the initial state of a few configurations. On the contrary, for the case of $\beta_{\rm exp}=0.001$ the competition between radiative and adiabatic timescales is more complex.
 The effect of this complex interplay among the cooling timescales is investigated in detail in Section \ref{sect:response}. We also notice that, for the parameter space investigated in the present analysis, the crossing time is always  shorter than the other timescales, making the approximation of ignoring its effect hereafter relatively  plausible.

Finally, we can express these relations ---which are valid in the adiabatic-dominated cooling regime--- in the observer frame:
\begin{eqnarray}
\label{eq:times_obs_t_R0}
\Delta t ^{\rm obs}_{\nu^{0,\rm obs}_{\rm SSA}\rightarrow
\nu^{*, \rm obs}_{\rm SSA}}&=&\frac{1+z}{\delta}\Big[t_{\rm exp}+\frac{R_0}{\beta_{\rm exp}c}\Big(\Big(\frac{\nu^0_{\rm SSA}}{\nu^{*}_{\rm SSA}}\Big)^{\psi}
-1\Big)\Big]\\  \nonumber
t^{\rm obs}_{\rm decay} &=& \frac{(1+z)}{\delta}
\frac{R_0}{m_B \beta_{\rm exp}c} \Big( \frac{\nu^0_{\rm SSA}}{\nu^*_{\rm SSA}}\Big)^{\psi}
 \\  \nonumber
t^{\rm obs}_{\rm peak} &=& \frac{(1+z)}{\delta}
\frac{R_0}{\beta_{\rm exp}c}\Big[ \Big( \frac{\nu^0_{\rm SSA}}{\nu^*_{\rm SSA}}\Big)^{\psi}  - 1 \Big],
\end{eqnarray}
where $\nu^{\rm obs}$=$\nu \frac{\delta}{z+1}$, and the $\delta$ is a beaming factor:
\begin{equation}
\delta=\frac{1}{\Gamma(1-\beta_{\Gamma} \cos(\theta))},
\end{equation}
where $\Gamma$ is the bulk Lorentz factor of the blob moving with velocity $\beta_{\Gamma} c$, and $\theta$ is the observing angle of the jet. 
   We can substitute the  observed timescale variability $t_{var}^{\rm obs}=\frac{(1+z)R_0 }{\delta c}$ in the equations above:
\begin{eqnarray}
\label{eq:times_obs_t_var}
\Delta t^{\rm obs}_{\nu^{\rm  0,obs}_{\rm SSA}\rightarrow \nu^{\rm *, obs}_{\rm SSA}} &=& t^{\rm obs}_{\rm exp} 
+\frac{t_{var}^{\rm  obs}}{\beta_{\rm exp}}\Big[\Big(\frac{\nu^0_{\rm SSA}}{\nu^*_{\rm SSA}}\Big)^{\psi} -1 \Big]\\ \nonumber
t^{ \rm obs}_{\rm peak} &=&\frac{t_{var}^{\rm obs}}{\beta_{\rm exp}}\Big[\Big(\frac{\nu^0_{\rm SSA} }{\nu^*_{\rm SSA}}\Big)^{\psi} -1\Big ]\\ \nonumber
t^{\rm obs}_{\rm decay} &=& \frac{t_{var}^{\rm obs}}{ m_B\beta_{\rm exp}} \Big(\frac{\nu^0_{\rm SSA}}{\nu^*_{\rm SSA}}\Big)^{\psi},\\ \nonumber
\end{eqnarray}
where $t^{\rm obs}_{\rm exp} =t_{\rm exp}(1+z)/\delta .$
The distance travelled along the jet during the time interval $\Delta t^{\rm obs}_{\nu^{\rm  0,obs}_{\rm SSA}}$  will be
\begin{eqnarray}
\label{eq:delat_r}
\Delta_{r}=\Gamma \Delta t_{\nu^0_{\rm SSA}\rightarrow \nu^*_{\rm SSA}}\beta_{\Gamma} c=\frac{\delta\Gamma \beta_{\Gamma} c \Delta t^{\rm obs}_{\nu^{\rm  0,obs}_{\rm SSA}}}{1+z}
,\end{eqnarray}
which is in agreement with the phenomenological derivations by \cite{Pushkarev_2010ApJ} and \cite{max-moerbeck_2014MNRAS.445..428M}, 
if $\nu^{0,\rm obs}_{\rm SSA}$ corresponds to the initial observed SSA frequency at the time of the flaring episode, and  $\nu^{*,\rm obs}_{\rm SSA}$ corresponds to the observed frequency of the delayed radio flare.

\section{Self-consistent temporal evolution of an expanding blob with the ~\jetset ~ code.}
\label{sect:fp-approac}
To follow the evolution of the emitting particle distribution, and the radiative fields, we use the {\ttfamily JetTimeEvol} class  from the {\ttfamily jet\_timedep} module of the open-source \jetset \footnote{ \url{https://github.com/andreatramacere/jetset}} framework \citep{jetset2020,Tramacere2011,Tramacere2009}. This class allows the user to evolve the particle distribution under the effects of radiative cooling, adiabatic expansion, and acceleration processes (both systematic and stochastic), and to extract SEDs and light curves at any given time (see Appendix \ref{sect:code} for further details on  code availability and reproducibility).
The code proceeds through the numerical solution of a kinetic equation, following the same approach as in \cite{Tramacere2011} based on the  employment of  the quasi-linear approximation with the inclusion of a
momentum diffusion term  \citep{Ramaty1979,Becker2006}. The equation governing the temporal evolution of the particle energy distribution $n(\gamma)=dN(\gamma)/d\gamma$ is the Fokker-Planck (FP) equation that reads 
\begin{eqnarray}
\label{eq:fp_eq}
\frac{\partial n(\gamma,t)}{\partial t}&=&
\frac{\partial }{\partial \gamma }
\Big\{-[S(\gamma,t) + D_A(\gamma,t) ]n(\gamma,t)\Big\}\\
&+&\frac{\partial }{\partial \gamma } \Big\{D_{p}(\gamma,t) \frac{\partial
n(\gamma,t)}{\partial \gamma }\nonumber \Big\}
-\frac{n(\gamma,t)}{T_{esc}(\gamma)} -\frac{n(\gamma,t)}{T_{ad}(t)}  +Q(\gamma,t).
\end{eqnarray}
The momentum diffusion coefficient $D_{p}(\gamma,t)\propto (\gamma^q)$  \citep{Becker2006}  and the average energy
change term resulting from the momentum-diffusion process
$D_A(\gamma,t)=(2/\gamma)D_{p}(\gamma,t)$   \citep{Becker2006}  represent the contribution from a
stochastic momentum-diffusion acceleration mechanism \citep{Kardashev1962,Melrose1969,Katarzynski2006,Stawarz2008}.     The systematic term
$S(\gamma,t)=-C(\gamma,t)+A(\gamma,t)$  describes systematic energy loss ($C$)
and/or gain ($A$), and $Q(\gamma,t)$ is the injection term.
A detailed description of the cooling terms and  the diffusion coefficient is provided in Appendix \ref{app:fp_eq}. The term $\frac{n(\gamma,t)}{T_{ad}}$ corresponds to the decrease in particle density due to the expansion process, with $T_{ad}(t)=\frac{1}{3} \frac{R(t)}{\beta_{\rm exp}c}$ \citep{Gould1975}. This term is connected to source geometry and should not be confused with the cooling term defined in Equation \ref{eq:t_ad_cool} and plugged in the $C(\gamma, t)$ term (see Appendix \ref{app:fp_eq}).
The term $\frac{n(\gamma,t)}{T_{esc}(\gamma)}$ represents the particle escape term. The injection function $Q(\gamma_{inj},t)$ is normalised according to\begin{equation} 
\label{eq:q_inj}
L_{inj}=V_{\rm acc}\int\gamma m_ec^2 Q(\gamma,t)d\gamma ~~~ (erg/s),
\end{equation}
where $V_{\rm acc}$ is the volume of the acceleration region, and the integration is performed over the numerical grid used to solve Equation \ref{eq:fp_eq} (see the following section for further details).
The numerical solution of the FP equation is obtained using the same approach as \cite{Tramacere2011}, which is based on the method proposed by \cite{Chang1970} as described in \cite{Park1996}.

\setcounter{table}{0}
\begin{table}[ht]
\begin{center}
\caption{Parameters for the flaring simulation}
\label{tab:flare_sim_par}
\begin{tabular}{ll|l|l}
\hline
&&rad. region& acc. region\\
\hline
\hline
$R$                  &(cm)         &$5\times10^{15}$  &  $5\times10^{15}$    \\
$\Delta_R^{acc}$     &(cm)         & -                &  $5\times10^{14}$    \\
$R_{H0}$             &(cm)         & -                &  $10^{17}$    \\
$B$                  &(G)          &0.2               &  0.2 \\
$\delta$             &             &30                &  30 \\
$z$                  &             &0.03              &  0.03 \\
$L_{inj}$            &(erg/s)      &$5\times10^{39}$  &  -  \\
$q$                  &             &  -               &  2               \\
$t_A$                & (s)         &  -               & $2.5\times 10^4$ \\
$t_{D_0 }=1/D_{P0}$  & (s)         &   -              & $1.5\times 10^5 $ \\
$T_{esc}$            & (s)         &$\infty $         & $5\times 10^4$    \\
{\it Duration}             & (s)         &$10^6$            & $10^6$      \\
{\it Duration acc.}        & (s)         &-                 & $10^5$    \\
{\it Duration inj.}       & (s)         &-                 & $10^5$    \\
$T_{\rm size}$       &             &$2\times 10^4$    & $2\times 10^4$    \\
$NUM_{SET}$          &             &200               & 200 \\
$\gamma_{inj}$       &             &-                  &  10.0    \\
\hline
\end{tabular}
\end{center}
\end{table}

\setcounter{table}{1}
\begin{table}[ht]
\begin{center}
\caption{Parameters for the long-term simulation with expansion}
\label{tab:exp_sim_par}
\begin{tabular}{ll|l}
\hline
&& expanding rad. region \\
\hline
\hline
$R_0$                &(cm)         & $5\times10^{15}$  \\
$B_0$                &(G)          & 0.2               \\
$\delta$             &             &30                 \\
$m_B$                &             & 1  \\
$\beta_{\rm exp}$    &(c)          & $[0.001-0.3]$                \\
$t_{ \rm exp}$        & (s)        &  $1\times 10^{7}$                \\
$T_{esc}$            & (s)         &$\infty $         \\
{\it Duration}             & (s)         &$[1.6\times 10^7  - 1.9\times 10^{9}]$             \\
$T_{\rm size}$       &             &$[1.6\times10^4 - 1.8\times 10^{6}]$       \\
$NUM_{SET}$          &             &[1000, 5000]               \\
\hline
\end{tabular}
\end{center}
\end{table}

\begin{figure*}
\centering
\begin{tabular}{lc}
\includegraphics[width=0.5\textwidth,angle=0]{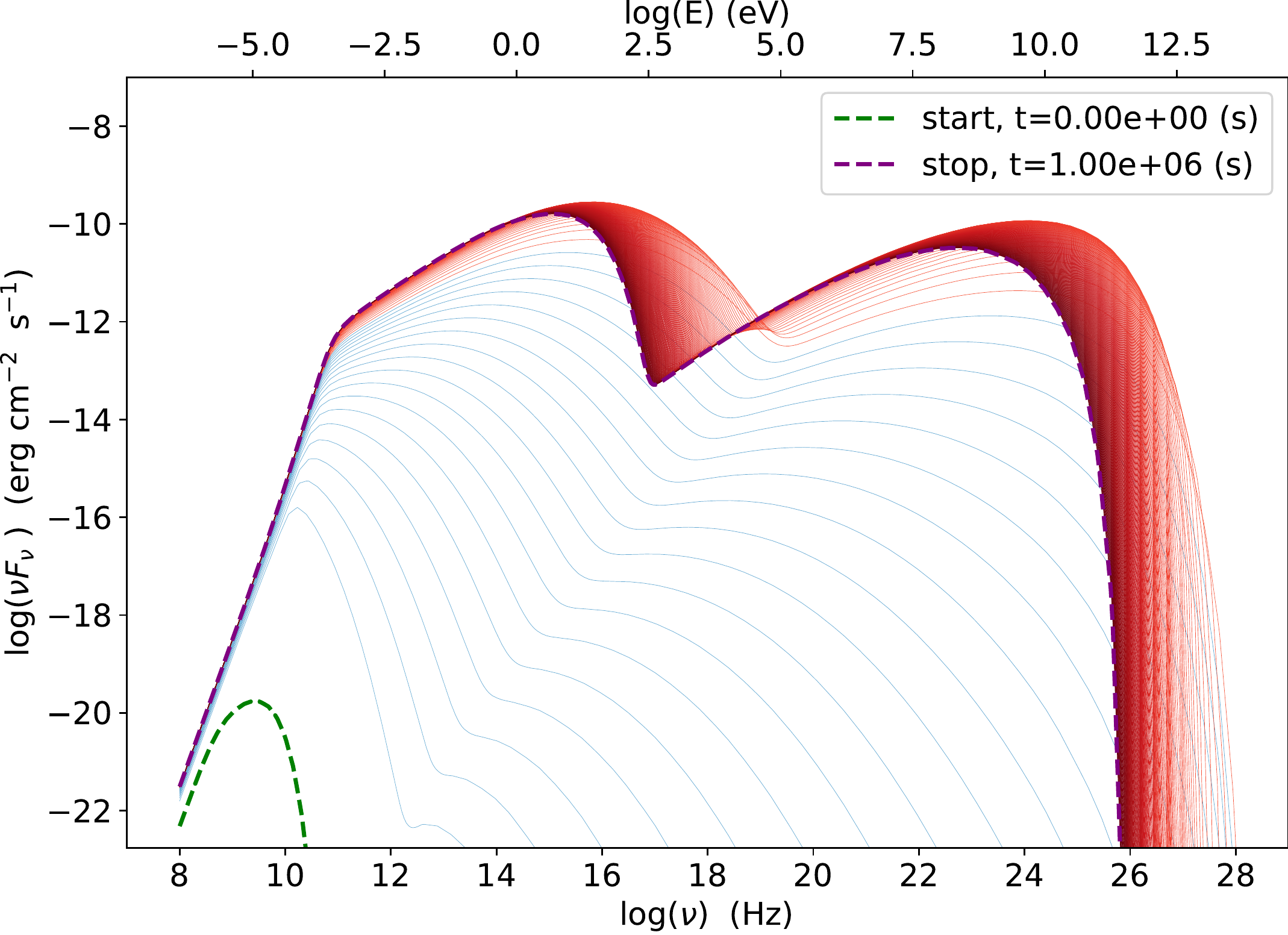}
&\includegraphics[width=0.5\textwidth,angle=0]{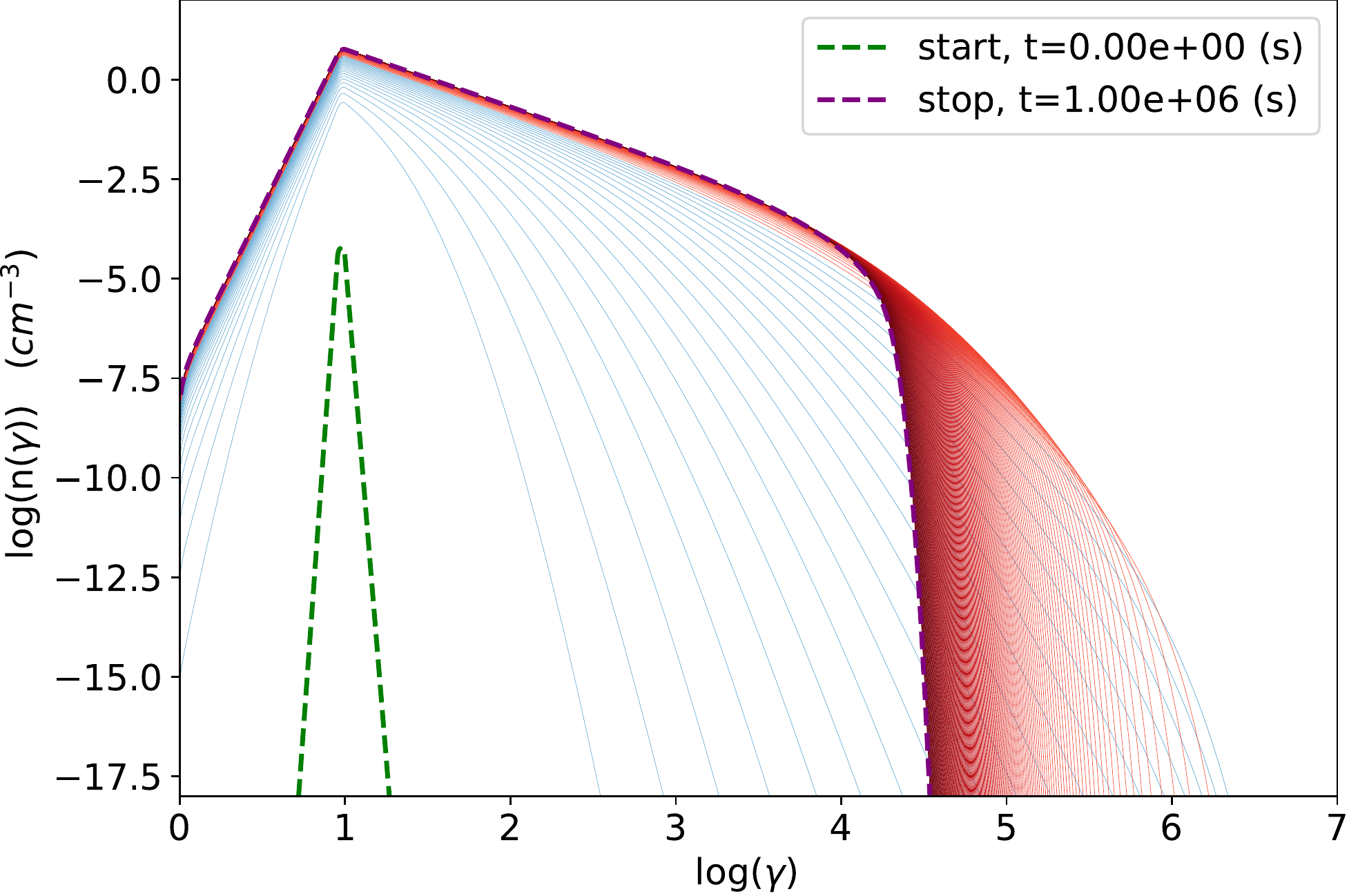}
\end{tabular}
\caption{
\label{fig:flare_sed}
\textit{Left panel:} SEDs corresponding to the simulation of the flaring state, for the radiative region. The dashed green line corresponds to the earliest of the  SEDs stored by the code, the blue lines correspond to the period when the injection, acceleration, and radiative  process are active, and the red lines correspond to the period when only the radiative processes are active. The times reported in the label are in the blob frame. \textit{Right panel:} Same as in left panel, but for the electron energy distribution in the radiative region.
}
\end{figure*}

\begin{figure*}
\centering
\begin{tabular}{lc}

\includegraphics[width=0.5\textwidth,angle=-0]{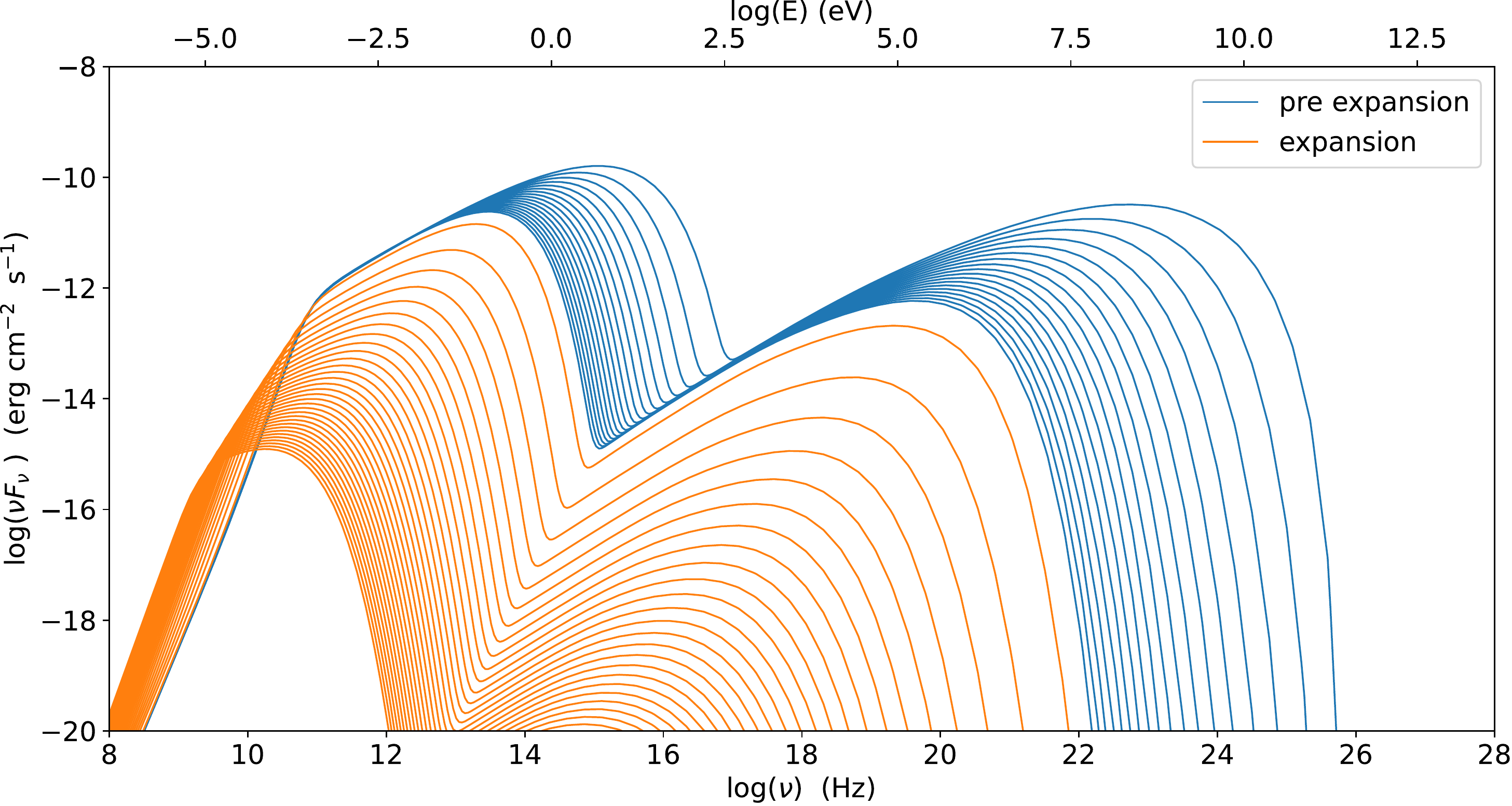}&
\includegraphics[width=0.5\textwidth,,angle=-0]{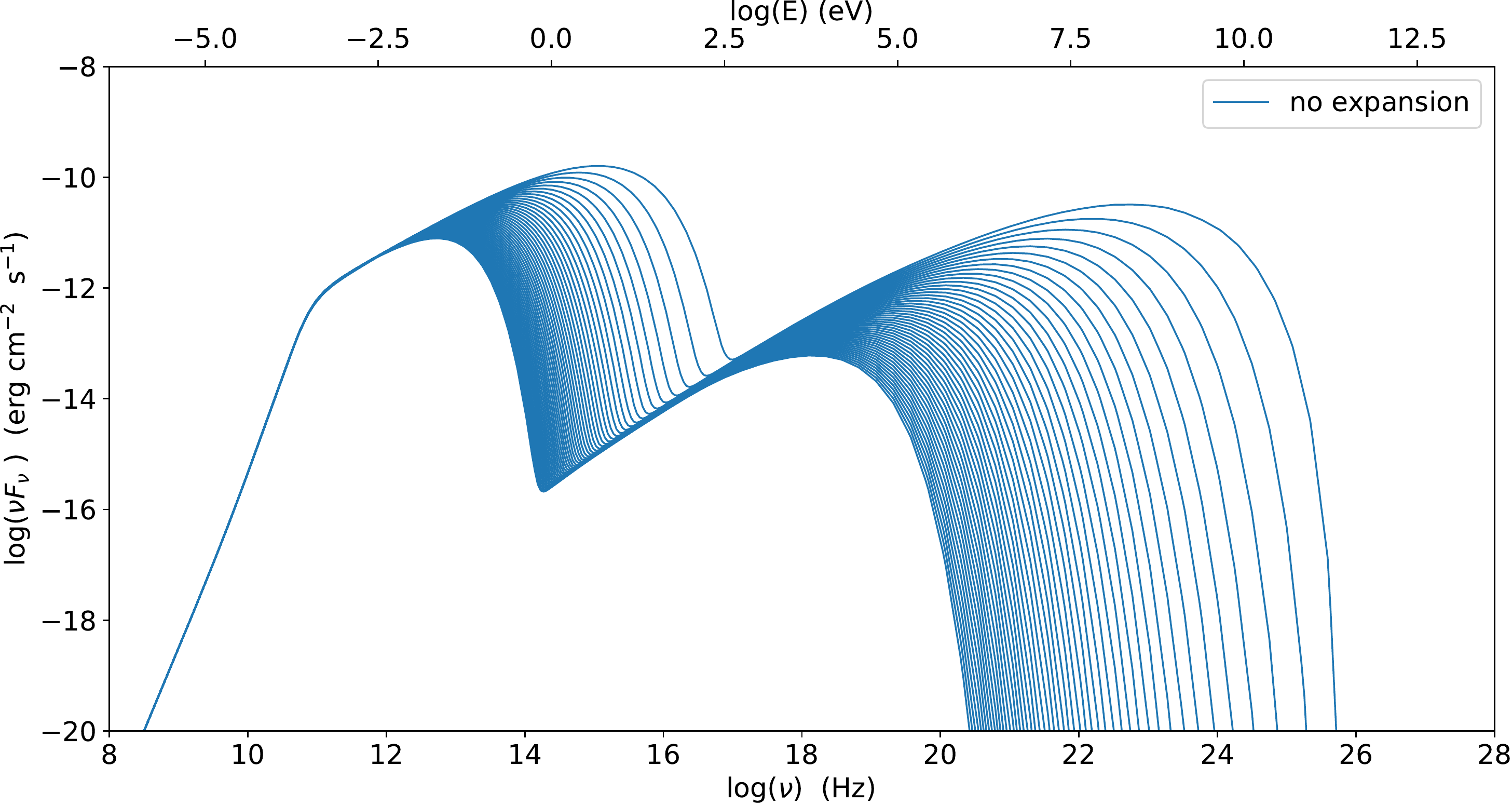}\\

\includegraphics[width=0.5\textwidth,angle=-0]{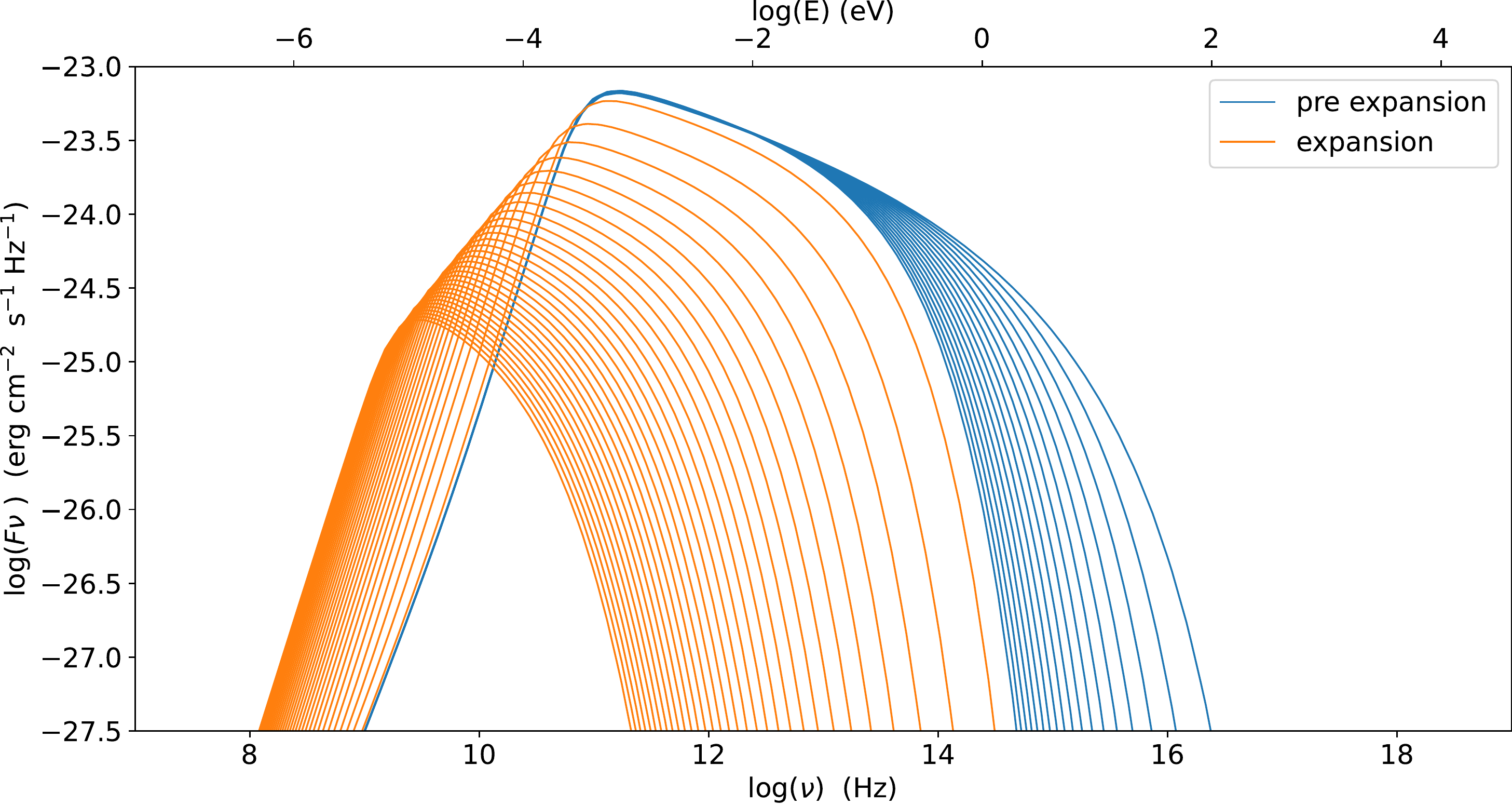}& 
\includegraphics[width=0.5\textwidth,,angle=-0]{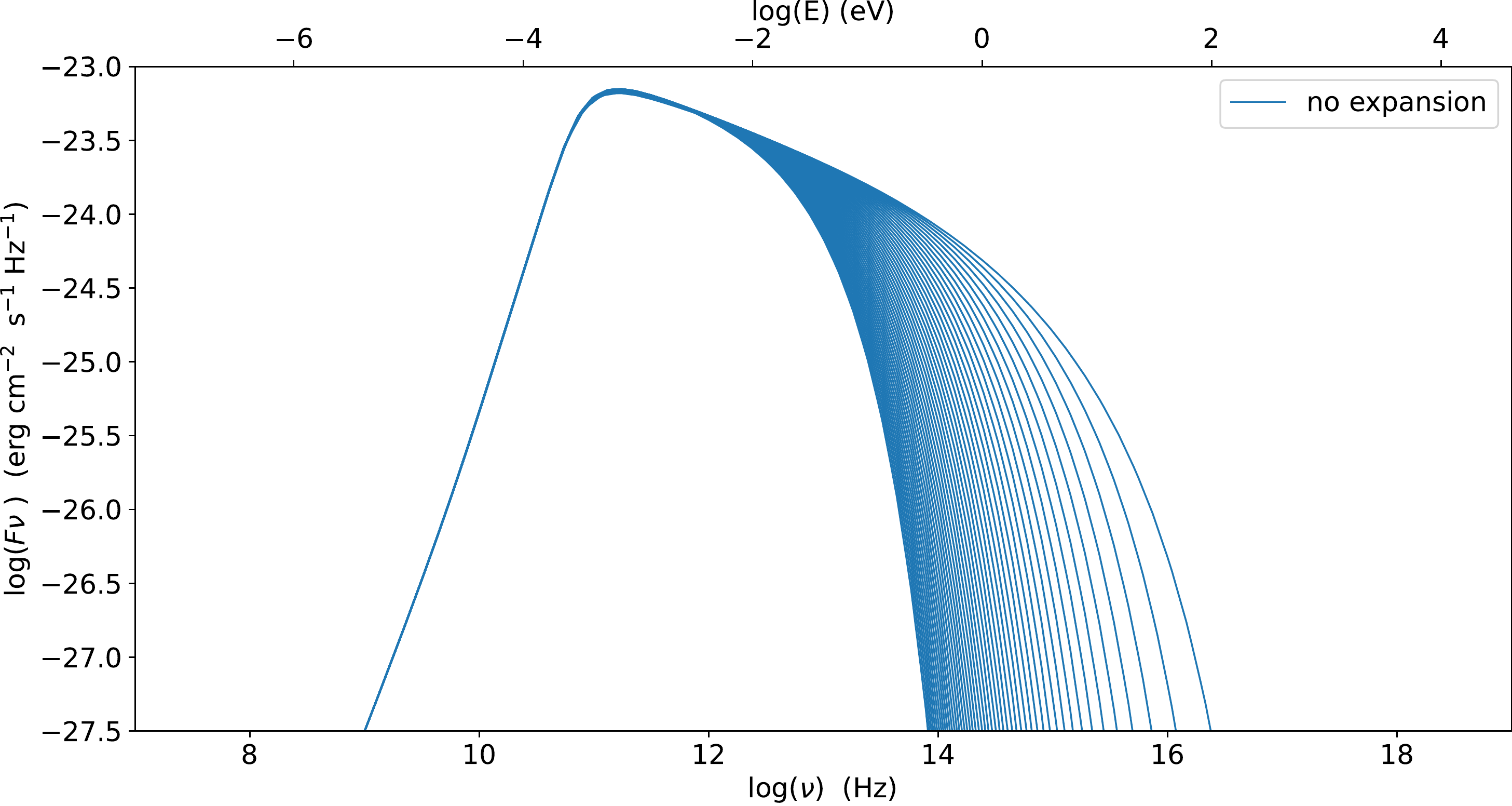}\\
      
\includegraphics[width=0.5\textwidth,angle=-0]{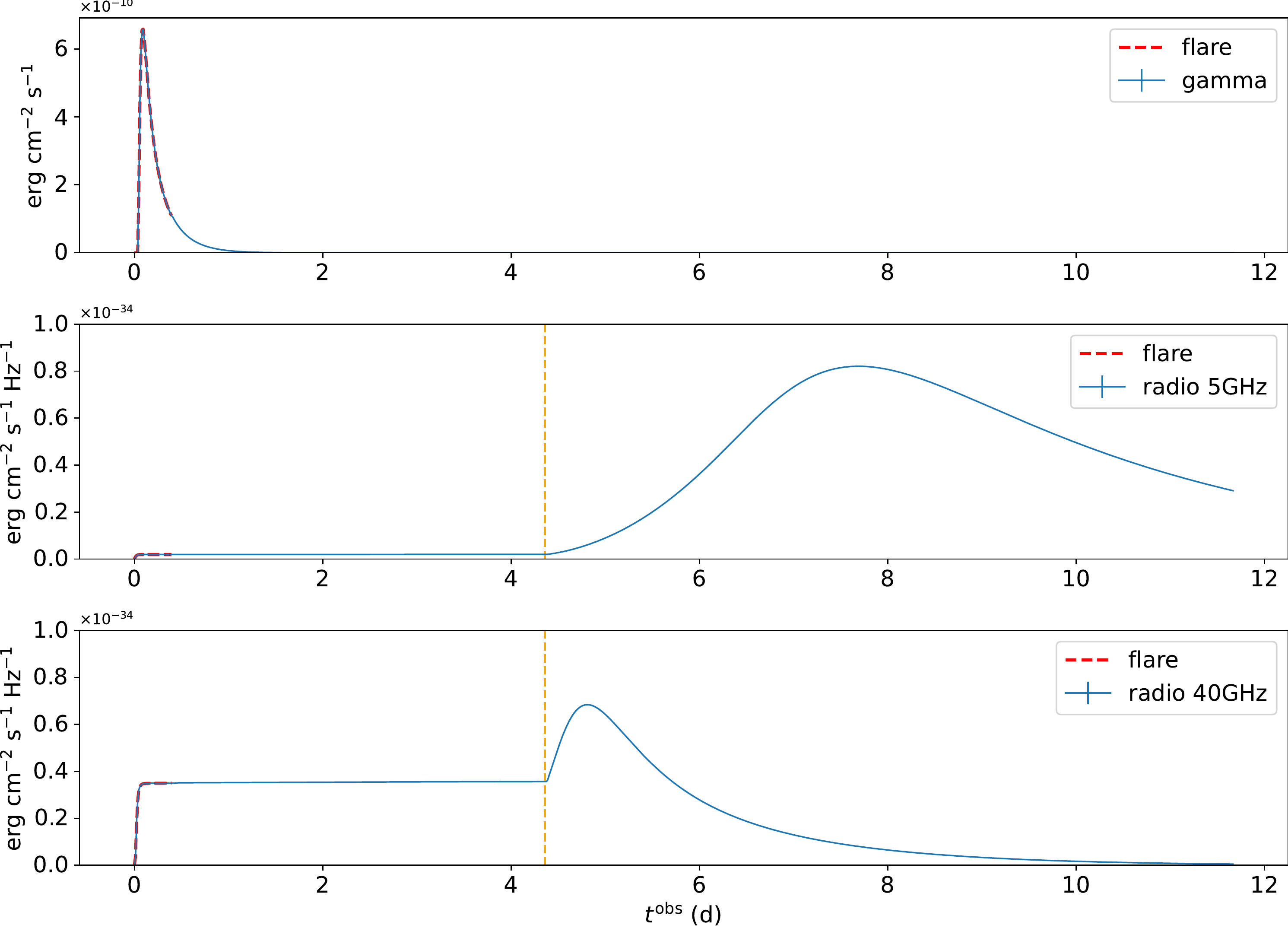}~ & 
\includegraphics[width=0.5\textwidth,,angle=-0]{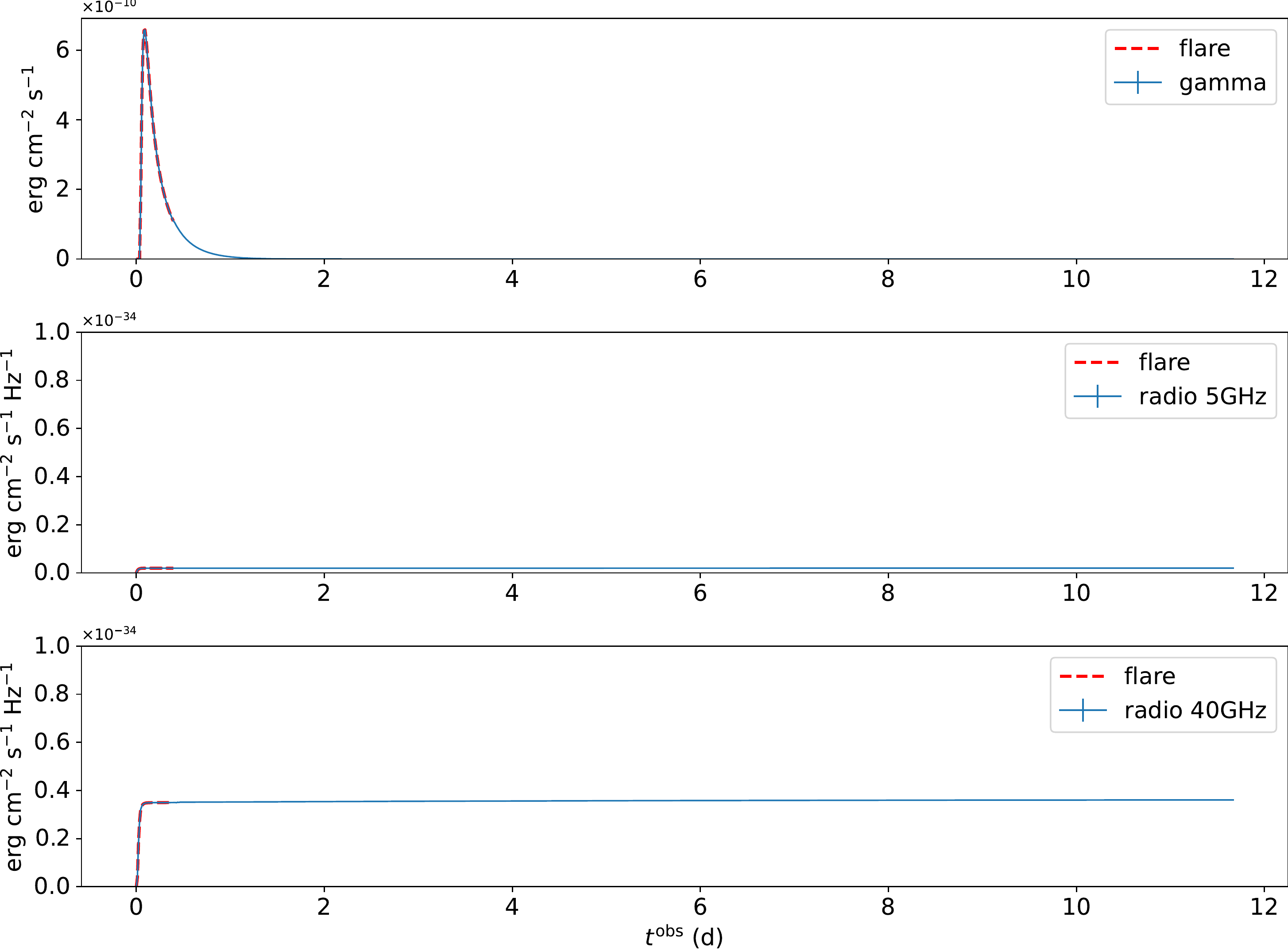}\\
\end{tabular}
\caption{Comparison of non-expanding ({\it right panels}) vs expanding ({\it left panels})
for $\beta_{\rm exp}=0.1$. The {\it top panels} show the evolution of the SEDs after the flaring stage, where the  blue 
colour indicates to the non/pre-expansion case, and orange indicates the expansion.
The second row of panels shows the evolution of the  flux density ($F_{\nu}$). 
The  three {\it bottom panels} show the merged light curves of both the flaring and the long-term simulation 
in the Fermi-LAT band, and at 5 and 40 GHz. The red dashed lines mark the light-curve segment belonging to the flaring stage and the orange vertical dashed lines mark the beginning of the expansion.}
\label{fig:exp_vs_no_exp}
\end{figure*}

\subsection{General setup of the simulation, assumptions, and limitations}
\label{sect:sim_setup}
We set up our simulation in order to reproduce an initial flaring episode, and a following expansion process within a leptonic synchrotron self-Compton (SSC) scenario. During the flare, particles  are injected and accelerated in the acceleration region (AR) where they undergo both cooling and acceleration processes, and diffuse toward the radiative region (RR), where only losses take place. After a time $t_{\rm exp}$ (measured in the blob frame), the expansion process takes place in the RR region. We follow the long-term evolution under the effects of radiative cooling and adiabatic expansion, setting the duration of the simulation  to be long enough  to follow the particle evolution due to the expansion process. A schematic representation of these processes is shown in Figure \ref{fig:jet_scheme}.

In the current approach, the values of the magnetic field ($B$) and the radius ($R$) in the RR during the flaring episode are taken from the typical values derived from MW modelling for HBLs, and these values coincide with the initial values at the beginning of the expansion ($B_0$ and $R_0$). Hence, we only extrapolate the evolution of $B$ according to $m_B$ and $R(t)$ from  the beginning of the expansion process. We adopte this approximation for the current approach because we are mostly interested in the determination of the radio-$\gamma$ response in terms of delay and expansion velocity, and are not interested in investigating the jet structure before the flaring site. Nevertheless, our model  can be easily generalised to a generic conical  jet geometry simply by replacing the temporal law $R(t)$ in order to follow the jet cross-section as a function of the jet opening angle and of the distance from the BH, setting a scaling parameter $z(t) = R_H(t)/R_{H0}$, and then expressing $ R(t)  = R_0 z(t)^{m_R}$, and $ B(t)  = B_0 z(t)^{-m_B m_R}$, where the expansion index of the jet $m_R$ is assumed to be $\in [0,1]$. In the ballistic case  \citep[$m_R=1$, ][]{Kaiser2006} the initial opening angle of the jet will be given by $\tan{\theta_0}=R_0/R_{H0}$, and will change with $z$ according to $\tan{(\theta(z))} =\tan{(\theta_0)}(R_H(t)/R_{H0})^{m_R-1}$, i.e. will be constant.

Both for the flaring and long-term (expansion) simulations, the time grid for the solution of the FP equation is tuned to have a temporal mesh  at least two orders of magnitude smaller than the shortest cooling and 
acceleration timescale.  We use an energy grid with 1500 points and $1 \leq \gamma \leq 10^{8}$. As the total number of time steps used in the FP numerical solution  ($T_{\rm size}$) can be very large,  
a subsample of the time steps of the simulation ($NUM_{SET}$) is stored in arrays, and can be used to build both light curves and SEDs. In the current simulation,  we use $NUM_{SET}=200$ for the flaring stage and $NUM_{SET} \in [1000,5000]$  for the long-term evolution,  depending on the duration of the simulation.
This guarantees an adequate time  sampling for light curves and spectral evolution.
SEDs are computed from the stored electron distributions, and from the blob parameters (according 
to their temporal evolution). In our case, the blob variable parameters are
the source radius ($R$) and magnetic field ($B$), which evolve according
to Equations \ref{eq:R(t)} and \ref{eq:B(t)}, respectively.
Light curves are obtained by integrating SEDs between two frequencies, or as monochromatic.
The code offers the possibility to convolve the light curves with the light-crossing time. In the present analysis, we skip this option because, as shown in section \ref{sect:phenom}, the light-crossing time is always shorter than the other competing timescales. This approximation used in the current approach will be removed in a forthcoming paper, where it will be treated accurately.
We also decided to use a constant bulk Lorentz factor. We tested and verified that, for the current scope of the simulations, the difference between enabling and disabling the IC cooling is negligible, and therefore to  speed up the computational time we use only synchrotron cooling for the radiative terms.

\subsection{Flare simulation}
\label{sect:flare_sim}
To generate  the flaring event, we use the {\ttfamily JetTimeEvol} configuration with 
a separated  acceleration and radiative region. With this configuration, particles are injected into the
acceleration region (AR), and then diffused toward the radiative region (RR) for
a timescale corresponding to the flare duration. We set the parameters for the flaring stage in order 
to reproduce the typical SED of HBLs, according to \cite{Tramacere2011}. We assume that both radiative  
and  first and second-order acceleration processes, occur in the AR, whilst in the RR region, we only 
take cooling processes into account. Particles are injected in the AR  with a quasi-monoenergetic distribution, normalised according to Equation \ref{eq:q_inj}. This initial distribution evolves under the effect of radiative and accelerative mechanisms, leading to the formation of a distribution with a low-energy power-law branch that bends close to the equilibrium energy. 
The high-energy branch exhibits a log-parabolic shape during the acceleration-dominated stage, and approaches a relativistic Maxwellian cut-off at the
equilibrium. The spectral index of the low-energy power law is dictated by 
the ratio of the first-order acceleration timescale to the escape time 
from the acceleration region, whilst the curvature during the acceleration-dominated 
stage is dictated by the momentum diffusion term. The acceleration region is modelled as a cylindrical shell
with a radius equal to the radiative region, and we assume a ten times
smaller width. 
Particles leaving the  acceleration region (shock front) enter the radiative region with a rate 
derived for the escape probability $P_{\rm escape}(\Delta t_{mesh})= 1 - \exp^{\Delta t_{mesh}/T_{esc}}$ \citep{Park1996},
where $\Delta t_{mesh}$ is the temporal mesh for the numerical solution of the FP equation.
The radiative region is modelled with a spherical geometry, where only the cooling processes are active
and where we assume that particles are confined $(T_{esc}\gg {\rm Duration})$.
The position along the jet of the flaring region  is placed at $R_{H0}=10^{17}$ cm. 
Particles are injected and accelerated in the AR for a  duration equal to {\it Duration acc.} and equal to {\it Duration inj.}, respectively. The total time-span of the flare simulation is given by the parameter {\it Duration.} The parameters for the acceleration and radiative region are reported in Table \ref{tab:flare_sim_par}.

\subsection{Long-term simulation of the expanding radiative region}
\label{sect:logn_term_sim}
The long-term simulation is an extension of the flaring event over a longer timescale, only for
the RR, and without injection or particle escape. The duration of the simulation for the long-term evolution is estimated according to $T_{\rm long} = \Delta t_{\nu_{\rm SSA}(0)\rightarrow \nu_{\rm SSA}(t)}+ 10\ t_{\rm decay}$. We set the expansion time $t_{\rm exp}=10^7$ s, and we evaluate ten realizations of the process, with $\beta_{\rm exp}$ evaluated on a ten-point logarithmic grid [0.001, 0.3]  to evaluate the trends as a function of $\beta_{\rm exp}$. We realise a further simulation  with $\beta_{\rm exp}=0.1$ to investigate the trends as a function of the radio frequency. The parameters for the acceleration and radiative region are reported in Table \ref{tab:exp_sim_par}.
{We stress that the initial position of the flaring region $R_{H0}$ can be changed without loss of generality, because  there is no dependence  of any relevant timescale on $R_{H0}$ in our model.}

\begin{figure}
   \includegraphics[width=0.45\textwidth,angle=-0]{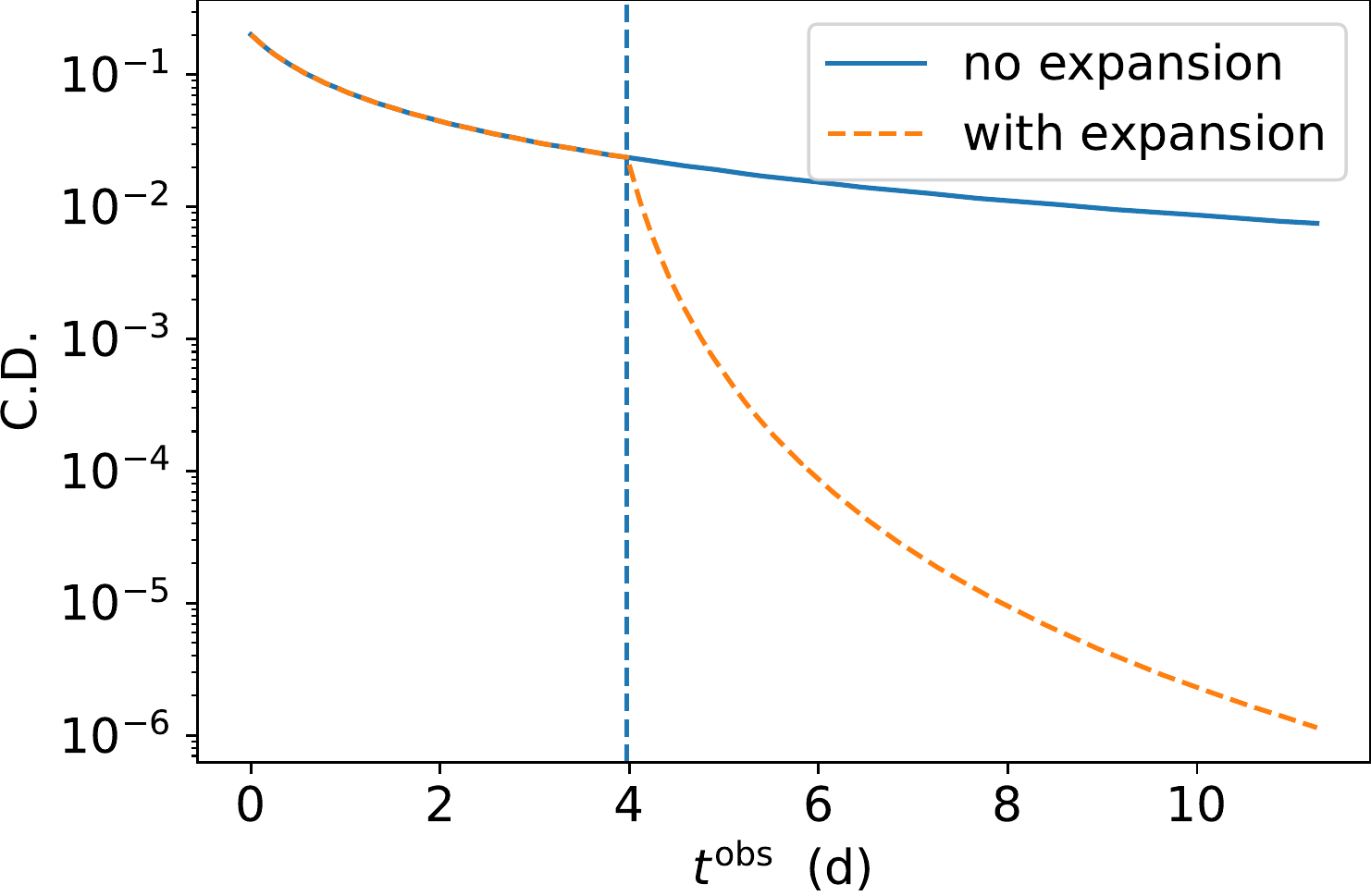}
   \caption{The CD versus the time of the simulation in the observer frame. The CD is evaluated as the ratio of the peak flux  of the IC component to the peak flux of the S component. The vertical dashed line marks the 
   beginning of the expansion (for the expanding case, orange line). It is clear that when the 
   adiabatic expansion begins, the IC starts to drop rapidly as a consequence of the larger 
   volume and lower seed photons density.}
   \label{fig:exp_vs_no_exp_CD}
   \end{figure}

\section{Comparison of non-expansion versus expansion}
\label{sect:exp_vs_no_exp}

In this section, we compare the different behaviours of  the spectral evolution when the
adiabatic expansion is active or switched off. For this purpose, we use the value of
$\beta_{\rm exp}=0.1$. The results are shown in Figure \ref{fig:exp_vs_no_exp}, where the plots in  
the left column refer to the expansion case and the plots in the  right column refer to the non-expansion
case. The top panels show the evolution of the SEDs, after the flaring stage, where the  blue 
flags mark the non-expanding and pre-expansion cases, and the orange flags mark the  case with expansion.
We note that, in the non-expanding simulation, the evolution follows the usual pattern
dictated by the radiative cooling timescales. On the contrary, in the expanding case, we notice 
that when the expansion starts, the patterns in both the synchrotron and IC components are 
different. The IC component is mainly affected by a significant drop in the Compton dominance
(CD). This can be better appreciated in Figure \ref{fig:exp_vs_no_exp_CD}, where we plot 
the CD versus the time of the simulation. The CD is evaluated as the ratio of the peak flux 
of the IC component to the peak flux of the S component. The vertical dashed line marks the 
beginning of the expansion (for the expanding case, orange line). It is clear that when the 
adiabatic expansion begins, the IC starts to drop rapidly, as a consequence of the larger 
volume and lower seed-photon density. This is a very interesting feature, and might already be 
visible during the flaring stage. The most integrating effect, for our analysis, is the evolution 
of the S component. On top of the flux decay dictated by the adiabatic losses, and decreased magnetic
field, we notice the shift in the SSA frequency, which is absent in the non-expanding case.
This effect can be better appreciated in the second row of panels in Figure \ref{fig:exp_vs_no_exp}, where we 
plot the evolution of the  flux density ($F_{\nu}$). Whilst in the non-expanding case the SSA is almost stable at the 
initial value of $\approx10^{11}$ Hz, in the expanding case the SSA decreases with 
time as predicted by Equation \ref{eq:nu_ssa_generic}. The actual trend will be investigated 
in detail in the following two sections. The  three bottom panels of Figure \ref{fig:exp_vs_no_exp}
show the light curves in the Fermi-LAT band, and at 5 and 40 GHz. We notice that, in the non-expanding
case, the temporal behaviour is again in agreement with a purely radiative cooling without particle 
escape. On the contrary, in the expanding case, we notice that for the radio light curves an increase
in the flux level happens after the beginning of the expansion, with the time of the maximum happening
earlier at larger frequencies. This can be quantitatively understood by looking at the flux density panel,
which shows that the SSA moves from the initial (non-expanding case) down to lower frequencies. We 
quantify these delays and the different rise and decay times in the following two sections.

\begin{figure}
\begin{tabular}{l}
\includegraphics[width=0.45\textwidth,angle=-0]{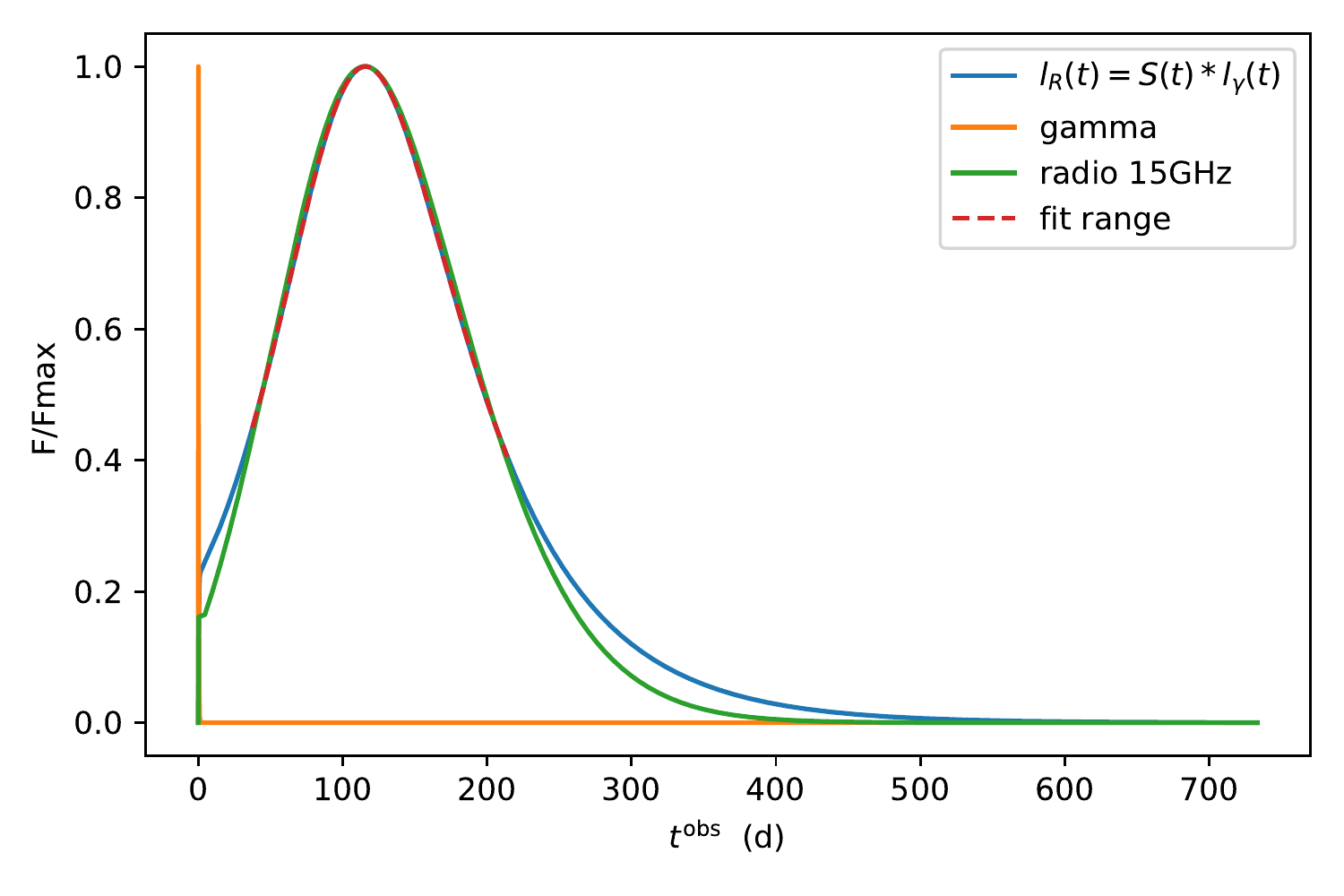}\\
\includegraphics[width=0.45\textwidth,angle=-0]{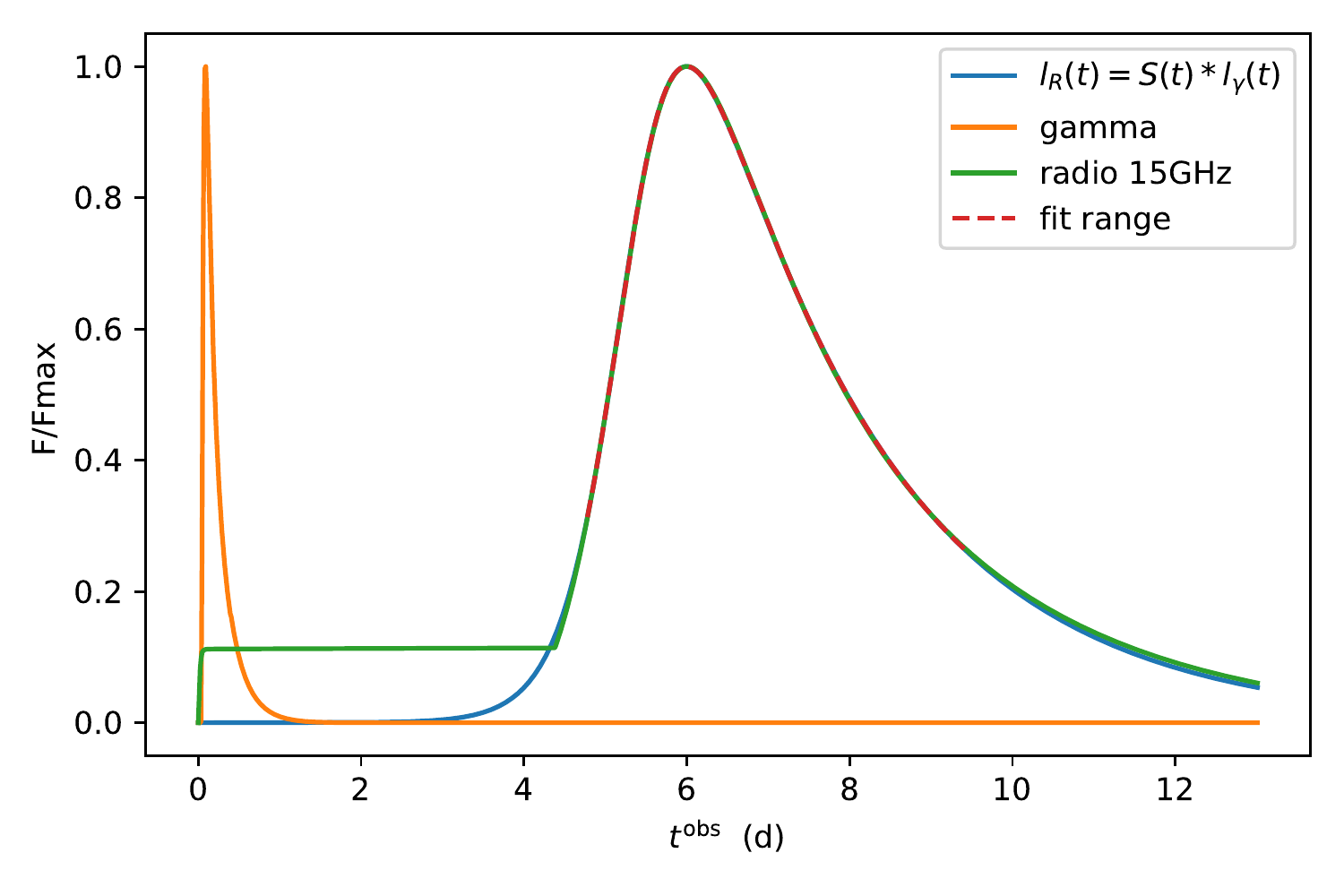}\\
\end{tabular}
\caption{Best fit for the radio-$\gamma$ response at 15 GHz for $\beta_{\rm exp}=0.001$ ({\it top panel}) 
and   at 15 GHz for $\beta_{\rm exp}=0.084$ ({\it bottom panel}), and $t_{\rm exp} =1\times 10^7$ s. All other parameters are the same as reported in Table \ref{tab:exp_sim_par}.  The light curves are in the observer frame. The red dashed line represents the actual fit interval, the orange line represents the simulated $\gamma$-ray light curve, the green line the simulated radio light curve, and the blue line is the best fit of the radio light curve obtained from the convolution of the  $\gamma$-ray light curve with the best-fit response.
The $\gamma$-ray light curve is the same in both panels, and appears different due to the different duration of the  simulations.}
\label{fig:conv_example}
\end{figure}

\section{Radio-$\gamma$ response and physical trends in the delay}
\label{sect:response}
In this section, we verify that the phenomenological trends derived in 
Section \ref{sect:phenom} are reproduced in our simulations and 
whether or not they can be  applied to observed data. The empirical determination of delays
is usually done by means of a discrete correlation function (DCF). A further and 
more interesting step is to determine a response function $S(t)$ such that radio
light curves ($l_R$) can be reproduced as a `response' of the $\gamma$-ray light curves ($l_{\gamma}$)  
as a convolution \citep{turler_1999A&A...349...45T,Sliusar2019A,Sliusar2019B} according to
\begin{equation}
\label{eq:S_conv}
l_R(t)=S(t)*l_{\gamma}(t),
\end{equation}
where the $S$ is an empirical function that depends on  parameters that can be related to the observable quantities investigated in Section \ref{sect:phenom}. 
We propose the following response function: 
\begin{equation}
\label{eq:Resp_func}
S(t)=A \frac{ \exp\frac{-(t-\Delta)}{t_{\rm f}}}{1+\exp\frac{-(t-\Delta  )}{t_{\rm f}}}
,\end{equation}
 where $t_{\rm f}$ is the decay time, and $t_{\rm f}$ is the rise time.

This is the combination of a logistic function for the rising part  and an exponential 
decay for the decaying part, with $A$ being a scaling factor.  The scaling factor depends mainly on the initial value of the Compton dominance, on the observed radio frequency, and on $m_B$. In the  present analysis, we are not investigating its impact. 
The peak of $S(t)$, corresponding to the radio-$\gamma$ delay,  reads
\begin{equation}
\label{eq:S_t_p}
\Delta t=\Delta-t_{\rm u}\ln\Big(\frac{t_{\rm u}}{t_{\rm f}-t_{\rm u}}\Big).   
\end{equation}
The actual determination of the rise and decay time is analytically complicated. We estimated $t_{\rm rise}$ and $t_{\rm decay}$ numerically by imposing the condition $S(t)=\frac{A}{2}$ for $t_{\rm rise}$ and $S(t)=\frac{A}{e}$ for $t_{\rm decay}$, and verified that within a maximum deviation of 5\% for $t_{\rm rise}$,  and of 0.2\% for $t_{\rm decay}$, these timescales  can be evaluated according to
\begin{eqnarray}
\label{eq:S_t_r_t_d}
t_{\rm rise} &=& t_u\Big(0.54+1.34 \Big(\frac{t_f}{t_u}\Big)^{1/4 } \Big) \\
t_{\rm decay}&=& t_f\Big(1.00+1.33 \Big(\frac{t_f}{t_u}\Big)^{-1.11} \Big) 
.\end{eqnarray}
For the propagation of the uncertainties we used the {\ttfamily Uncertainties}\footnote{Uncertainties: a Python package for calculations with uncertainties, Eric O. LEBIGOT,  \url{http://pythonhosted.org/uncertainties/}}  Python package.

\subsection{Validation of phenomenological relations}
\label{sect:validatio}
Before investigating the phenomenological trends, we validate the relations in  Section \ref{sect:phenom} using long-term simulations with ten different values of $\beta_{\rm exp}$ evaluated on a logarithmic grid  ranging $[0.001,0.3]$. We use two scenarios: one where we disable only the radiative cooling term  in the FP equation, and one with both radiative and adiabatic  cooling terms  enabled. As the phenomenological relations derived in Section \ref{sect:phenom} are valid  when the adiabatic cooling is dominant, the deviations in the trends with the radiative cooling enabled will highlight the effect of the competition between the synchrotron cooling and time the adiabatic time already discussed in   Section \ref{sect:phenom}.
In order to estimate the trends, we minimise the right-hand side of Equation \ref{eq:S_conv}
with respect to the left-hand side, where $l_{\gamma}$ and $l_{R}$ are the light curves produced in the simulations, leaving as free parameters $A$, $\Delta $, $t_{\rm u}$, and $t_{\rm f}$.
To  perform the analysis in the observer frame, we  express Equations \ref{eq:times_obs_t_R0} in terms of  $R^{\rm 0}_{\rm obs}=R_0\frac{1+z}{\delta}$ and of the observed radio frequencies: 

\begin{eqnarray}
\label{eq:fit_trends}
t^{\rm obs}_{\rm decay} &=&\frac{R_{\rm 0}^{\rm obs}}{m_B\beta_{\rm exp}c}\Big(\frac{\nu^{0,\rm obs}_{\rm SSA}}{\nu^{*,\rm obs}_{\rm SSA}}\Big)^{\phi} \\  
t^{\rm obs}_{\rm rise} &=& \frac{1}{2}t^{\rm obs}_{\rm peak}= \left \{
\begin{aligned}
& \frac{1}{2} \frac{R_{\rm 0}^{\rm obs}}{ \beta_{\rm exp}c}  \Big[ \Big( \frac{\nu^{0,\rm obs}_{\rm SSA}}{\nu^{*,\rm obs}_{\rm SSA}}\Big)^{\phi}  - 1 \Big]\nonumber && \text{if}\ \nu^{0,\rm obs}_{\rm SSA}>\nu^{*,\rm obs}_{\rm SSA} \\
&0 && \text{otherwise}
\end{aligned} \right. \\
\Delta t^{\rm obs}&=&t^{\rm obs}_{\rm exp}+t^{\rm obs}_{\rm peak}= t^{\rm obs}_{\rm exp}+ \frac{R_{\rm 0}^{\rm obs}}{  \beta_{\rm exp}c}  \Big[ \Big( \frac{\nu^{0,\rm obs}_{\rm SSA}}{\nu^{*,\rm obs}_{\rm SSA}}\Big)^{\phi}  - 1 \Big].\nonumber
\end{eqnarray}

\begin{figure}[!h]
\begin{tabular}{c}
\includegraphics[width=0.45\textwidth,angle=-0]{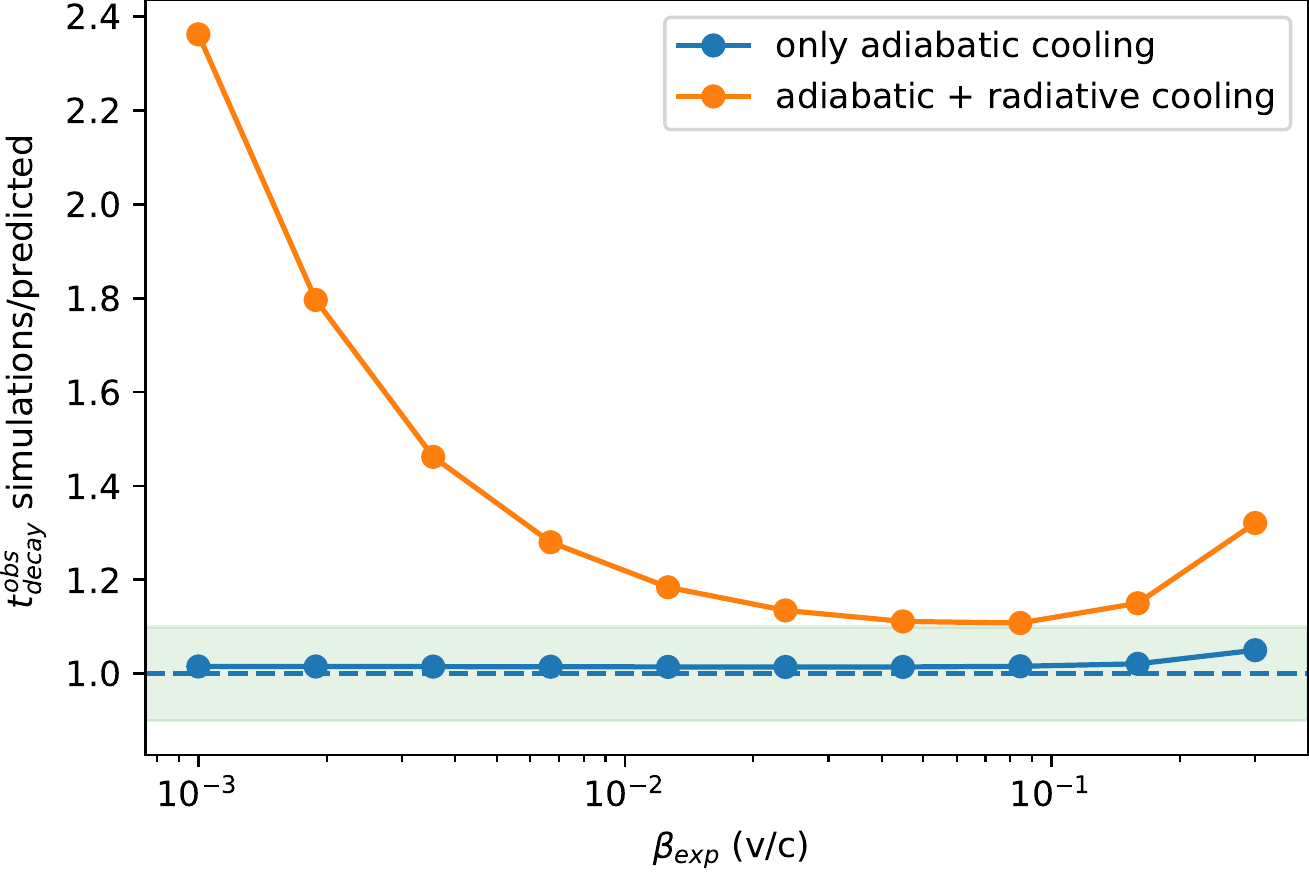}\\
\includegraphics[width=0.45\textwidth,angle=-0]{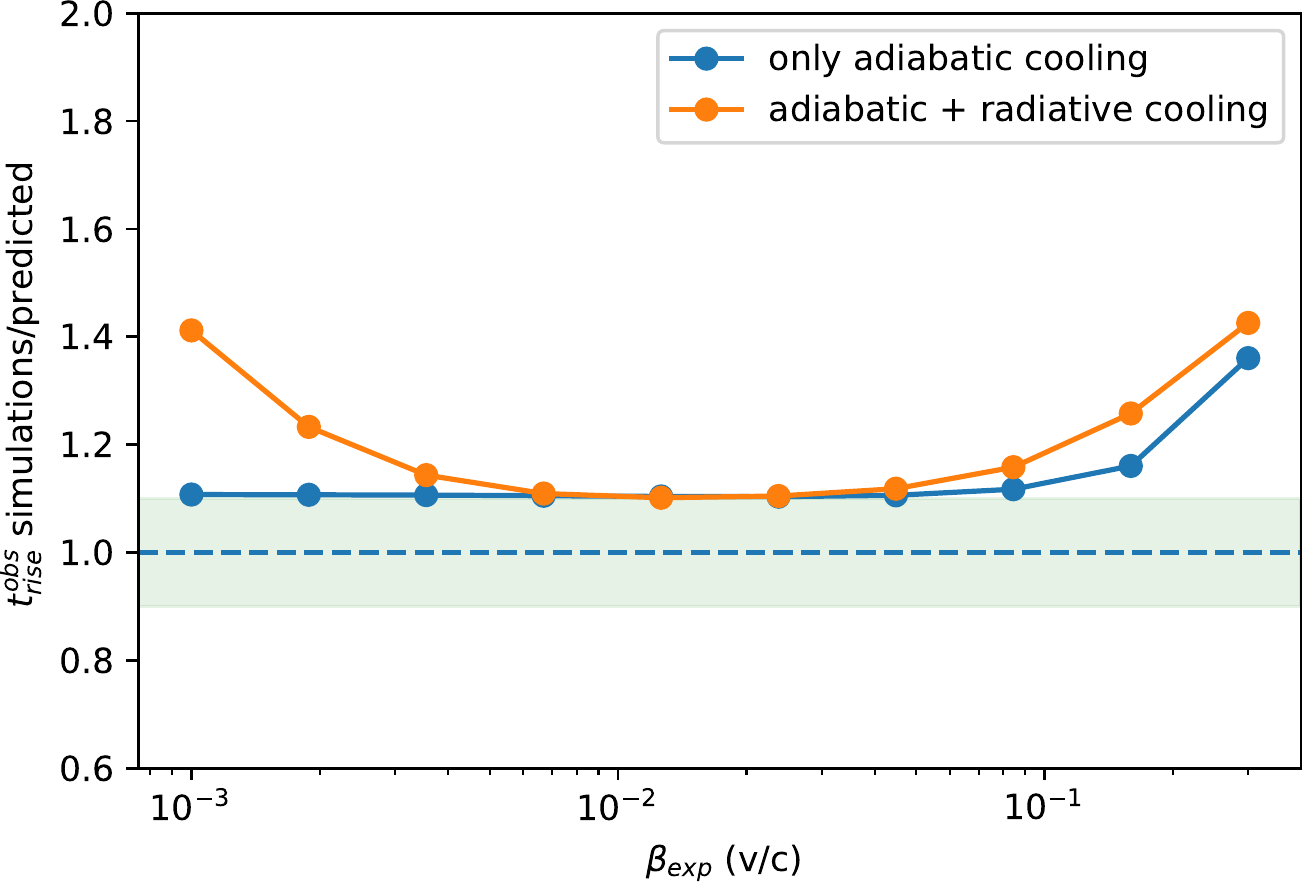}\\
\includegraphics[width=0.45\textwidth,angle=-0]{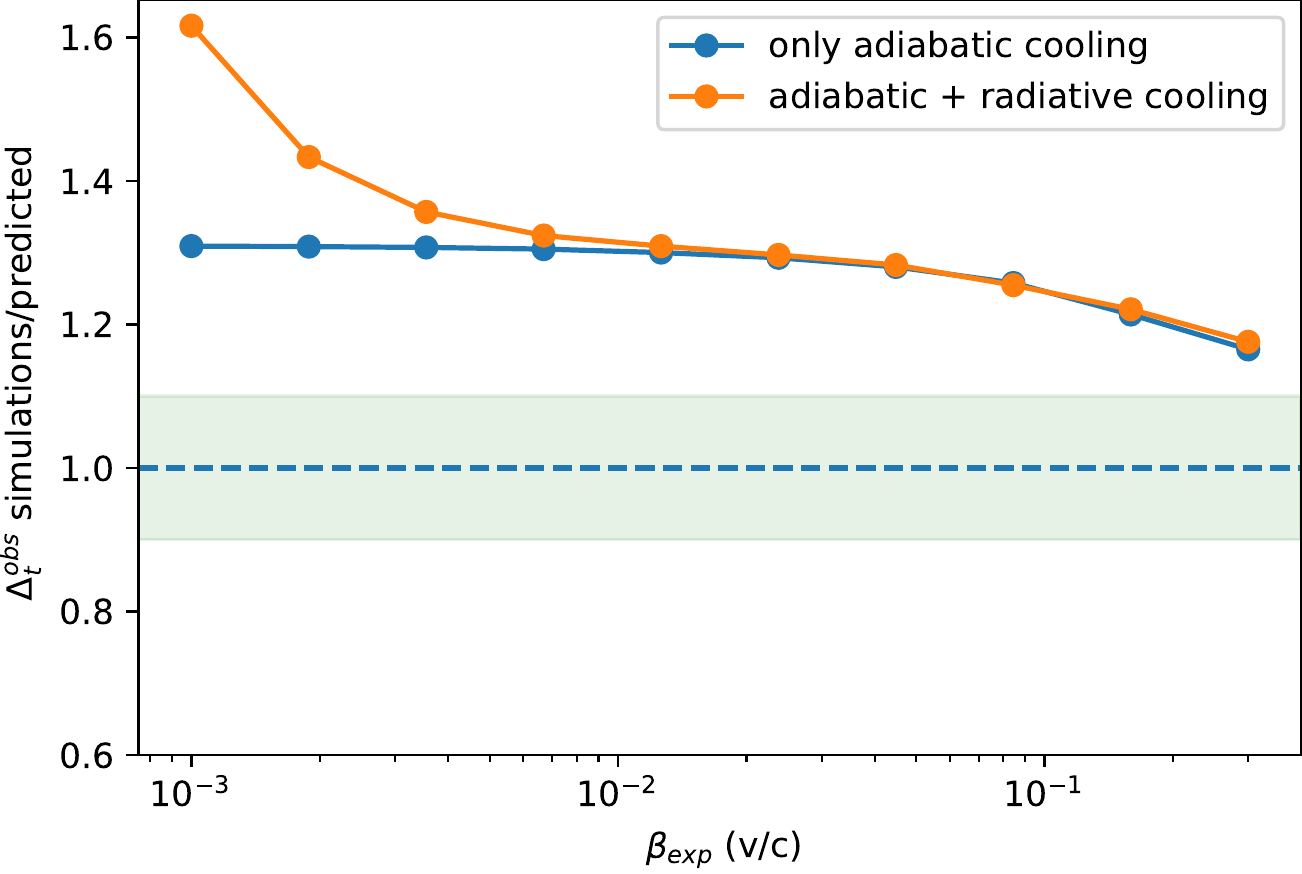}
\end{tabular}
\caption{
   Ratio of the timescales predicted by Equation \ref{eq:fit_trends} to  the results obtained by  the best fit of the radio-$\gamma$ response applied to the  numerical simulations. All the other parameters are the same as reported in Table \ref{tab:exp_sim_par}. The blue lines correspond to the case of only adiabatic cooling, and the orange lines to the case of radiative plus adiabatic cooling. The green shaded areas represent 
   the $\pm 10\%$  region with respect to the prediction from  Equation \ref{eq:fit_trends}, and the dashed horizontal lines indicate unity.
\textit{Top panel}: Radio$-\gamma$ delay. 
\textit{Middle panel}: Decay time. \textit{Bottom panel}: Rise time.
}
\label{fig:validation}
\end{figure}

\begin{figure}
\begin{tabular}{c}
\includegraphics[width=0.45\textwidth,angle=-0]{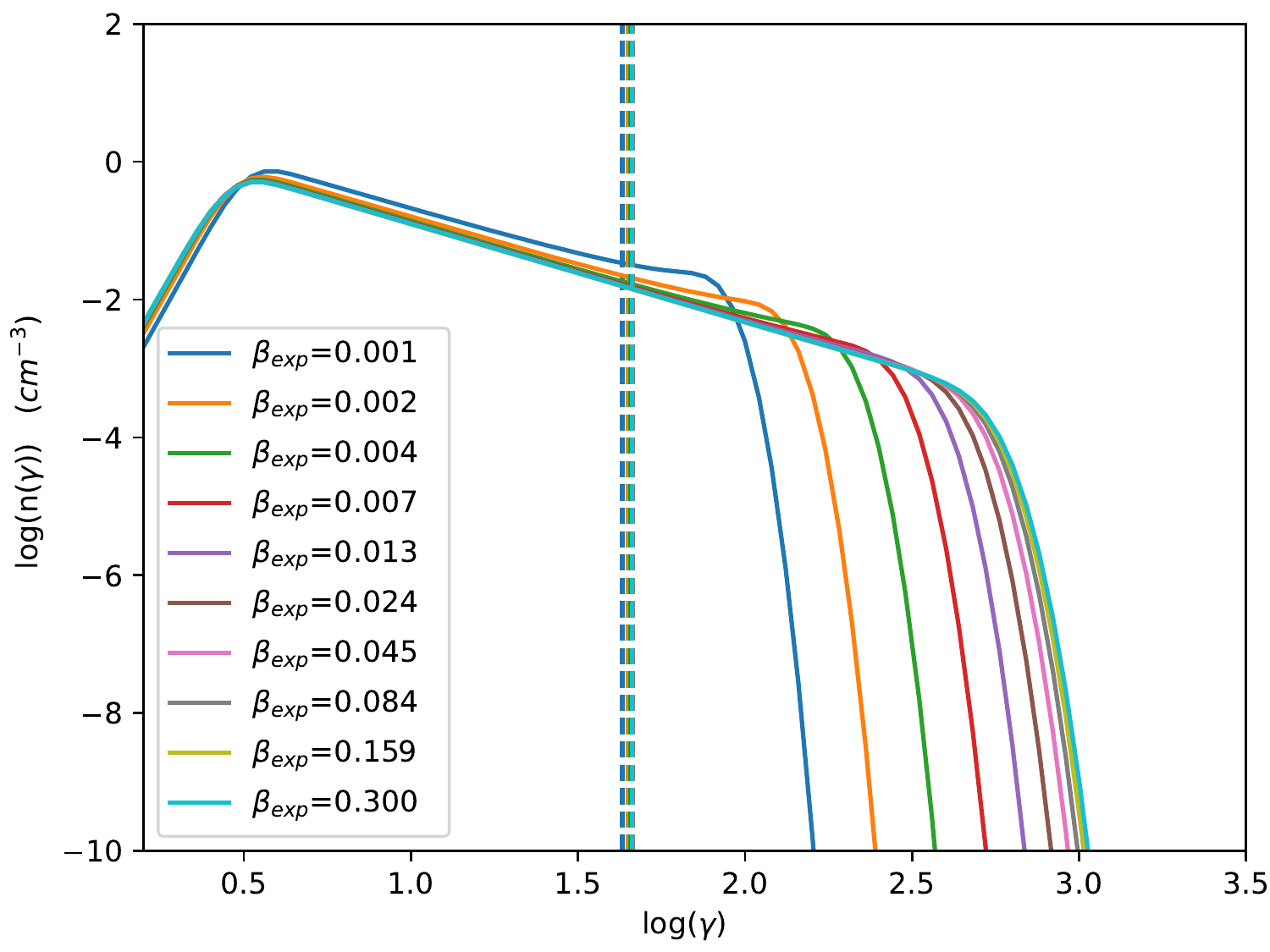}\\
\end{tabular}
\caption{  
State of the electron distributions at the time corresponding to the peak of the $\nu_{\rm obs}=15$ GHz light curves for the expansion simulations with  both radiative and adiabatic cooling enabled, and $\beta_{\rm exp}$ ranging $[0.001,0.3]$. The vertical dashed lines correspond to the Lorentz factor of the electrons most contributing to the observed 15 GHz frequency. All the other parameters are the same as reported in Table \ref{tab:exp_sim_par}}
\label{fig:emitters_evol}
\end{figure}

\begin{figure}
\begin{tabular}{c}
\includegraphics[width=0.45\textwidth,angle=-0]{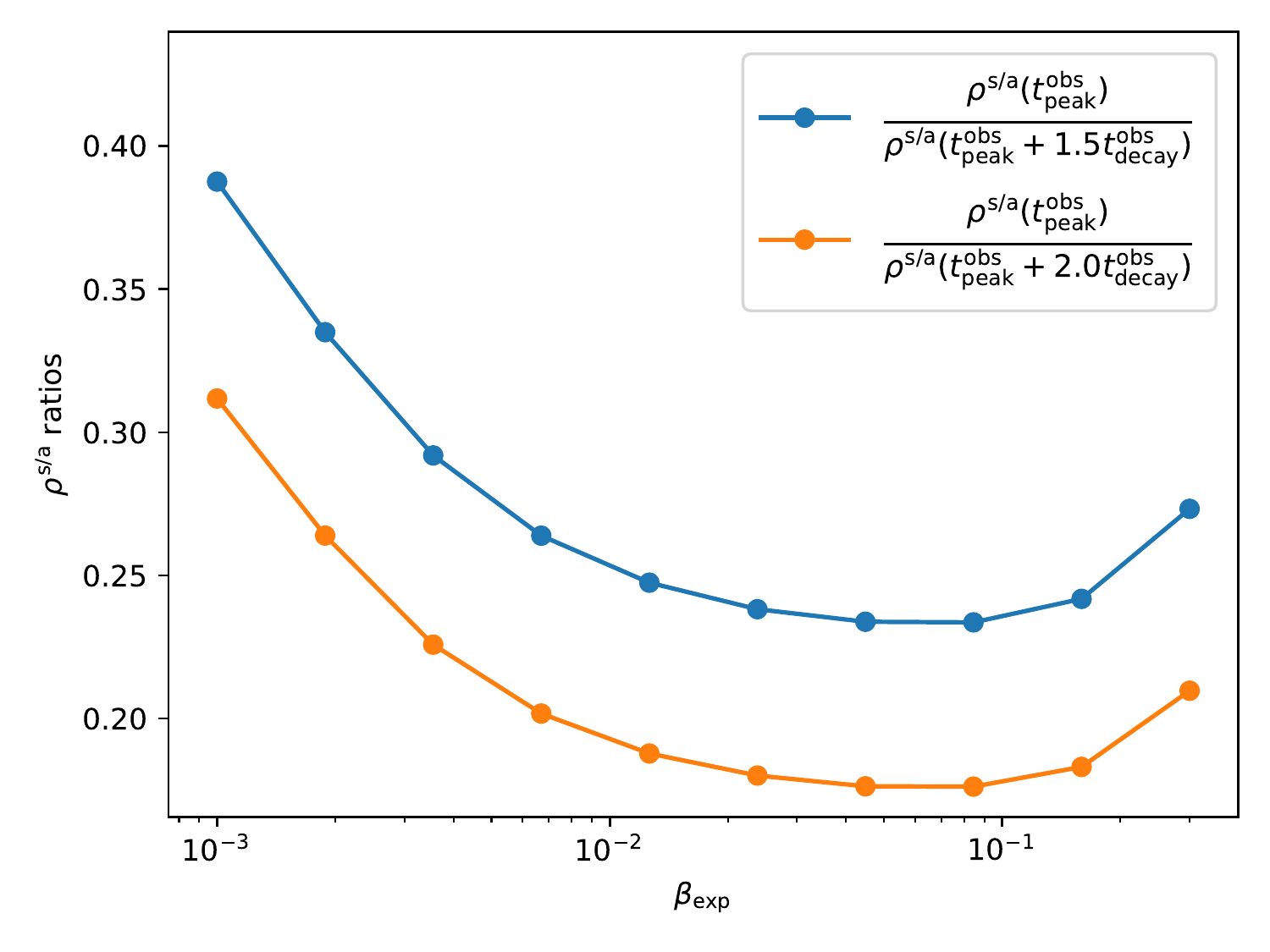}
\end{tabular}
\caption{  
Ratio of the synchrotron to adiabatic cooling timescales $\rho^{\rm s/a}(t)=t_{\rm sync}(t)/t_{\rm ad}(t)$ for the Lorentz factor of the electrons most contributing to the observed 15 GHz frequency, evaluated at $t=t^{\rm obs}_{\rm peak}$ and $t=t^{\rm obs}_{\rm peak} + 1.5 t^{\rm obs}_{\rm decay}$ (blue line), and evaluated at  $t=t^{\rm obs}_{\rm peak}$ and $t=t^{\rm obs}_{\rm peak} + 2.0 t^{\rm obs}_{\rm decay}$ (orange line). The same values  of $\beta_{\rm exp}$ are used as reported in Figure \ref{fig:emitters_evol}.
}
\label{fig:cooling_ratios}
\end{figure}
\begin{figure}[!h]
   \begin{tabular}{c}
   \includegraphics[width=0.45\textwidth,angle=-0]{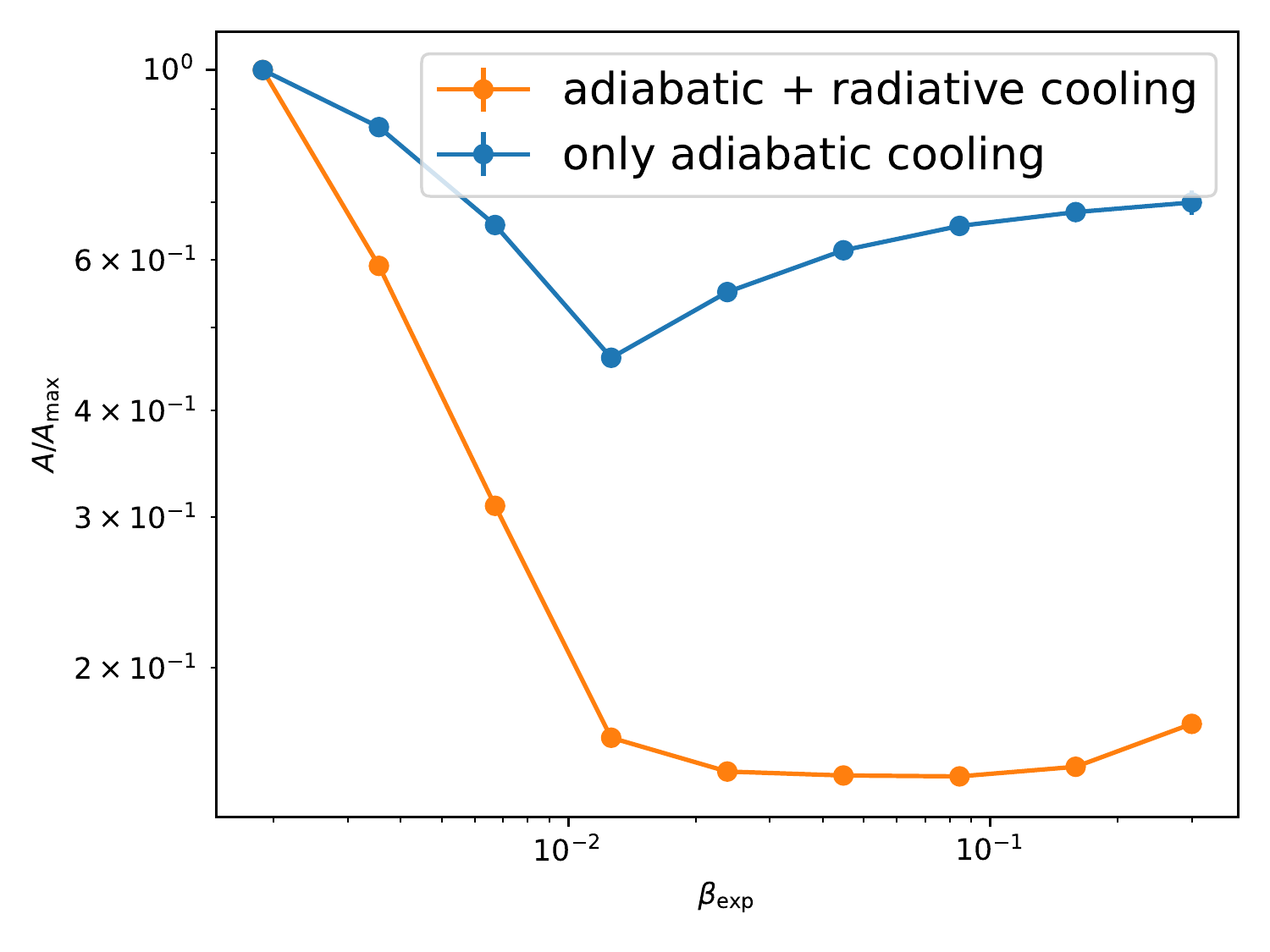}\\
   \includegraphics[width=0.45\textwidth,angle=-0]{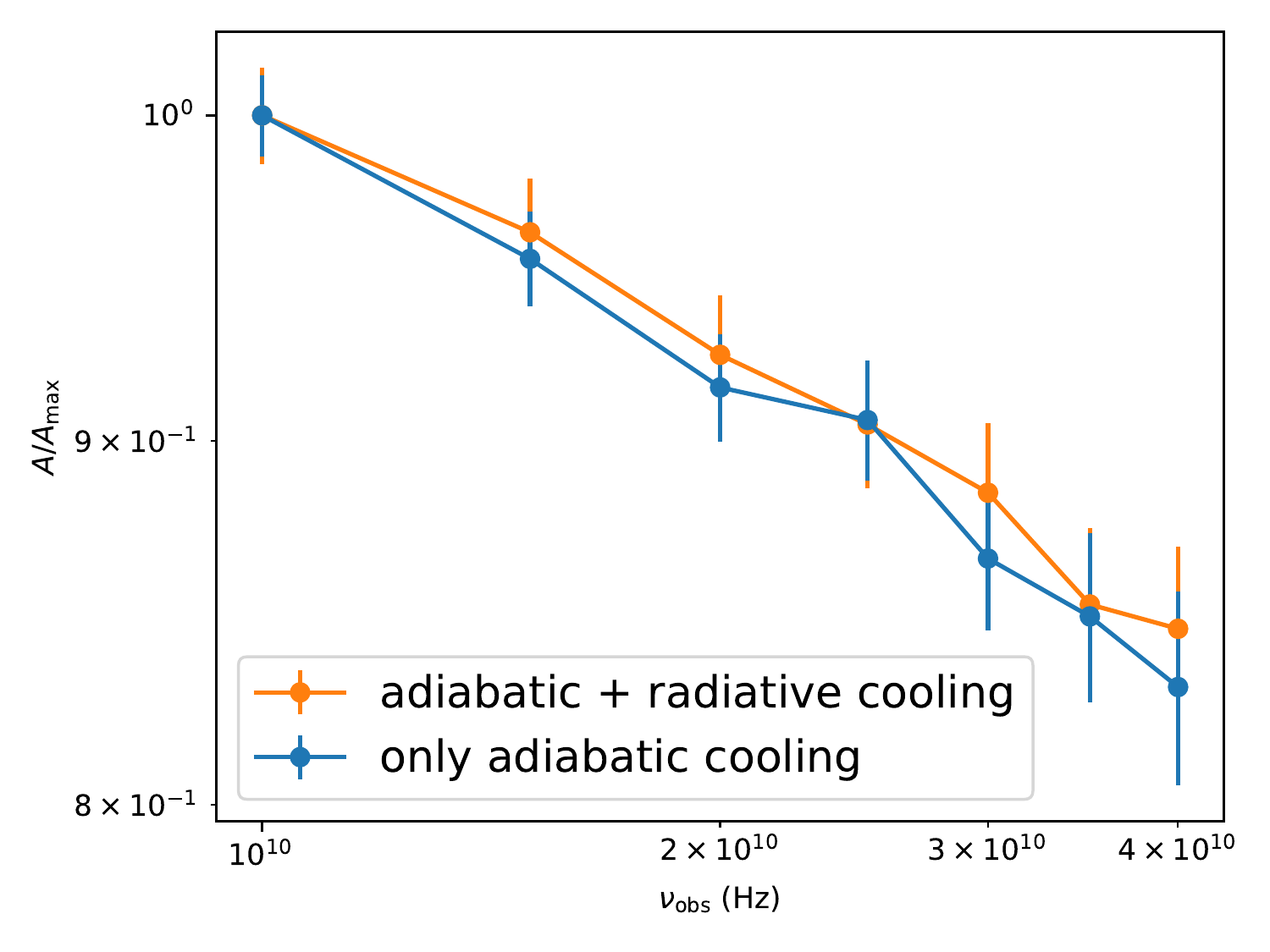}
   \end{tabular}
   \caption{  
   \textit{Top panel:}  Trends for the response amplitude ($A$) normalised to its maximum value for  $\beta_{\rm exp}\in[2\times10^{-3},0.3]$ and $\nu_{\rm obs} =15$ GHz. The blue lines correspond to the case of only adiabatic cooling, and the orange lines to the case of radiative plus adiabatic cooling. \textit{Bottom panel:} Same as in top panel but  for $\nu_{\rm obs} \in [10,40]$ GHz and $\beta_{\rm exp}=0.1$. All the other parameters are the same as reported in Table \ref{tab:exp_sim_par}.
   }
\label{fig:A_trend}
\end{figure}

As we show later, as the value of the electron index $p$  evolves with time,   the use  of $\psi$ as a function of a constant $p$  during the fit is  inappropriate. Hence, we use a generic index $\phi$  that is not related directly to  $p$ during the fit but still preserves the behaviour of the trends. In any case, we can still estimate the value of $p$ from the best-fit parameters using the second Equation \ref{eq:R_transp}.

An example of best-fit convolution is reported in Figure \ref{fig:conv_example}, where we show the results of 
the best fit for the response at 15 GHz for the case with the cooling terms active, for $\beta_{\rm exp}=0.084$ (top panel) and $\beta_{\rm exp}=0.001$ (bottom panel), and $t_{\rm exp} =1\times 10^7$ s;  all the other parameters are the same as reported in Table \ref{tab:exp_sim_par}.
The light curves are in the observer frame. 

The results of the validation of Equations \ref{eq:fit_trends} are summarised in Figure \ref{fig:validation}, where we show the ratio of the timescales predicted by Equations \ref{eq:fit_trends} to the actual results obtained by the best fit of the radio-$\gamma$ response applied to the  numerical simulations. The blue lines correspond to the case of only adiabatic cooling, and the orange lines to the case of only adiabatic  plus radiative cooling. The green shaded area corresponds to the $\pm 10\%$ region with respect to the prediction from  Equation \ref{eq:fit_trends}. For the decay time, we note that in the case of only adiabatic cooling, the trends in Equation \ref{eq:fit_trends} are valid within a maximum derivation of time $\lesssim 1\%$ for the delay time. This is consistent with our expectations, because Equation \ref{eq:fit_trends} takes into account only the contribution from adiabatic cooling and from flux variations related to the geometrical expansion. When the radiative cooling is also enabled, the deviations are larger (by up to a factor of 2), with a trend that decreases for larger values of  $\beta_{\rm exp}$. This trend in the deviations is due to the different interplay between radiative and adiabatic cooling timescales for different expansion velocities, which we investigate in more detail  below. For the rise time we observe a deviation  by up to $\approx 40\%$ for the cases of only adiabatic cooling and radiative plus adiabatic cooling.
For the delay time, the deviations are $\approx 20\%$ to  $\approx 30\%$ for the case of only adiabatic cooling, and $\approx 20\%$ to  $\approx 160\%$ for the case of radiative plus adiabatic cooling. The variations in the decay are larger than those observed in the rise time, because the peak of the response is affected both by decay and rise times (see Equation \ref{eq:S_t_p}).
The validation shows that the phenomenological trends predict the timescales of the response when the adiabatic cooling is dominant  with good accuracy, and, as anticipated in Section \ref{sect:beta_exp_trends}, can be biased  by  the competition between the  radiative cooling time and adiabatic cooling time. A different balance between adiabatic and radiative cooling will cause not only deviation in the adiabatic-dominated trends but also  significant changes in the   electron energy distribution at the radio peak time for different values of $\beta_{\rm exp}$.
For the purpose of illustration, we show in Figure \ref{fig:emitters_evol} the different states of the electron distributions at the time corresponding to the peak of the $\nu_{\rm obs}=15$ GHz light curves, for  the values of $\beta_{\rm exp}$ ranging $[0.001,0.3]$ used in our simulations. We note that, for smaller values of $\beta_{\rm exp}$, the expansion process lasts longer, and the slower  temporal decrease of the magnetic field produces a steeper distribution with a lower cut-off of the electron distributions. On the contrary, for larger values of $\beta_{\rm exp}$, the shorter duration and the faster decrease in $B$   results in a higher cut-off of the electron distribution.  In Figure \ref{fig:emitters_evol}, the dashed vertical lines represent the values of the electron Lorentz factor most contributing to the emission at $\nu_{\rm obs}=15$ GHz, $\gamma_t \simeq [3\nu_{\rm SSA}/(4\nu_{L})]^{1/2}$ \citep{Ghisellini2013b}, for the different values of $\beta_{\rm exp}$. 
Clearly, the corresponding value of $p$ is different for different states of the evolution, meaning that the use of $\psi$ as a function of a constant $p$  is inappropriate. For these reasons we use a generic index $\phi$ that is not related directly to $p$, but still preserves the behaviour of the trends.

A further effect due to the complex interplay between the cooling timescales is shown in   Figure  \ref{fig:cooling_ratios}, where we plot the ratio of the synchrotron to adiabatic cooling timescales $\rho^{\rm s/a}(t)=t_{\rm sync}(t)/t_{\rm ad}(t)$, for the Lorentz factor of the electrons most contributing to the observed 15 GHz frequency, evaluated at $t=t^{\rm obs}_{\rm peak}$ and $t=t^{\rm obs}_{\rm peak} + 1.5 t^{\rm obs}_{\rm decay}$ (blue line), and evaluated at  $t=t^{\rm obs}_{\rm peak}$ and $t=t^{\rm obs}_{\rm peak} + 2.0 t^{\rm obs}_{\rm decay}$ (orange line). This complex interplay between the cooling times can produce deviations in the  $t_{\rm decay}$ trend. We investigate these deviations in Section \ref{sect:beta_exp_trends}.

As a final comment, we investigate   the impact of the complex interplay between the cooling timescales on the response function amplitude $A$. Even if in the current analysis we are not investigating the dependence of $A$ on the physical parameters, it is worth showing the impact of the cooling times. In the top panel of Figure \ref{fig:A_trend}, we plot the trends for the response amplitude normalised to its maximum value, for  $\beta_{\rm exp}\in[2\times10^{-3},0.3]$ and $\nu_{\rm obs} =15$ GHz.  The blue lines correspond to the case of only adiabatic cooling, and the orange lines to the case of radiative plus adiabatic cooling. In the bottom panel of the same figure, we  plot the trends for $\nu_{\rm obs} \in [10,40]$ GHz and  $\beta_{\rm exp}=0.1$. All the other parameters are the same as reported in Table \ref{tab:exp_sim_par}. We notice that changes in $\beta_{\rm exp}$ introduce variations of up to one order of magnitude for the case of radiative plus adiabatic cooling, and  up to $\approx 40\%$ for the case of only adiabatic cooling. For the case of $\nu_{\rm obs}$ trends, the variation in the response is up to $\approx 20\%$, both for only adiabatic and adiabatic plus radiative cooling. Therefore,  the response amplitude is also affected mostly by the  balance between radiative and adiabatic cooling, which depends strongly on the expansion velocity. 

   \begin{figure*}
   \centering
   \begin{tabular}{lr}
   \includegraphics[width=0.5\textwidth,angle=-0]{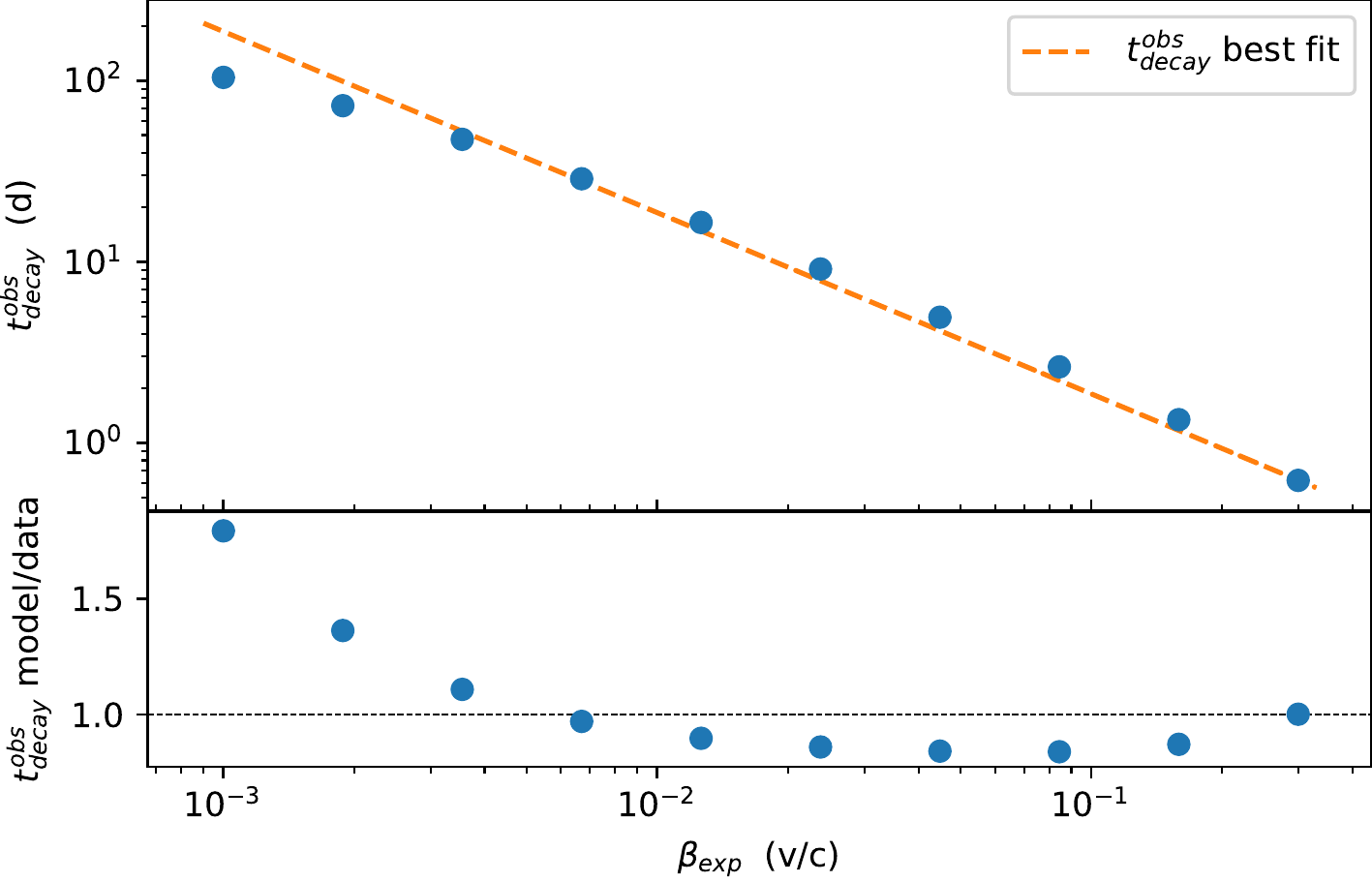}&
   \includegraphics[width=0.5\textwidth,angle=-0]{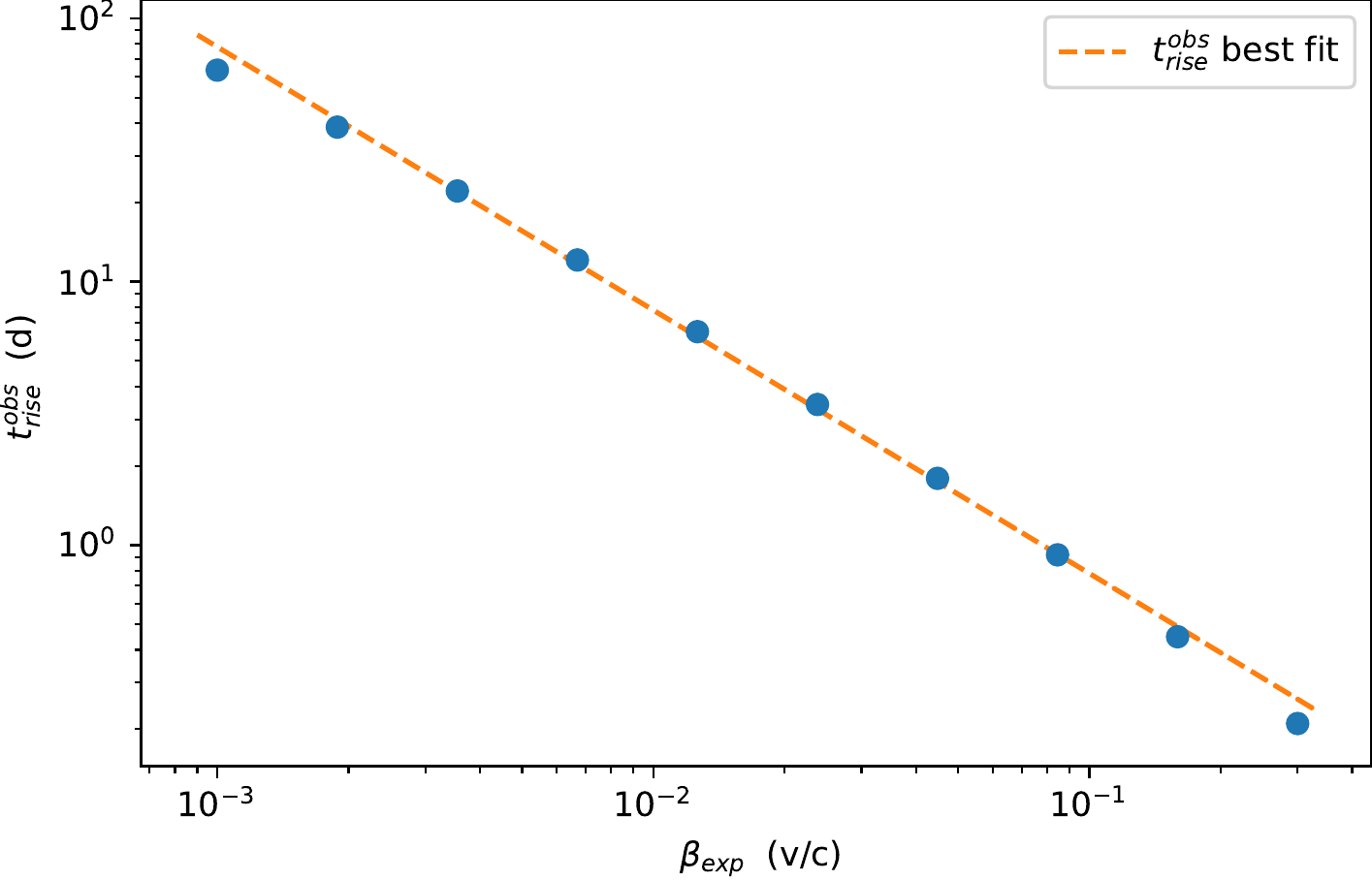}\\
   \includegraphics[width=0.5\textwidth,angle=-0]{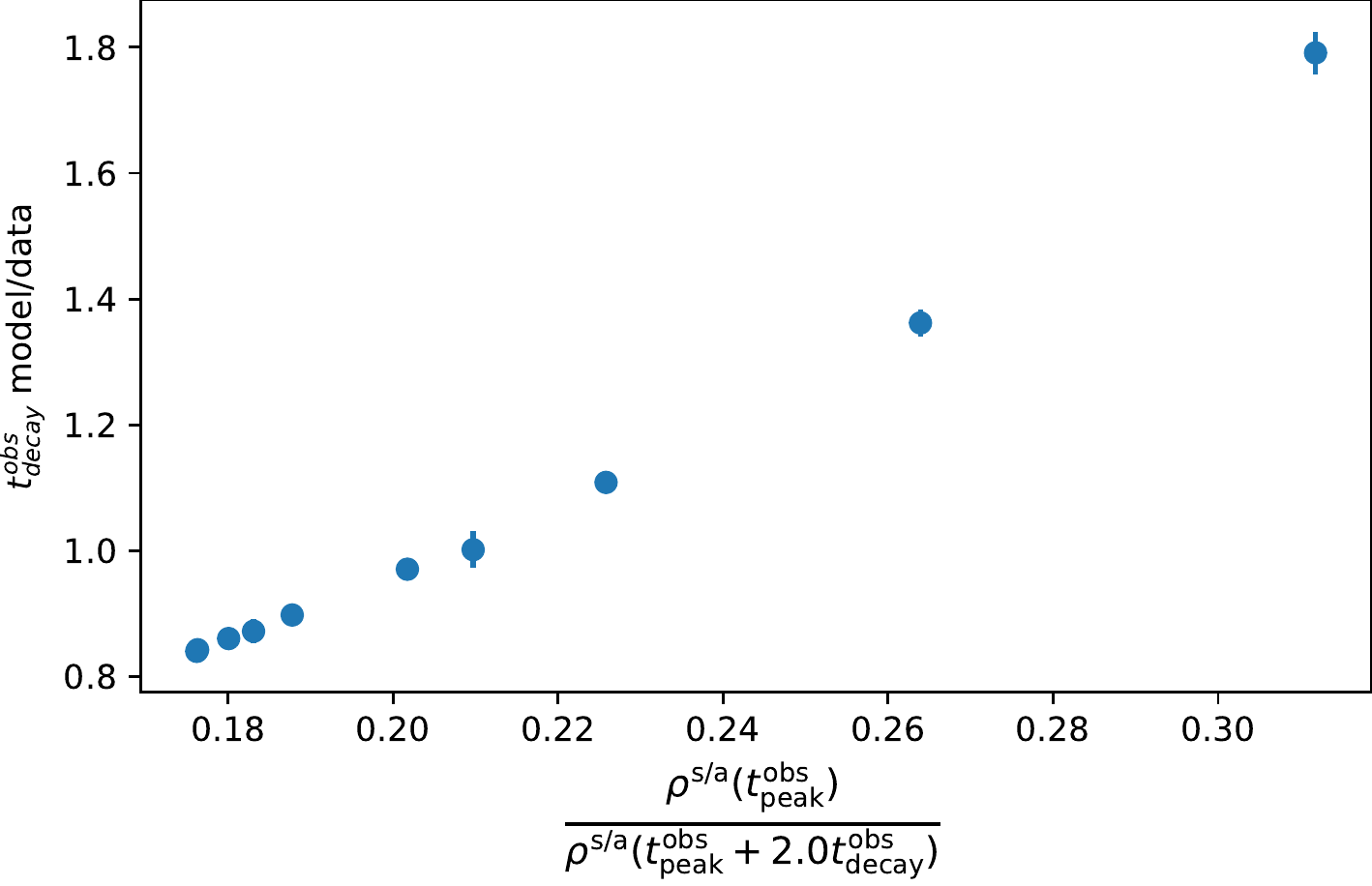}&
   \includegraphics[width=0.5\textwidth,angle=-0]{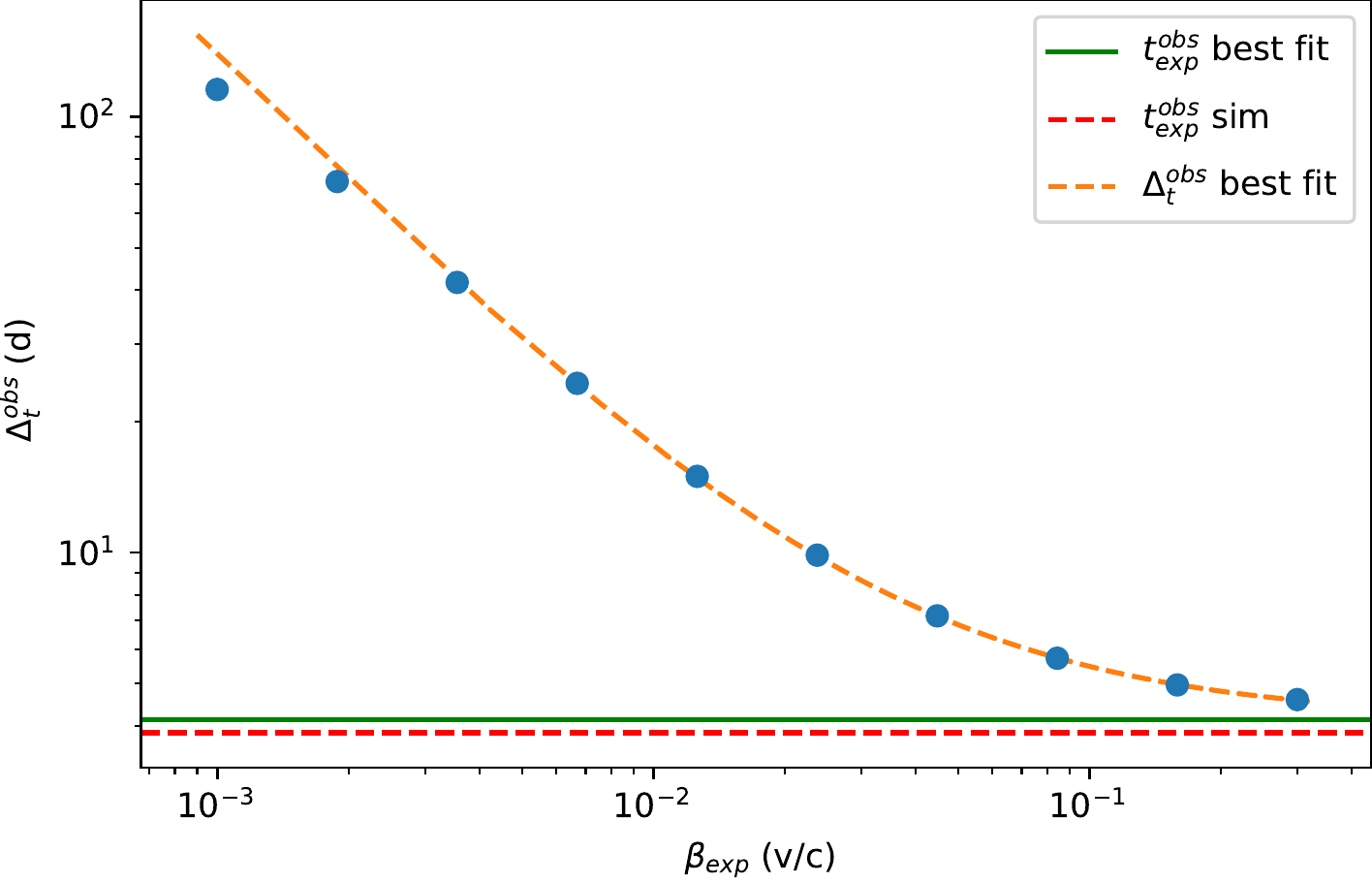}\\
   \end{tabular}
   \captionof{figure}{Expanding trends for $\beta_{\rm exp}$ obtained from the simulations. \textit{Top left panel}:  $t_{\rm decay}^{\rm obs}$ (blue solid points) obtained from the best fit for the radio-$\gamma$ response for the ten simulations with $\beta_{\rm exp}$ ranging 
    $[0.001,0.3]$. The dashed line represents the best fit with the first equation of Equation \ref{eq:fit_trends}. \textit{Top right panel}: Same as in the top left panel, but for the case of $t_{\rm rise}^{\rm obs}$. The dashed line corresponds to the best fit with the second equation of Equation \ref{eq:fit_trends}. \textit{Bottom right panel}: Same as in the top left panel, but for $\Delta_t^{\rm obs}$. The dashed line corresponds to the best fit with the third equation of Equation \ref{eq:fit_trends}.  \textit{Bottom left panel}:  Strong correlation between the fit model-to-data ratios and the cooling ratios in Figure \ref{fig:cooling_ratios}.
   }
   \label{fig:beta_exp_trends}
   \end{figure*}
   \begin{table*}
   \caption{Best-fit results, for the $\beta_{\rm exp}$ simulations}
   \centering  
   \begin{tabular}{ll|l|l|l|l|l|}
      \cline{3-7}
                                                   &    & \multicolumn{2}{c|}{actual values}                    & \multicolumn{3}{c|}{values from  $\beta$ trend best fit}                                                                                  \\ \cline{3-7} 
                                                   &    & \multicolumn{1}{c|}{blob}  & \multicolumn{1}{c|}{obs} & \multicolumn{1}{c|}{$t^{\rm obs}_{\rm rise}$} & \multicolumn{1}{c|}{$t^{\rm obs}_{\rm decay}$}& \multicolumn{1}{c|}{$\Delta t^{\rm obs}$} \\ \hline
      \multicolumn{1}{|l|}{$R_{0}$}                & cm & $5\times 10^{15}$          & $1.66\times 10^{14}$     & $(1.9\pm0.5) \times 10^{14}$                  & $(1.7\pm0.1) \times 10^{14}$                  & $(1.8\pm0.1) \times 10^{14}$              \\
      \multicolumn{1}{|l|}{$\nu_{\rm SSA}^{0}$}    & GHz&    3                       &    90                    & $110\pm 40 $                                  & $ 100\pm10$                          & $100\pm5$                                 \\
      \multicolumn{1}{|l|}{$t_{\rm exp}$}          & s  & $1\times 10^7$             & $3.3\times 10^5$         &                                               &                                               & $(3.57\pm0.01) \times 10^{5}$             \\
      \multicolumn{1}{|l|}{$m_B$}                  &    & 1                          &                          &                                               & $0.96\pm0.06$                                 &                                           \\
      \multicolumn{1}{|l|}{$\phi$}                 &    &                            &                          & $0.6\pm0.1 $                                  & $0.52\pm0.04$                                 & $0.54\pm0.02$                             \\ 
      \multicolumn{1}{|l|}{$p$}                    &   & 1.46                        &                          & $1.6\pm0.3 $                                  & $1.5\pm0.01$                                  & $1.57\pm0.05$                             \\ 
      \hline
      \end{tabular}%
   \tablefoot{Best-fit results, for the $\beta_{\rm exp}$ simulations of the trends reported in Equation \ref{eq:fit_trends} for $t_{\rm rise}^{\rm obs}$, $t_{\rm decay}^{\rm obs}$, and $\Delta t^{\rm obs}$, and shown in the top left, top right, and bottom left panels of Figure \ref{fig:beta_exp_trends}, respectively. The parameter  $p$, i.e. the electron distribution spectral index, is evaluated from the best-fit parameters using the second equation of Equation \ref{eq:R_transp}.}
   
   \label{tab:trends_best_fit_beta}
   \end{table*}

\subsection{$\beta_{\rm exp}$ trends}
\label{sect:beta_exp_trends}
To investigate how the system responds to changes in $\beta_{\rm exp}$, we use the long-term simulations, with ten different values of $\beta_{\rm exp}$ evaluated on a  logarithmic grid ranging $[0.001, 0.3]$, performing the analysis in the observer frame, and we select the 15 GHz frequency. For each value of $\beta_{\rm exp}$, we find the best fit response from  Equation \ref{eq:Resp_func}, and investigate the trends of the parameters $\Delta t^{\rm obs}$,  $t_{\rm rise}^{\rm obs}$,  $t_{\rm decay}^{\rm obs}$ 
with the predictions from Equations \ref{eq:fit_trends}. 


\begin{figure*}
\centering
\begin{tabular}{lr}
\includegraphics[width=0.5\textwidth,angle=-0]{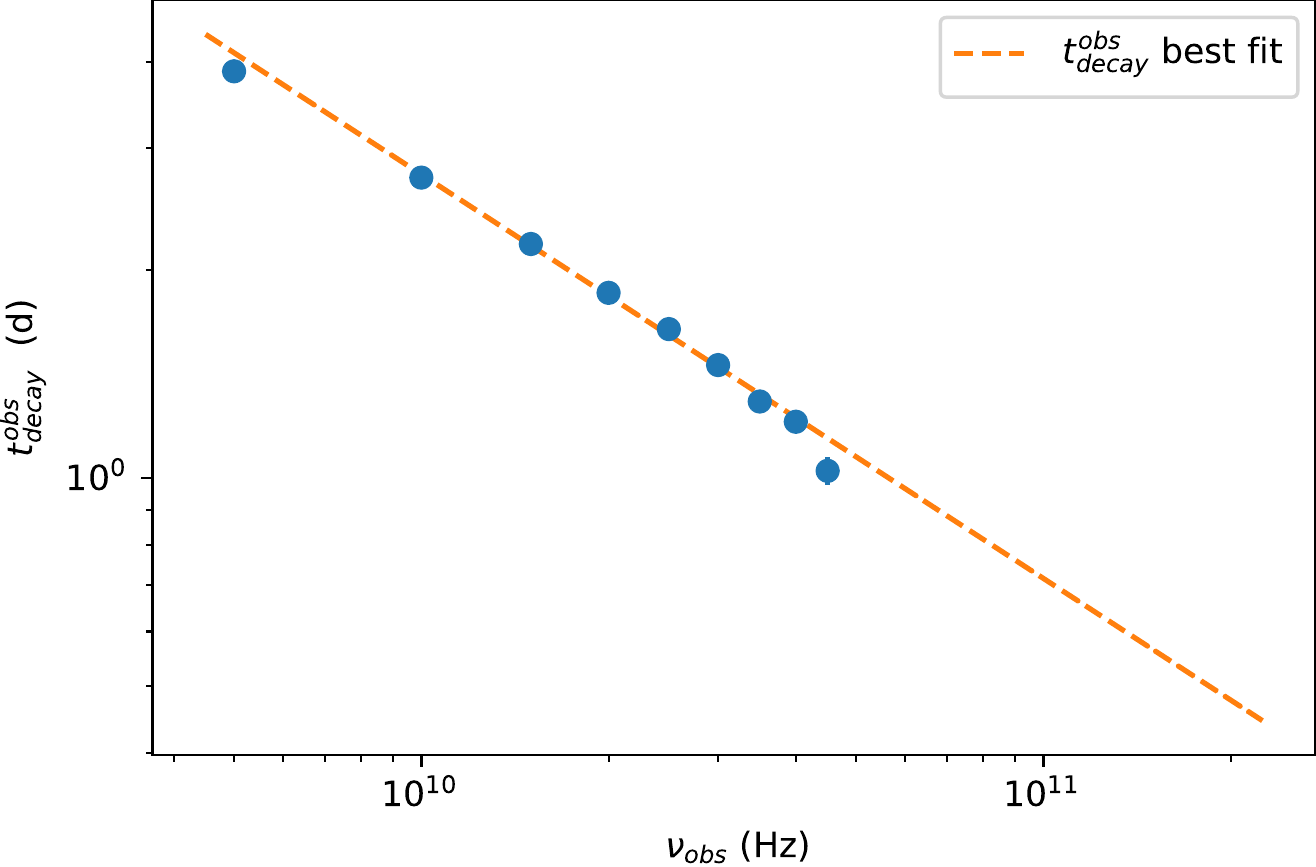}&
\includegraphics[width=0.5\textwidth,angle=-0]{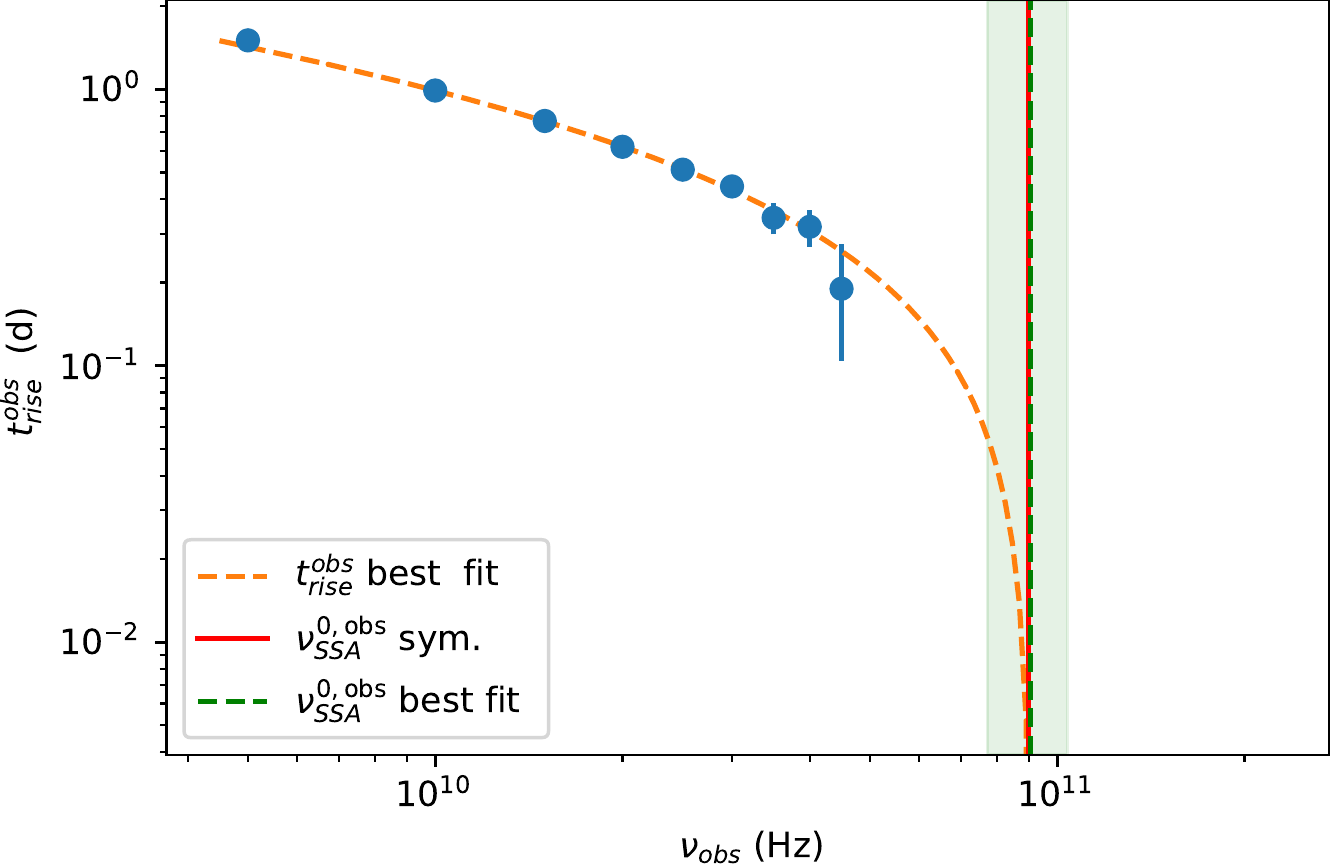}\\
\includegraphics[width=0.5\textwidth,angle=-0]{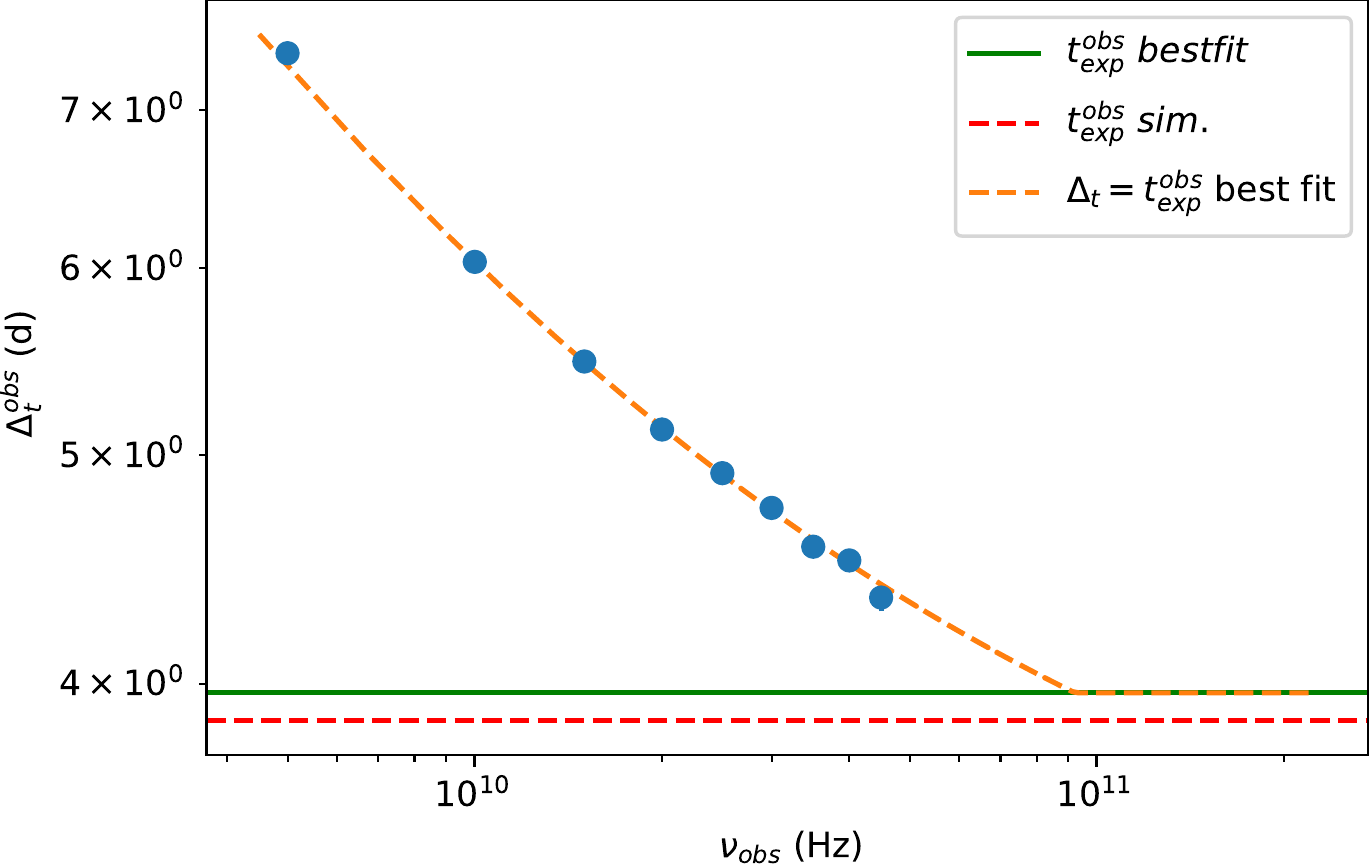}&
\includegraphics[width=0.5\textwidth,angle=-0]{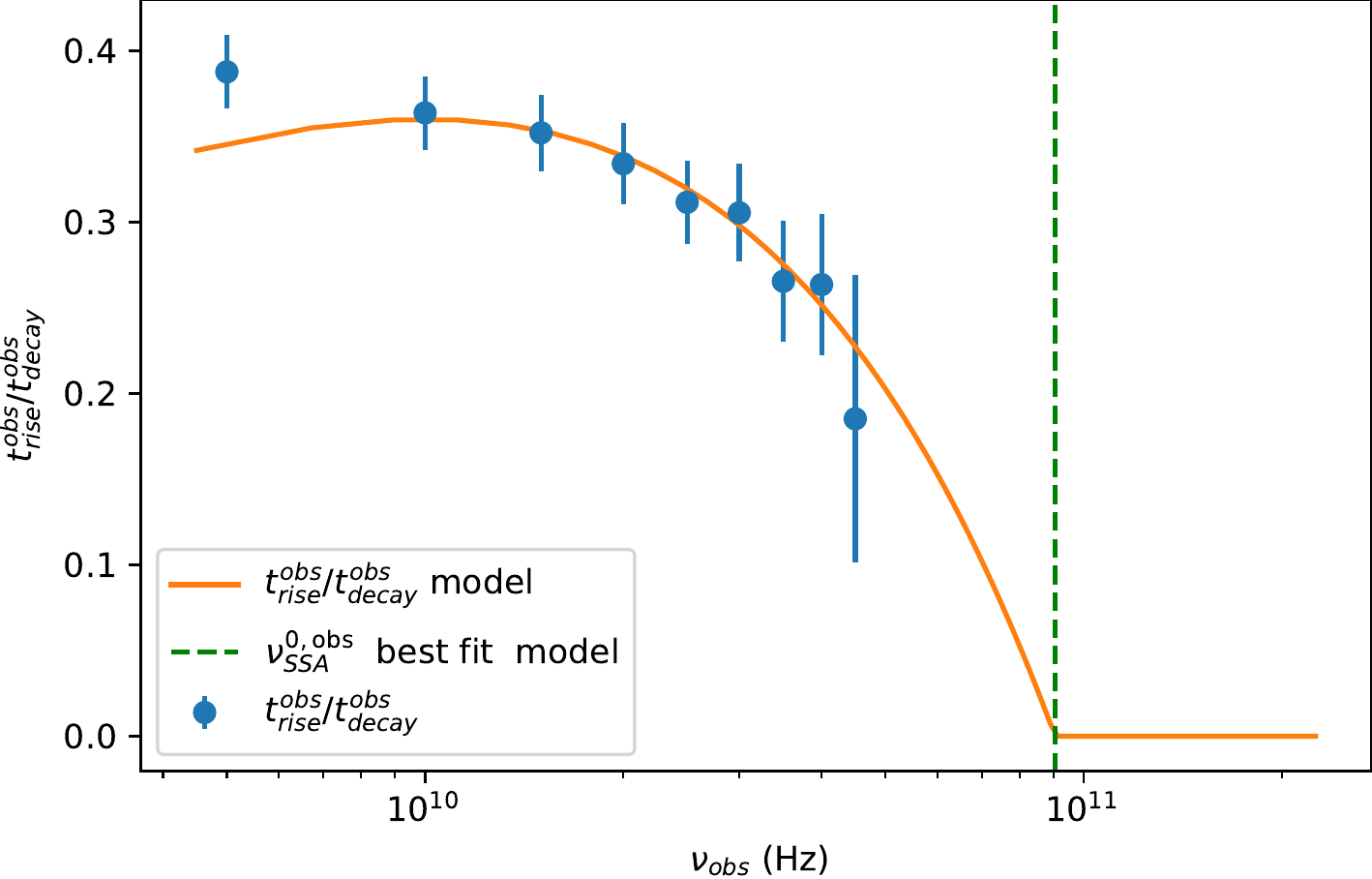}\\
\end{tabular}
\caption{Expanding trends for $\nu$  obtained from the simulations. \textit{Top left panel}: Decay times (blue solid points) 
obtained from the best fit for the radio-$\gamma$ response, for the  simulation with $\beta_{\rm exp}=0.1$ and ranging $\nu_{\rm obs}=[5,45]$ GHz. The orange dashed line represents the best fit with first equation of Equation \ref{eq:fit_trends}. \textit{Top right panel}: Same as in the top left panel, but for the case of $t_{\rm rise}^{\rm obs}$. The dashed line corresponds to the best fit with the second equation of Equation \ref{eq:fit_trends}. \textit{Bottom left panel}: Same as in the top left panel, but for $\Delta t^{\rm obs}$. The dashed line corresponds to the best fit with the third equation of Equation \ref{eq:fit_trends}.  \textit{Bottom right panel}:  Trend of  $t_{\rm rise}^{\rm obs}/t_{\rm decay}^{\rm obs}$ as observed 
in the simulations (solid blue points) compared to the expectation from the individual best-fit trends of  $t_{\rm rise}^{\rm obs}$ and $t_{\rm decay}^{\rm obs}$ (dashed line).}
\label{fig:nu_trends}
\end{figure*}
\begin{table*}
\caption{Best fit results, for the $\nu_{\rm obs}$ simulations}
\centering
\begin{tabular}{ll|l|l|l|l|l|}
\cline{3-7}
                                             &    & \multicolumn{2}{c|}{actual values}                   & \multicolumn{3}{c|}{values from  $\nu$ trend best fit}                                                                 \\ \cline{3-7} 
                                             &    & \multicolumn{1}{c|}{blob} & \multicolumn{1}{c|}{obs} & \multicolumn{1}{c|}{$t^{\rm obs}_{\rm rise}$} & \multicolumn{1}{c|}{$t^{\rm obs}_{\rm decay}$} & \multicolumn{1}{c|}{$\Delta t^{\rm obs}$} \\ \hline
\multicolumn{1}{|l|}{$R_{0}$}                & cm & $5\times 10^{15}$         & $1.66\times 10^{14}$     & $(2.4\pm1.0) \times 10^{14}$                  & $(1.7\pm0.2) \times 10^{14}$                     & $(1.6\pm0.1) \times 10^{14}$              \\
\multicolumn{1}{|l|}{$\nu_{\rm SSA}^{0}$}    & GHz&     3                    &     90                   & $90\pm 10 $                                   & $100\pm20$                                       & $90\pm10$                                 \\
\multicolumn{1}{|l|}{$t_{\rm exp}$}          & s  & $1\times 10^7$            & $3.3\times 10^5$         &                                               &                                                  & $(3.4\pm0.1) \times 10^{5}$               \\
\multicolumn{1}{|l|}{$m_B$}                  &    &  1                        &                          &                                               & $1.0\pm0.1$                                      &                                          \\
\multicolumn{1}{|l|}{$\beta_{\rm exp}$}      & c  &  0.1                      &                          & $0.03\pm0.01$                                 & $0.09\pm0.01$                                    & $0.06\pm0.01$                              \\
\multicolumn{1}{|l|}{$\phi$}                 &    &                           &                          & $0.24\pm0.07$                                 & $0.58\pm0.02$                                    & $0.50\pm0.02$                             \\
\multicolumn{1}{|l|}{$p$}                    &    &1.46                       &                          & $0.6\pm0.2$                                   & $1.7\pm0.1$                                      & $1.4\pm0.1$                             \\ 

\hline
\end{tabular}%
\tablefoot{Best fit results, for the $\nu_{\rm obs}$ simulations, of the trends reported in Equation \ref{eq:fit_trends} for $t_{\rm rise}^{\rm obs}$, $t_{\rm decay}^{\rm obs}$, and $\Delta t^{\rm obs}$, and shown in the top left, top right, and bottom left panels of Figure \ref{fig:nu_trends},  respectively. The parameter  $p$, i.e. the electron distribution spectral index, is evaluated from the best-fit parameters using the second equation of Equations \ref{eq:R_transp}.}
\label{tab:trends_best_fit_nu}
\end{table*}

The results for the trends are shown in Figure \ref{fig:beta_exp_trends}, and the corresponding best-fit values 
are reported in Table \ref{tab:trends_best_fit_beta}. We start from the decay trend (top left panel, orange dashed line). The best-fit value for the parameter  $ R_{0}^{\rm obs}=  (1.7\pm 0.1) \times 10^{14}$ cm, which corresponds to a blob frame value of $R_{0}  \simeq 5.2\times 10^{15} $ cm and is in very good agreement with the value of $R_0=5\times 10^{15}$ cm used in our simulation. The best-fit estimated value $\nu^{0, \rm obs}_{\rm SSA}= 100 \pm 10 $ GHz is compatible within one sigma with the simulation value of 90 GHz. Also,  the value of $m_B$  returned by  the delay trend $m_B= 0.96\pm0.06$ is compatible with the simulation value within one sigma.
Nevertheless, we notice systematic deviations from the  trend. These are better visible in the model-to-data ratios in the top left panel of Figure \ref{fig:beta_exp_trends}. This behaviour is due to the complex interplay between the radiative and adiabatic cooling  timescales discussed in the previous section, and we notice that the shape of the model-to-data ratios matches the shape of the cooling ratios in Figure \ref{fig:cooling_ratios}. This agreement can be appreciated better in the bottom-left panel of Figure \ref{fig:beta_exp_trends},  which shows the strong correlation between the fit model-to-data ratios and the cooling ratios in Figure \ref{fig:cooling_ratios}.

For the rising time trend (top right panel of Figure \ref{fig:beta_exp_trends}), the best fit value of $R_{0}^{\rm obs}=  (1.9 \pm 0.5) \times 10^{14}$ cm  corresponds to a blob rest-frame value of $R_{0}\simeq 5.7\times 10^{15}$ cm,  which is within a $\approx15\%$ deviation from the simulated value. The best-fit value of   $\nu^{0, \rm obs}_{\rm SSA}=  110\pm 40$  is compatible with the simulated value  of 90 GHz but with a significant dispersion.
For the delay trend (bottom right panel of Figure \ref{fig:beta_exp_trends}),  we again find values of $R_{0}$ and $\nu^{0, \rm obs}_{\rm SSA}$, close to the simulation values. We are also able to estimate the value of $t_{\rm exp}^{\rm obs}=(3.57\pm0.01)\times 10^5$s  with good accuracy, which is an agreement with the actual one within a few percent. Finally, we estimate the electron distribution index  $p$ from the best-fit parameters $\phi$ and $m_B$ using the second equation of Equation \ref{eq:R_transp}. The estimated values of $p$ for all the three trends are in a very good agreement with the simulation value of $p\approx 1.5$.

\subsection{$\nu$ trends}
\label{sect:nu_trends}
To investigate the trends as a function of $\nu^{\rm obs}$,  we use the simulation with $\beta_{\rm exp }=0.1$, where $\nu^{\rm obs}$ is the frequency of the observed radio light curves. We stress that the ratio $\nu^0_{\rm SSA}/\nu^*_{\rm SSA}$ can be expressed using either the $blob$ or the $observed$ frequencies,
because the ratio will cancel out the beaming transformation.
We follow the same approach as in Section \ref{sect:beta_exp_trends}, with the only difference being that we use $\nu$ as  an independent variable. The results are shown in Figure \ref{fig:nu_trends}, and  the best-fit results are reported in Table \ref{tab:trends_best_fit_nu}. 
For the decay trend (top left panel), we find $R_{0}^{\rm obs}\simeq  1.7 \times 10^{14}$ cm, corresponding to a blob frame value of $R_{0}\simeq 5.1\times 10^{15}$ cm, which is in good agreement with the simulation value. The best-fit value of $\nu^{0, \rm obs}_{\rm SSA}=  100\pm 20$ GHz is compatible with that measured in the simulated SEDs. Both the estimate of $\beta_{\rm exp}= 0.09\pm0.01$  and  the estimate of $m_B 1.0\pm0.1$ are in excellent agreement with the simulation values.
The $t_{\rm rise}$ trend (top right panel of Figure \ref{fig:nu_trends}) returns a value of $R_{0}^{\rm obs}= 2.4 \pm 1.0 \times 10^{14}$, which is $\approx 60\%$ larger than the simulation value, but still compatible within one sigma. The $\beta_{\rm exp}= 0.03\pm0.01$ is  significantly lower than the simulated one. The value of $\nu^{0,\rm  obs}_{\rm SSA} \simeq  90 \pm 10$  provides a very good estimate of the simulation value $\nu^{0, \rm obs}_{\rm SSA} =  90$ GHz, but, as mentioned in the previous section, this value represents a lower bound. In the top left panel of Figure \ref{fig:nu_trends},  the green shaded area shows the 1-$\sigma$  the interval from the best fit, and the vertical red dashed line represents the simulation value.
The delay trend (bottom left panel of Figure \ref{fig:nu_trends}) returns an estimate of $ \beta_{\rm exp}=0.06\pm0.01$, which is $\approx 40\%$  lower than the simulation value, and an estimate of $\nu^{0,\rm  obs}_{\rm SSA} \simeq  90 \pm 10$ GHz, which is in agreement with the simulation value. In this case,  we also estimate the value of $t_{\rm exp}^{\rm obs}= (3.4\pm0.1)\times 10^5$s with good accuracy,  which is in agreement with the simulation value
within a few percent. 
Finally, we comment on the effect of the initial SSA frequency on the rising time. As already noted, the rising time decreases to zero as  $\nu^*_{\rm SSA}$ approaches $\nu^0_{\rm SSA}$. This implies that even if we obtain a long decay time because of the low expansion rate, we might expect a short rising time if $\nu^*_{\rm SSA}$ is close to $\nu^0_{\rm SSA}$. As in the case of the $\beta_{\rm exp}$ trends, we estimate the electron distribution index  $p$ from the best-fit parameters using the second equation of Equation \ref{eq:R_transp}. The agreement with the simulation value is lower than in the  case of the $\beta_{\rm exp}$ trends, in particular  for the case of the rise time trend, but nevertheless both delay and decay times provide an estimate that is consistent within one sigma with the simulation value.

\section{Comparison with observational data}
\label{sect:data_comparison}
Here we show how we derive the radio-$\gamma$-ray response using real data from {\it Fermi}-LAT \citep{2009ApJ...697.1071A} and from the OVRO radio telescope \citep{2011ApJS..194...29R} for three well-known sources. The results are discussed in Sect. \ref{sect:discussion} in the framework of the model outlined above.

To reconstruct the radio light curve, we convolved the GeV light curve with the response profile given by Eq.~\ref{eq:Resp_func} and optimised the response parameters to obtain a match with the observed radio data.
The response is based on a single flare profile, even if the radio light curves consist of many overlapping  consecutive flares in addition to a background flux. As, in reality, the response parameters might change from flare to flare, the response we derive should be considered as an average, possibly driven by the most prominent flares.  As a consequence,
short-term features, such as spiky structures, could be smoothed and suppressed. The main aim of this section is to show that
the physical mechanism investigated in the previous section is responsible for the systematic  delayed radio emission, and to understand  whether or not these average timescales are compatible with those predicted by the model, assuming physical parameters within the range of those used in the simulations

Delayed responses were already observed by \cite{max-moerbeck_2014MNRAS.445..428M} for specific flares of Mrk\,421 and Mrk\,501 and interpreted as the propagation of shocks through conical jets. The same interpretation was proposed \citep{turler_1999A&A...349...45T} to explain the long-term light curves of 3C 273, and in particular the radio flares corresponding to overlapping stretched and delayed GeV flares \citep{esposito_2015A&A...576A.122E}. 

The results indicated below show that a constant response profile is adequate for Mrk\,421 and Mrk\,501, while the response amplitude appears variable from flare to flare in the case of 3C\,273. We also find that the background flux could be considered as constant in Mrk\,421 and Mrk\,501, possibly accounting for the core emission, while it is slowly variable and firmly associated to the jet in 3C\,273.

\begin{figure*}
   \centering
   \includegraphics[width=0.85\linewidth]{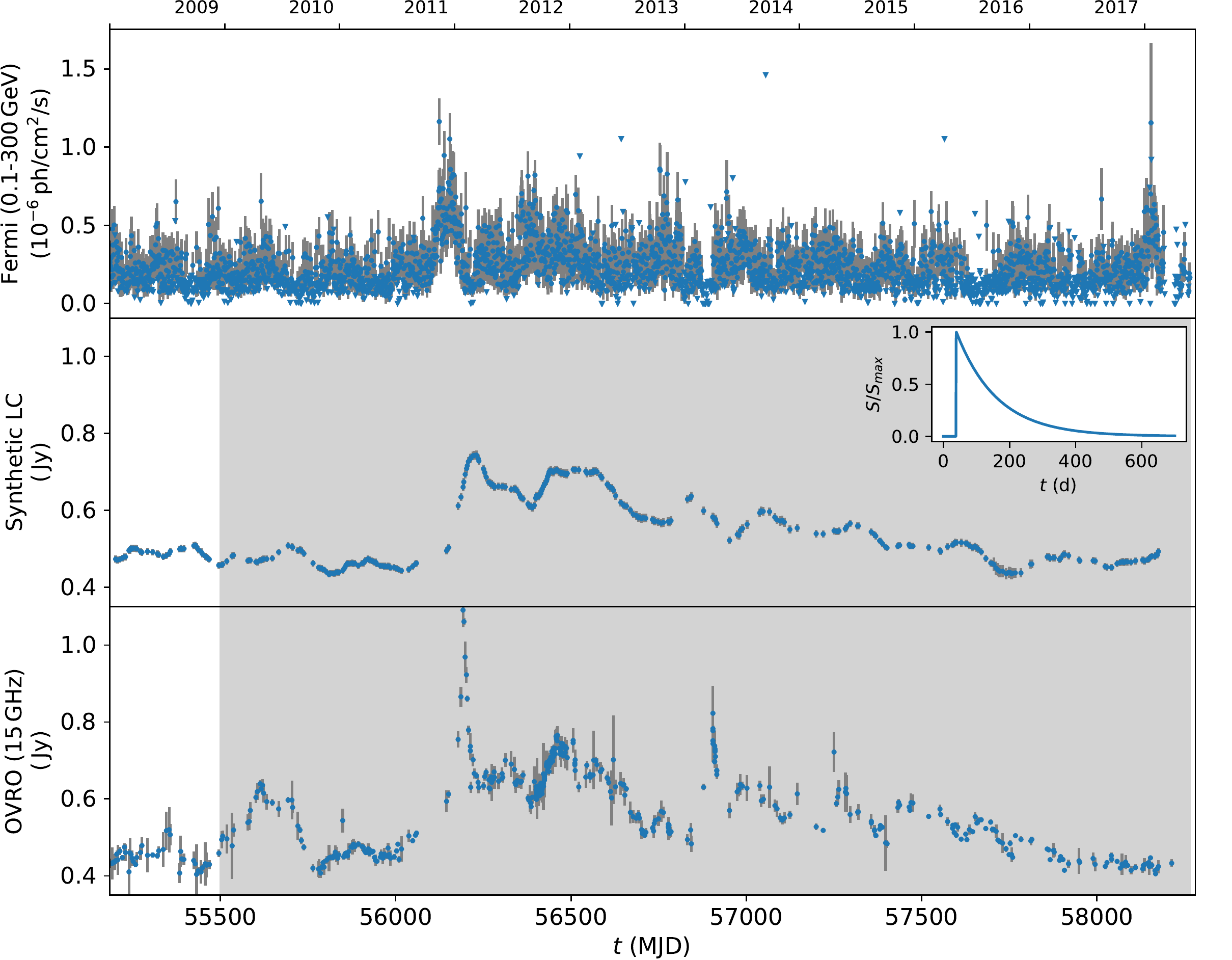}
   \caption{Synthetic radio light curve for Mrk 421 (middle) created as a convolution of the day-binned Fermi-LAT 0.1-300\,GeV light curve (top) and of the radio response (inset panel), compared with the OVRO 15\,GHz radio light curve (bottom). Fitting time range is highlighted in grey.}
   \label{fig:fermi_radio_conv_mrk421}
   \end{figure*}
\subsection{Data\label{subsec:data}}
The Large Area Telescope on board the Fermi Gamma-ray Space Telescope ({\it Fermi}-LAT) is the most sensitive $\gamma$-ray telescope in the 20 MeV $<$ E $<$ 300\,GeV energy range. {\it Fermi}-LAT  uses a charged particle tracker and a calorimeter to detect photons. The point spread function (PSF) depends on energy, reaching a $1\sigma$-equivalent containment radius of $\sim 0.1^{\circ}$ at 40\,GeV \citep{2009ApJ...697.1071A}. Despite {\it Fermi}-LAT being sensitive to $\gamma$-rays of about 20\,MeV, because of the energy-dependent PSF we can only reliably consider photons with energies between $100$\,MeV and 300\,GeV. Data reduction was performed using the PASS8 pipeline and the Fermi Science Tool v10r0p5 package. Sources from the {\it Fermi}-LAT four-year point-source catalogue were used for the fitting model. More details on the performed analysis can be found in \cite{2021A&A...647A..88A}. Because of their different $\gamma$-ray fluxes, we used different time bins for each source: 3 days, 1 day, and 7 days for 3C 273, Mrk 421, and Mrk 501, respectively. In the case of Mrk 501, we also reduced the considered energy range to 1-300\,GeV leading to a reduction in the flux uncertainties thanks to the better PSF and a lower background in that energy range. 

Regular radio observations of Mrk\,421, Mrk\,501, and 3C\,273 were performed at 15\,GHz by the 40\,m radio telescope of the Owens Valley Radio Observatory (OVRO) \citep{2011ApJS..194...29R}. These observations were conducted as part of the {\it Fermi} blazars monitoring campaign. The light curves (with twice-per-week cadence) were publicly available (when we started the analysis) from the OVRO archive \footnote{http://www.astro.caltech.edu/ovroblazars/}. We removed all the data points with a significance of less than $5\sigma$, because these  observations were performed during unfavourable observing conditions.

\subsection{Mrk\,421}
\label{subsec:crosscor_421}
The radio light curve of Mrk\,421 is broadly correlated to the GeV light curve, with radio lagging behind the GeV variations by $30-100$ days at the maximum of the discrete correlation function \citep{2021A&A...647A..88A}.

\begin{table}[b]
\centering
\caption{Best-fit parameters for the $\gamma$-ray to radio response profile for Mrk\,421.}
\label{tab:profile_mrk421}
\begin{tabular}[t]{lc}
Parameter & Value\\
\hline
\hline
$A$ & $12.5_{-0.013}^{+0.5}\times 10^3$ Jy cm$^2$ s/ph\\ 
$t_{rise}$ & $\lesssim 1$ day \\ 
$t_{decay}$ & $126.5_{-1.3}^{+1.3}$ days \\
$\Delta t$ & $37.58_{-0.13}^{+0.13}$ days \\
$F_{background}$ & $0.18_{-0.0004}^{+0.008}$ Jy \\
\hline
\end{tabular}
\tablefoot{Best-fit parameters for the $\gamma$-ray to radio response profile (Eq.~\ref{eq:Resp_func}) with the addition of a background radio flux. }
\end{table}

\begin{figure*}
\centering
\includegraphics[width=0.85\linewidth]{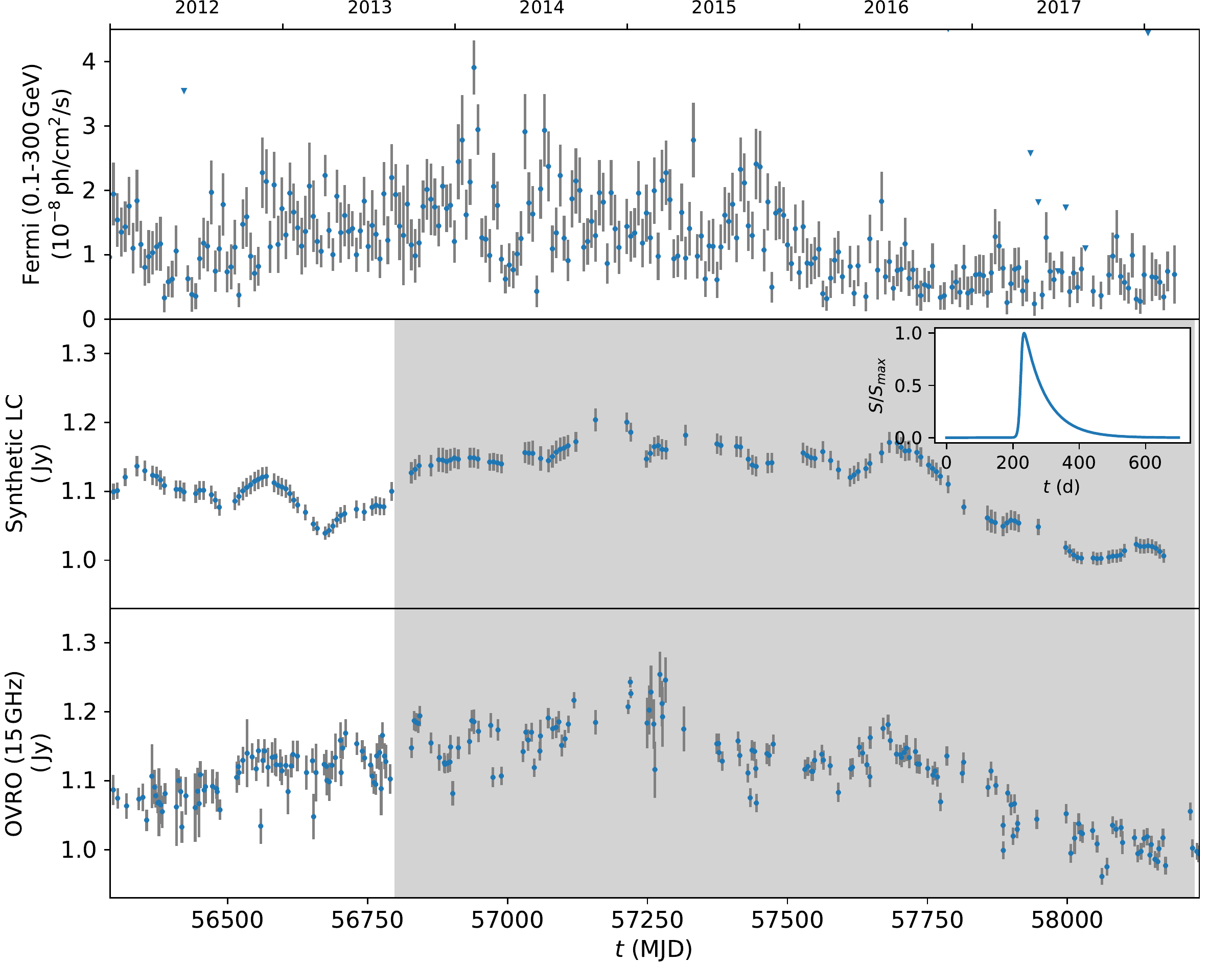}
\caption{Mrk 501 synthetic radio light curve (middle), created as a convolution of the weekly binned Fermi-LAT 1-300\,GeV light curve (top) and of the radio response (inset of the middle panel), compared with the OVRO 15\,GHz radio light curve (bottom). The fitting time range is highlighted in grey.}
\label{fig:fermi_radio_conv_mrk501}
\end{figure*}

The best-fit response was obtained by minimising, over a 7.5-year period (MJD 55500-58226), the deviations between the observed radio light curve and the synthetic light curve obtained by the convolution of the Fermi LAT daily binned light curve with a response (Eq.~\ref{eq:Resp_func}). The Fermi LAT light curve starts about two years before the period used for the minimisation to account for the long-lasting effect of the response, particularly $\Delta t$.

The observed GeV and radio and the resulting synthetic radio light curves are shown in Fig. \ref{fig:fermi_radio_conv_mrk421}, and the best-fit parameters are listed in Table~\ref{tab:profile_mrk421}.
The uncertainties on the synthetic light curve were derived using direct uncertainty propagation from the Fermi LAT light curve and response profile through the convolution with the response profile. The best-fit synthetic light curve is similar, but does not perfectly match the observed radio light curve, possibly indicating that the intensity of the response (e.g. the constant A) might be variable with time. Also, a fast radio flare near MJD 56897 and a wider flare at about MJD 55600 (see Fig. \ref{fig:fermi_radio_conv_mrk421}) could not be reproduced. These flares could have a different origin or may need a different response \citep[such adjustments were also used by][and could indicate different conditions in various shocks]{esposito_2015A&A...576A.122E}. As the uncertainties on the GeV light curve and on the response profile are not Gaussian nor independent, the goodness of fit $ (\chi^2/\nu= 1243/218=5.9 )$ between the observed and synthetic radio data is only indicative.
As the response rise time is similar to the binning time of the GeV light curve, its value 
indicates a rising time of shorter than one day.

\subsection{Mrk\,501\label{subsec:crosscor_501}}
The correlation between the GeV and radio light curves is also strong in Mrk\,501 with radio variations lagging behind the GeV ones by 170-250 days \citep{2021FACTSubmitted}. To further probe the connection between these bands, we searched for the delayed response profile, which, when convolved with the GeV light curve, can mimic the radio variations. 

Adopting the analytical response profile defined by Eq.~\ref{eq:Resp_func} together with a constant background emission ($\sim 0.9$ Jy) and minimising the deviations between the observed and synthetic radio data ($\chi^2/\nu = 258/112 = 2.3$) led to a best-fit response decaying in 66 days after a delay of about 224 days as displayed in Fig.~\ref{fig:fermi_radio_conv_mrk501} (inset panel) and parametrised in Table~\ref{tab:profile_mrk501}.
The minimisation was performed for the period starting on MJD 56800, because prior to that the GeV-radio correlation is weak, indicating that additional noise or emission components are present. 
 
\begin{figure*}
   \centering
   \includegraphics[width=0.85\linewidth]{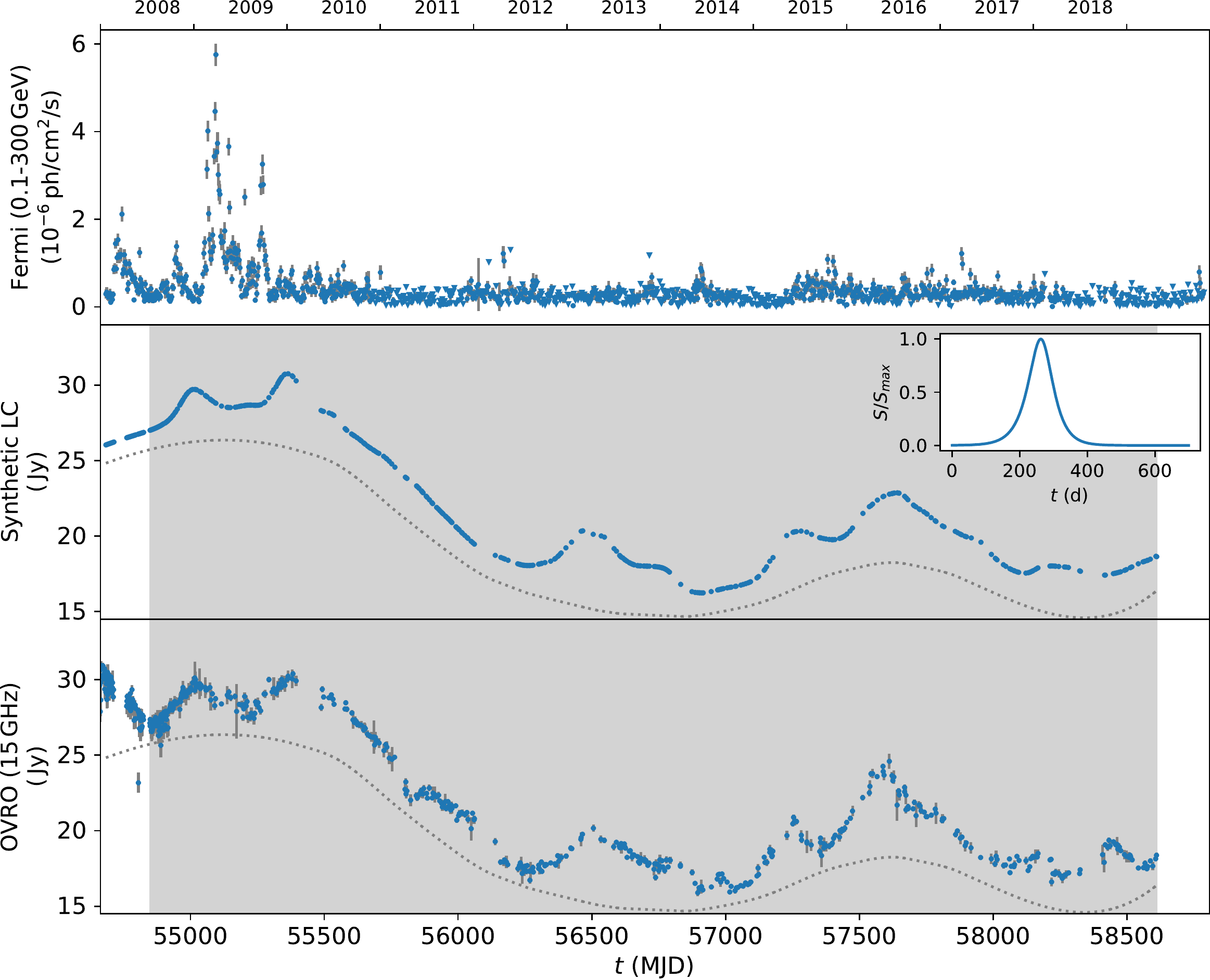}
   \caption{3C 273 synthetic radio light curve (middle), created as a convolution of the 3-days-binned Fermi-LAT 0.1-300\,GeV light curve (top) and of the radio response (inset panel), compared with the OVRO 15\,GHz radio light curve (bottom). Amplitude fitting time range is highlighted in grey. The shape of the response profile (Table~\ref{tab:profile_3c273}) was determined using the $\gamma$-ray flaring period 54710 - 54890 MJD. The dotted line shows a slowly varying background radio flux, which cannot be reproduced by the response convolution approach. The response amplitude was adjusted for different time intervals. 
   }
   \label{fig:fermi_radio_conv_3c273}
   \end{figure*}

\begin{table}[b]
\centering
\caption{Best-fit parameters for the $\gamma$-ray-to-radio response  for Mrk\,501.}
\label{tab:profile_mrk501}
\begin{tabular}[t]{lc}
Parameter & Value \\
\hline
\hline
$A$ & $166^{+5}_{-3}\times 10^4$ Jy cm$^2$ s/ph \\

$t_{rise}$ & $12^{+4}_{-4}$ days \\
$t_{decay}$ & $73^{+3.6}_{-3.6}$ days\\ 
$\Delta t$ & $234^{+10}_{-10}$ days \\ 
$F_{background}$ & $0.915^{+0.004}_{-0.004}$ Jy \\ \hline
\end{tabular}
\tablefoot{Best-fit parameters for the $\gamma$-ray-to-radio response profile (Eq.~\ref{eq:Resp_func}) with the addition of a background radio flux.
}
\end{table}

\subsection{3C~ 273\label{subsec:crosscor_3c273}}
The GeV-radio response for the FSRQ 3C 273 was investigated and discussed in \cite{esposito_2015A&A...576A.122E} using light curves lasting for about 6 years. The radio data could be reproduced using a convolution of the GeV light curve through a response profile varying only in amplitude from flare to flare.
 
 We performed a similar analysis including additional data and found that for 3C 273 a single response profile cannot reproduce the complete radio light curve, unlike for Mrk 421 and Mrk 501. The addition of a slowly changing background radio emission was required, which was estimated with a third-order Savitzky-Golay filter of the observed radio emission with a windows size of 501 days, which is long enough to prevent fitting of individual radio flares (dotted line in Fig.~\ref{fig:fermi_radio_conv_3c273}). This variable background component needs to be bright (about 90\% of the total emission) and cannot be accounted for by the core emission of 3C\,273. This emission is linked to the jet itself. The variability of that component cannot be well constrained.
 
Figure~\ref{fig:fermi_radio_conv_3c273} shows the original {\it Fermi}-LAT and radio light curves, and the synthetic radio light curve derived as described above. We had to adjust the amplitude of the response for different individual flaring periods while the other response parameters could remain unchanged, as found by \cite{esposito_2015A&A...576A.122E}.
The response profile (best-fit parameters listed in Table~\ref{tab:profile_3c273}) was derived using a single flare ($\gamma$-ray time range [54710, 54890] MJD) with a goodness of fit $\chi^2/\nu=88.5/59 = 1.5$. The amplitudes of the response for the individual flaring periods (listed in Tab.~\ref{tab:profile_3c273_amplitudes}) were determined on the full range of radio observations.
The variations of the response amplitude  during different flares can be explained by two possible factors:
\begin{itemize}
   \item A transition from a SSC- to an  external Compton (EC)-dominated IC emission regime. Indeed, the amplitude of the response strongly depends  on the initial radiative output of both the IC and S components at the flaring state. Hence,  a different contribution from the EC emission can impact the response amplitude.
   
   \item As shown in Section \ref{sect:response}, and in particular in Figure \ref{fig:A_trend}, there is a strong effect of the competition between synchrotron and adiabatic cooling times on the modulation of the response amplitude. In particular,  we notice that, for changes in  $\beta_{\rm exp}$, variations on the response amplitude of up to one order of  magnitude are possible, whilst for changes in $\nu_{\rm obs}$ the variations can be up to $40\%$. As for 3C 373, the modulation of the amplitude ranges form $\approx 0.5$ to $\approx 5$  (see Table \ref{tab:profile_3c273_amplitudes}), we can assume that the observed modulation can also be explained as a change in  $\beta_{\rm exp}$. 
   Moreover, a change in the EC contribution can impact the radiative--adiabatic balance, producing effects similar to those produced by  changes  in $\beta_{\rm exp}.$
\end{itemize}

We also note that the slowly changing background (dotted line in the middle and bottom panels of Figure \ref{fig:fermi_radio_conv_3c273}) needed for the radio emission does not affect the $\gamma-$ray component. This discrepancy seems to be unrelated to the EC/SSC transition, because the radio-$\gamma$ response reproduces the detrended radio light curves. A possible explanation could be provided by a change in the  baseline of the radio emission at the expansion site, possibly related to a change in the beaming factor and therefore related to the jet--blob geometry. In any case, the fact that the modulated baseline alone allows us to reproduce the trend suggests that the physical cause of the modulation is not affecting the physical mechanism of the response.

\begin{table}[ht]
\centering
\caption{Best-fit parameters of the $\gamma$-ray({\it Fermi}-LAT)-to-radio response profile  for 3C 273. }
\label{tab:profile_3c273}
\begin{tabular}[t]{lc}
Parameter & Value\\
\hline
\hline
$A_0$ & $174_{-11}^{+12}\times 10^3$  Jy cm$^2$ s/ph\\
$t_{rise}$ & $37_{-2}^{+2}$ days \\
$t_{decay}$ & $69_{-3}^{+3}$ days \\ 
$\Delta t$ & $276_{-10}^{+10}$ days \\
\hline
\end{tabular}
\tablefoot{Best-fit parameters of the $\gamma$-ray({\it Fermi}-LAT)-to-radio response profile (see Eq.~\ref{eq:Resp_func}).}
\end{table}

\begin{table}[ht]
\centering
\caption{Response amplitude multiplicative factor for different flaring periods of 3C 273.}
\label{tab:profile_3c273_amplitudes}
\begin{tabular}[t]{lc}
Period, MJD & Value\\
\hline
\hline
54684-54890 & $1.07 \pm 0.11$ \\
54890-55340 & $0.69 \pm 0.04$ \\
55340-56125 & $1.39 \pm 0.08$ \\
56125-56530 & $5.31 \pm 0.16$ \\
56530-56920 & $0.46 \pm 0.15$ \\
56920-57250 & $5.3 \pm 0.4$ \\
57250-57710 & $2.71 \pm 0.12$ \\
57710-57910 & $0.88 \pm 0.17$ \\
57910-58110 & $3.5 \pm 0.3$ \\
58110-58610 & $2.4 \pm 0.3$ \\ \hline
\end{tabular}
\end{table}

\section{Discussion}
\label{sect:discussion}
The results presented in the present analysis provide a self-consistent framework with which to explain the observed delay in the radio emission with respect to the $\gamma$-ray activity observed in several blazars. The development of a flaring episode, followed taking into account radiative and acceleration processes, results in an initial self-absorption frequency of above $10^{11}$ Hz (according to our initial setup), and therefore the radio delay will occur under the hypothesis of an expanding blob when the source size is large enough to move the SSA at frequencies comparable to or lower than the observed radio light curve. The particles need to be confined in order to observe a delayed peak in the radio light curve. In our analysis, we
decoupled the acceleration region from the radiative region, and this is why escape timescales acting in the radiative region do not impact the accelerated electrons. Moreover, because of the expansion, it is natural to expect the escape time of the particles to increase. 

Similar analyses and comparable results have been presented by \cite{Boula2018} and  \cite{Potter2018} but only a qualitative comparison could be performed with our results as these latter works did not present a quantitative analysis of the delays or the phenomenological relations linking the delay to the physical parameters of the jet, and assume an arbitrary electron distribution in the flaring region. The time lags versus the expansion velocity presented in \cite{Boula2018} follows a similar  trend to that reported in the bottom left panel of Figure \ref{fig:beta_exp_trends}, with a delay going asymptotically to $\Delta t=0$, as predicted by our model and confirmed by our simulations. The  time lags versus the initial source size $R_0$ \citep[Fig. 3 of][]{Boula2018} is the result of the change in the SSA frequency with the source size and can therefore be compared with
our result shown in the bottom left panel of Figure \ref{fig:nu_trends}. \cite{Potter2018}  used a large-scale parabolic to conical jet structure in qualitative agreement with our results.
Even if the agreement with these models can only be qualitative ---because the analyses were performed with different methods and numerical codes \citep{Potter2018,Boula2018}---, the comparison provides orthogonal checks, the results of which appear to support the hypothesis of an expanding blob related to the radio-$\gamma$ connection in blazars.

One of the main novelties of our analysis, and one that is missing in previous studies, is the determination  of the single flare response, and its  verification via a self-consistent numerical model, taking into account both acceleration and radiation process, and the determination of the phenomenological relations that link not only the delay but also the rise and decay time to the expansion velocity, magnetic field index, and initial SSA frequency. The proposed single-flare response is able to reproduce the radio light curve as a convolution of the $\gamma-$ray light curve, and we verified that the timescales of the response follow the phenomenological trends. This allowed us to establish a link between some physical parameters, such as the emitting region initial  size, $R_0$, the jet magnetic field index $m_B$, and the observed response. In particular, the derivation of the initial source size can provide an orthogonal  method compared to the determination based on MW SED fitting or variability timescales.

We also investigated other effects of the adiabatic expansion. In particular, we analysed the impact on the CD, verifying that as the source size increases, the consequent decrease in photon and electron density leads to a drop in the CD. This effect, if present at the time of the flare ($t_{\exp}=0$), can provide a hint for 
the blob expansion before the observation of the radio delay. It is extremely interesting that, very recently,  \cite{MAGIC_Mrk421_2021}  presented an analysis of the correlation patterns of Mrk 421 in 2017, finding that adiabatic expansion without significant particle losses can be invoked to explain the pattern of the CD evolution. The authors also verified that the adiabatic cooling timescales should be longer than those necessary to explain a cooling break compatible with their MW data, requiring adiabatic timescales of the order of weeks to months in the observer frame that are in good agreement with our proposed scenario. Moreover, the authors also state that the expanding blob can explain $\gamma$-ray orphan flares.  The scenario proposed in  \cite{MAGIC_Mrk421_2021} is highly compatible with our scenario, as we can set $t_{\rm exp}=0$  without any loss of generality, and proves that, for Mrk 421, both the radio delay and the CD effect are confirmed by the data. 
Also, comparisons with the data shown in Section \ref{sect:data_comparison} indicate that our model can accurately reproduce the radio light curve as a response to the $\gamma$-ray  light curve over a time-span of years, and the relevant timescales derived from the response function are in agreement with those derived from the self-consistent modelling.  In particular, for Mrk 421, the analysis in Section \ref{subsec:crosscor_421}  returns a best-fit value of $\Delta_t \approx 37$ days with a decay time of $\approx 126$ days, which corresponds to $\beta_{\rm exp}\lesssim 0.01$ (see Sect. \ref{sect:beta_exp_trends}). The value of the decay time, which according to our model is a proxy for the adiabatic cooling time, is also in nice agreement with the decay time of the order of weeks to months reported in \cite{MAGIC_Mrk421_2021}. 

\begin{figure*}[!h]

   \includegraphics[width=0.9\textwidth,angle=-0]{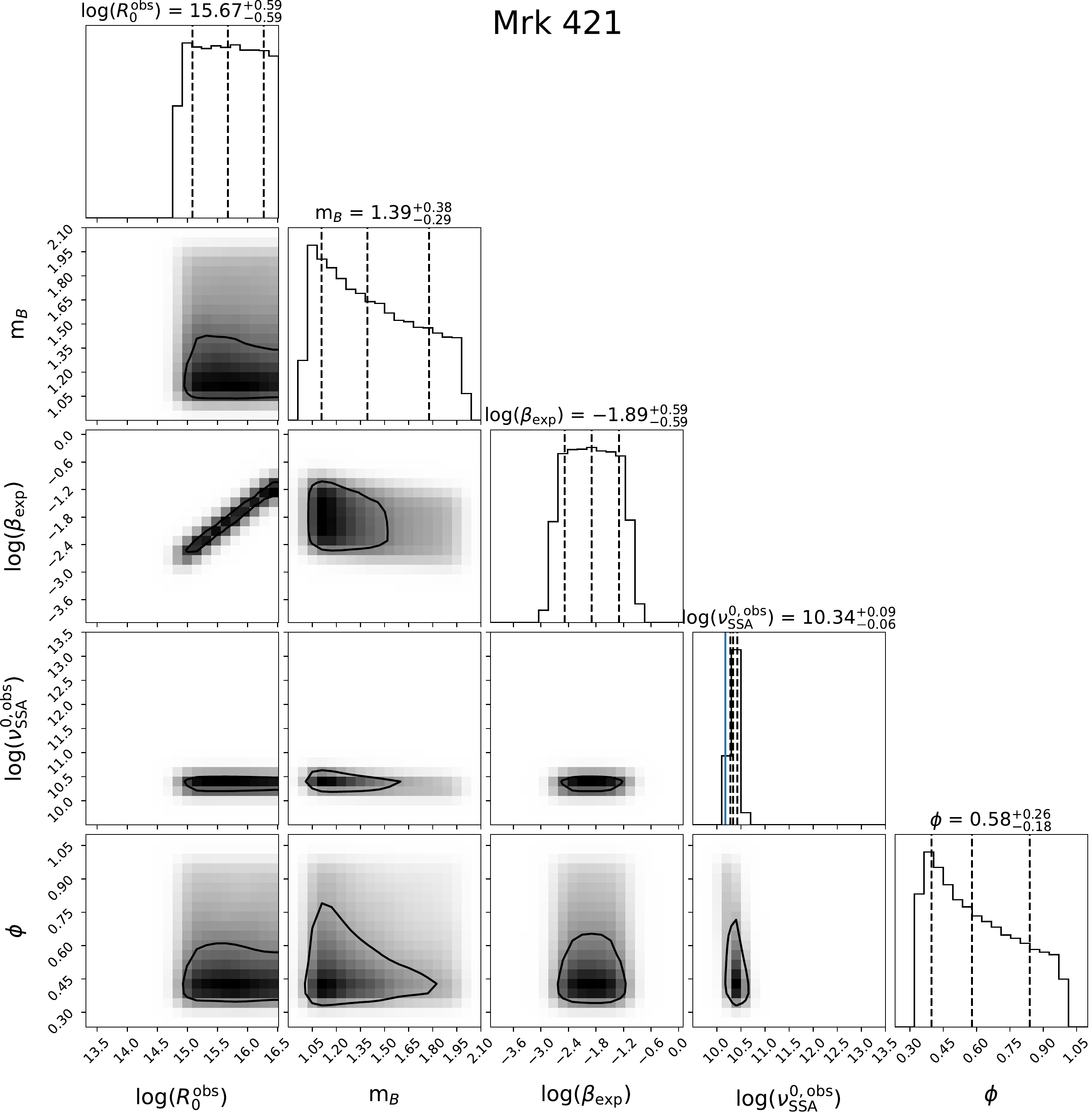}
 
   \caption{
      Posterior distribution for a MCMC sampling  of the composite log-likelihood reported in Equation \ref{eq:log-like} for rise, decay, and delay time in Equation \ref{eq:fit_trends}. To sample the parameter space, we use uninformative flat priors, with  $m_{B}\in[1,2]$, $\phi\in [1/3,1]$, $\nu^{0,\rm obs}_{\rm SSA} \in [10,10^4]$ GHz,  $\beta_{\rm exp} \in [10^{-4}, 1]$. The range of $R_{0}^{\rm  obs}$ is determined by setting a flat range for the observed $\gamma$-ray variability timescale  $t_{\gamma}^{\rm var}\in [0.25,14]$ days, and setting $R_{0}^{\rm  obs}=t_{\gamma}^{\rm var} c$, corresponding to $R_{0}^{\rm  obs} \in [6.5\times10^{13},3.6\times10^{17}]$ cm. We plot the posterior contour maps (where the solid black identifies the 1-$\sigma$ containment for a bivariate Gaussian distribution). On the diagonal, we plot the marginalised posterior distributions, and with the  vertical dashed black lines we indicate the $0.16,0.5,$ and $0.84$ quantiles. The blue vertical line in the $\log(\nu^{0,\rm obs}_{\rm SSA})$ histogram identifies the 15 GHz observed OVRO frequency. On top of each marginalised histogram we report the confidence level corresponding to the quantiles.
   }
   \label{fig:mcmc_mrk_421}
\end{figure*}

\begin{figure*}[!h]
  
   \includegraphics[width=0.9\textwidth,angle=-0]{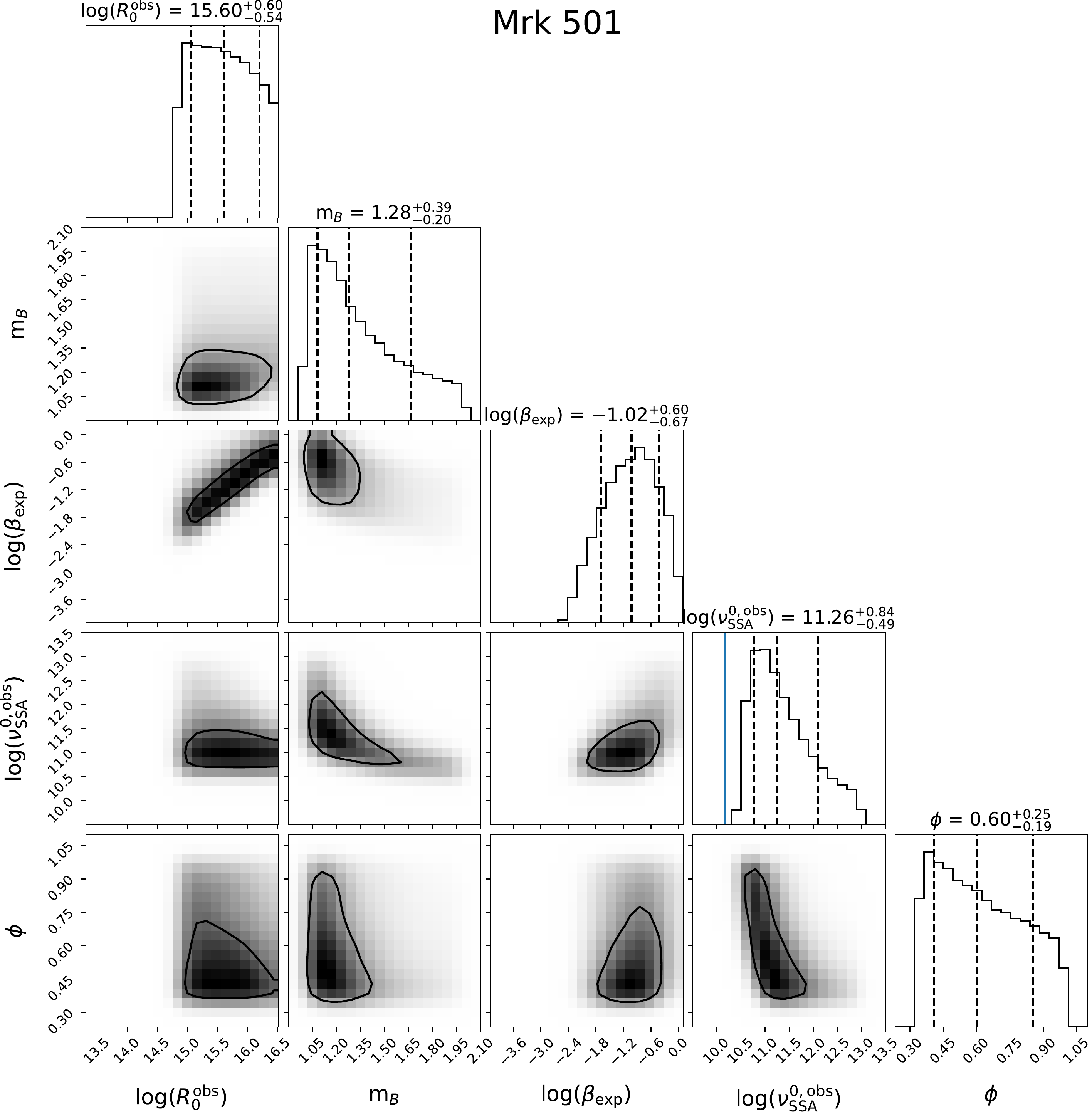}
   
   \caption{ Same as in Figure \ref{fig:mcmc_mrk_421}, but for the case of Mrk 501.}
   \label{fig:mcmc_mrk_501}
\end{figure*}

\begin{figure*}[!h]
  
   \includegraphics[width=0.9\textwidth,angle=-0]{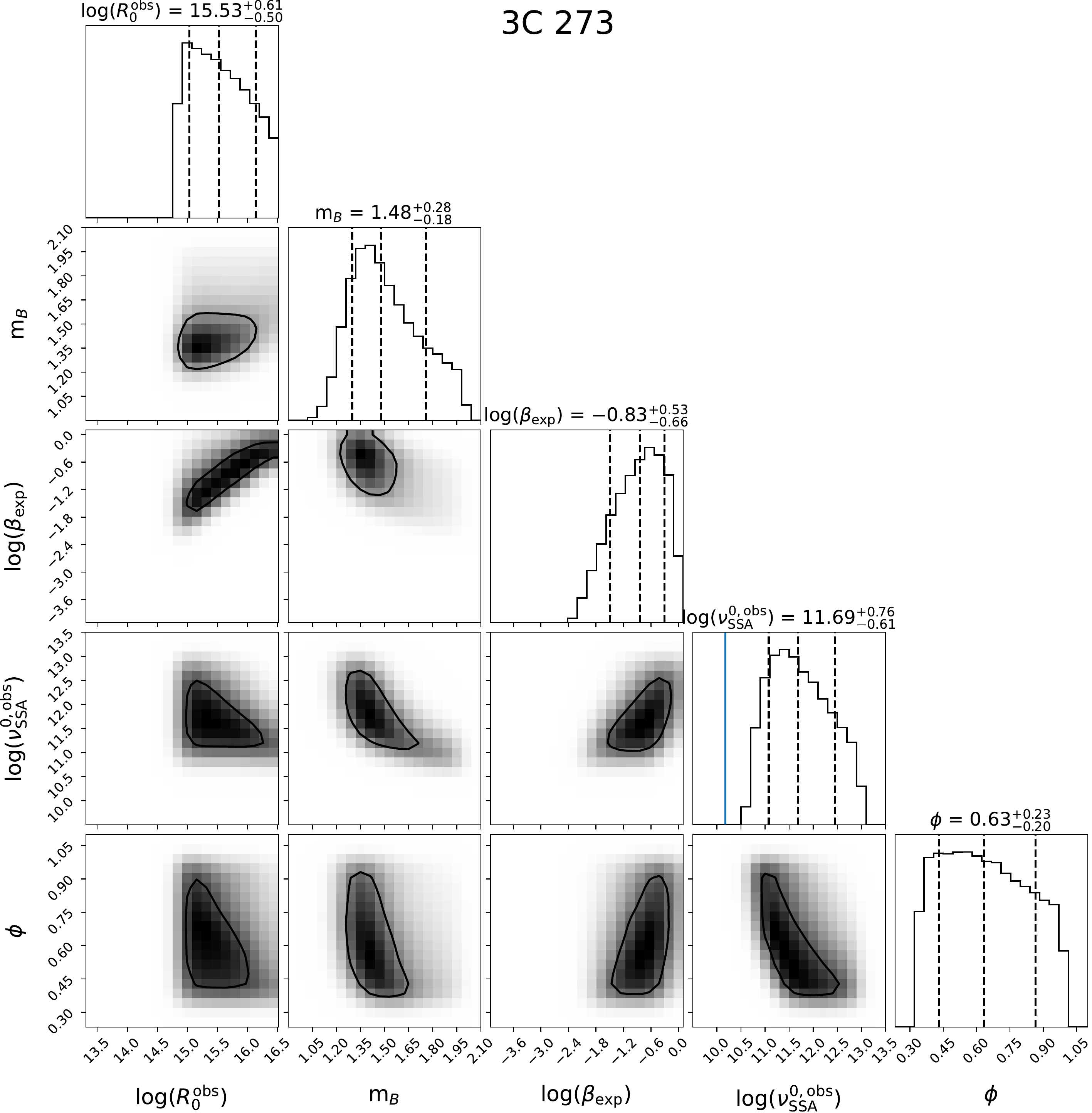}
   
   \caption{  Same as in Figure \ref{fig:mcmc_mrk_421}, but for the case of Mrk 3C 273.}
   \label{fig:mcmc_3C723}
\end{figure*}

\begin{figure}
   \begin{tabular}{c}
   \includegraphics[width=0.4\textwidth,angle=-0]{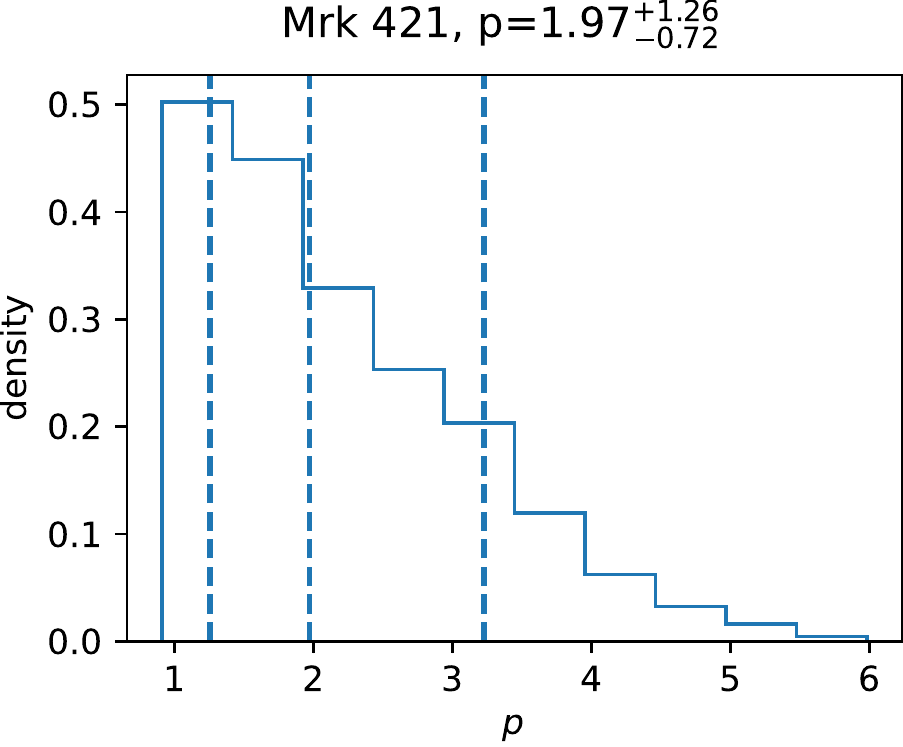}\\
   \includegraphics[width=0.4\textwidth,angle=-0]{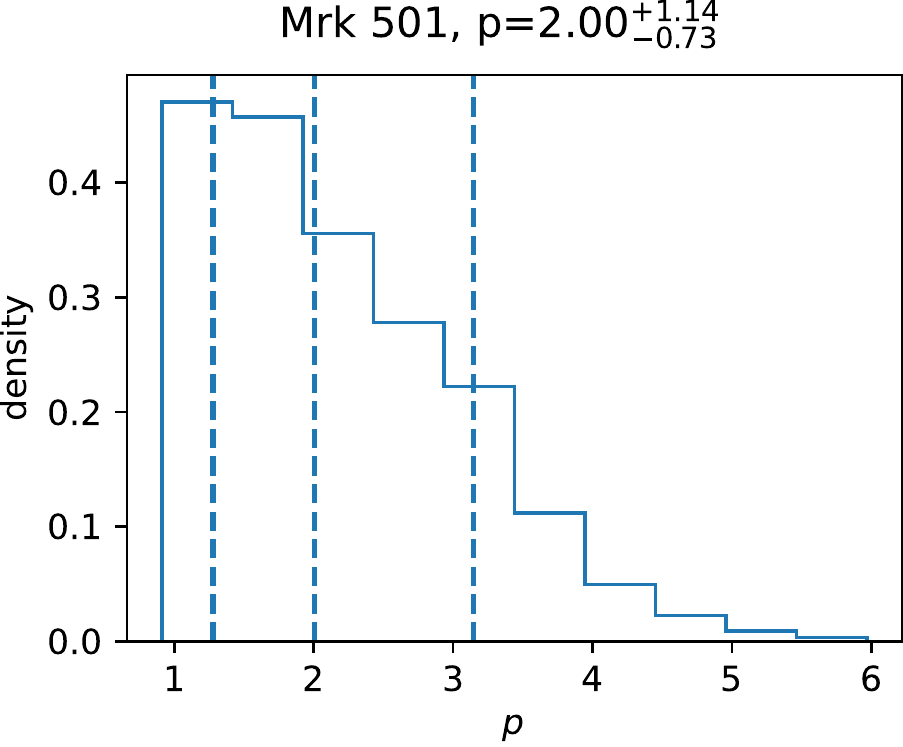}\\
   \includegraphics[width=0.4\textwidth,angle=-0]{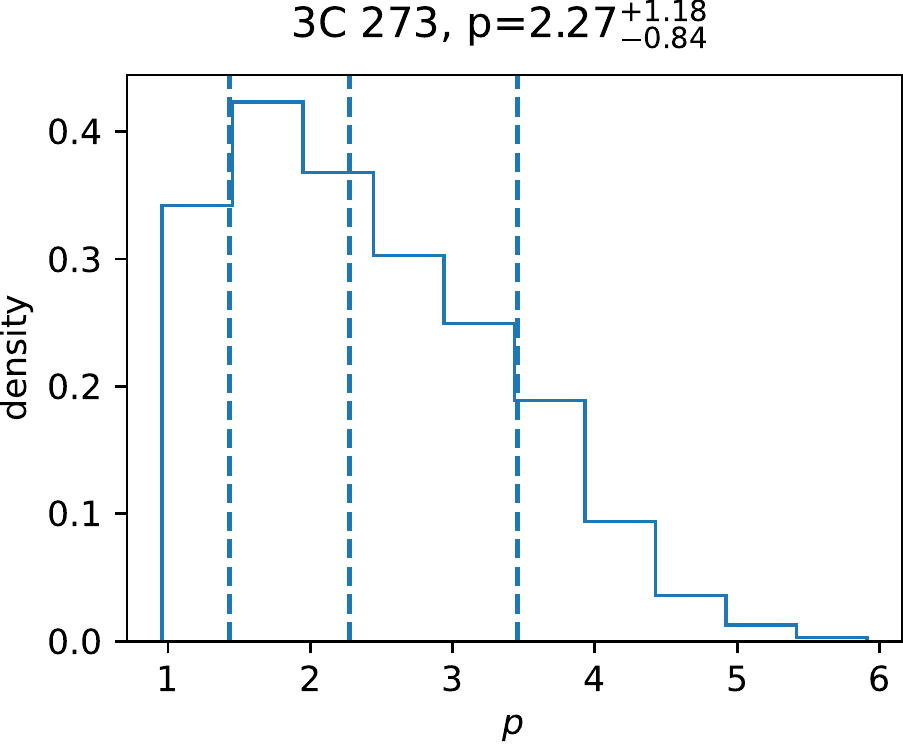}

   \end{tabular}
   \caption{  
   Posterior distributions of the electron distribution index $p$ obtained form the posterior of $m_B$ and $\phi$ 
   returned by the MCMC sampler.
   \textit{Top panel:} Mrk 421,  \textit{middle panel:} Mrk 501, \textit{Bottom panel:} 3C273.
   }
\label{fig:mcmc_p}
\end{figure}

To get a deeper understanding of the physics embedded in the convolution analysis, we used a MCMC approach using the \texttt{emcee}\footnote{ \url{https://emcee.readthedocs.io/en/stable/}}  package \citep{emcee}. We define a composite log-likelihood $\mathcal{L}= \mathcal{L}_{\rm rise} + \mathcal{L}_{\rm decay} + \mathcal{L}_{\rm delay}$, where $\mathcal{L}_{\rm rise}$,  $\mathcal{L}_{\rm decay}$, and   $\mathcal{L}_{\rm delay}$ represent the log-likelihood functions corresponding to rise, decay, and delay time in Equation \ref{eq:fit_trends}. Each likelihood is evaluated assuming that the parameters returned by the convolution analysis are distributed according to a Gaussian PDF:
\begin{equation}
\mathcal{L}\propto \sum_{i=[1,2,3]}-\frac{1}{2} \frac{(x_{i}-\mu_{i})^2}{2\sigma_{i}^2} - \frac{1}{2}\ln(\sigma_{i}^2),
\label{eq:log-like}
\end{equation}
where $\mu_i$ and $\sigma_i$ represent  best-fit parameter values and the 1-$\sigma$ errors, respectively,  for $\Delta_t$, $t^{\rm obs}_{\rm rise}$, and  $t^{\rm obs}_{\rm decay}$ obtained in the convolution analysis, and $x_i$ represents the corresponding parameter evaluated from Equation \ref{eq:fit_trends}. In order to sample the parameter space, we ran a chain with  $10^4$ steps, and a burn-in length of 1000 steps, checking that the chain was always converging. We use uninformative flat priors, with  $m_{B}\in[1,2]$, $\phi\in [1/3,1]$, $\nu^{0,\rm obs}_{\rm SSA} \in [10,10^4]$ GHz,  $\beta_{\rm exp} \in [10^{-4}, 1]$. To determine the range on $R_{0}^{\rm  obs}$, we started from setting a flat range for the observed $\gamma$-ray variability timescale  $t_{\gamma}^{\rm var}\in [0.25,14]$ days, and we set $R_{0}^{\rm  obs}=t_{\gamma}^{\rm var} c$, leading to $R_{0}^{\rm  obs} \in [6.5\times10^{13},3.6\times10^{17}]$ cm. As the phenomenological relations have a bias with respect to the case where radiative cooling is taken into account (see Section \ref{sect:response}, Figure \ref{fig:validation}) we add $5\%$ systematic error to the convolution analysis results. In Appendix \ref{sect:mcmc_validation} we provide a validation of the method against the simulation for $\beta_{exp}=0.1$. In Figures \ref{fig:mcmc_mrk_421}, \ref{fig:mcmc_mrk_501}, and \ref{fig:mcmc_3C723}, we plot the posterior contour maps (where the solid black line identifies the 1-$\sigma$ containment for a bivariate Gaussian distribution). On the diagonal, we plot the marginalised posterior distributions. 
The blue vertical line in the $\log(\nu^{0,\rm obs}_{\rm SSA})$ histogram identifies the 15 GHz observed OVRO frequency.
In figure \ref{fig:mcmc_p}, we plot the histogram of the values of the electron distribution index $p$ obtained from the posterior values of $m_B$ and $\phi$, and using the second equation of Equation \ref{eq:R_transp}.
We notice that the MCMC is able to provide informative confidence regions for the  parameters of interest, except for $\log(R_{0})$  estimated for Mrk 421, where we notice a flat posterior for $\log(R_{0}^{\rm  obs})=15.67^{+0.59}_{-0.59}$. In all the other cases, we get  informative posteriors. The magnetic index, for Mrk 421, $m_B=1.39^{+0.38}_{-0.29}$ has the peak of the PDF at $m_B=1$. For the same source, we notice the low value of   $\log(\nu^{0,\rm obs}_{\rm SSA})=10.34^{+0.09}_{-0.06}$, corresponding to $\nu^{0,\rm obs}_{\rm SSA}\approx 22$ GHz, driven by the short $t^{\rm obs}_{rise}$ returned by the convolution analysis, and very close the observed OVRO frequency of 15 GHz. For the case of Mrk 501, we obtain $\log(R_{0}^{\rm  obs})=15.60^{+0.60}_{-0.54}$, $m_B=1.31^{+0.36}_{0.22}$. In this case, the PDF is also peaking at  $m_B\approx1.0$. The initial SSA frequency $\log(\nu^{0,\rm obs}_{\rm SSA})=11.26^{+0.84}_{-0.49}$, which is larger than in the case of Mrk5 421, is compatible with the longer rise time.  Regarding 3C 273, we obtain $\log(R_{0}^{\rm obs})=15.53^{+0.61}_{-0.50}$, $m_B=1.48^{+0.28}_{-0.18}$, but in this case the PDF is not peaking at $m_B=1$. In this case, as in the case of Mrk 501, we also obtain a larger value of $\log(\nu^{0,\rm obs}_{\rm SSA})=11.69^{+0.76}_{-0.61}$ compared to the case of Mrk 421. Regarding the index $\phi$, we find it interesting to discuss the estimate on the electron index $p$. Indeed, as $\phi$ plays the same role as $\psi$, we can use the second equation of Equation \ref{eq:R_transp} to estimate the posterior distribution of $p$ from the  posterior values of $m_B$ and $\phi$. The result is shown in Figure \ref{fig:mcmc_p}. For the case of Mrk 421, we find a confidence level of $p=1.97^{+1.26}_{-0.72}$, for Mrk 501 $p=2.00^{+1.14}_{-0.73}$, and for 3C 273 $p=2.27^{+1.18}_{-0.84}$. The value of 3C 273 is compatible with a stronger cooling regime, and is therefore  in agreement with the presence of  an EC radiative component. For all the objects, the values of $p$ are compatible with the predictions from Fermi first-order acceleration plus a  stochastic component, which is in agreement with previous theoretical and observational analyses \citep{Tramacere2009,Tramacere2011}.

Nevertheless, the convolution analysis performed with the observed data can be affected by numerous sources of bias; in particular, our use of a single response for the full time-span,  encompassing several flaring episodes, with the possibility that the response from flare to flare might change due to different physical conditions in each single flare. 
There are further  effects that we do  not take into account in the present work and that can have an impact: 
\begin{itemize}
   \item  Changes in the Bulk Lorentz factor, or in the magnetic index $m_B$. These effects might actually be connected, and could lead to deviations in the trends presented in the previous section. For example, a decrease in the Bulk Lorentz factor after the beginning of the expansion  would produce observed decay times that are significantly longer than those derived in the blob frame.
   \item External photon fields can impact  the  response amplitude, and indeed the evolution of the radio flare if the EC field is dominant at the scale where the expansion takes place. In this latter case, we would expect a lower response amplitude compared to the case of pure SSC. Moreover, the transition from SSC- to EC-dominated regimes can impact the competition between the adiabatic and radiative timescales, with effects similar to those investigated in the case of the synchrotron cooling.
   \item A further bias is introduced by the sampling of the IC window for different classes of objects. For example, in the case of FSRQs, the Fermi window   (0.1-300\,GeV)  is sampling the EC bump close to the SED peak, whilst in the case of HBLs, Fermi is sampling the rising part of the IC bump. We therefore expect a different dynamics in the light curves.
   \item  Regarding the interplay between escape times and source size, because particle escape time is $\propto{R/c}$, we could expect significant escape at the time of the flare, with increasing escape times as long as the source expands. This, again, might lead to modulation of the amplitude and also the SSA frequency. The combination of these effects could impact the slopes of the trends in Equation \ref{eq:fit_trends}
   \item In general,  changes to the initial SSA frequency, $\nu^{0,\rm obs}_{\rm SSA}$, related to different combinations of $N^{\rm tot}$ and $B_0$ will  lead to the same trends as those in Equation \ref{eq:fit_trends}. Given  the functional form of the term $\left(\nu^{0,\rm obs}_{\rm SSA}/\nu^{*,\rm obs}_{\rm SSA}\right)^{\phi}$, these trends with respect to $\nu^{*,\rm obs}_{\rm SSA}$ will  have a changed sing in the exponent.
   
\end{itemize}

Nevertheless, the agreement between the observed radio data and the results obtained from the convolution with a single response is satisfactory and supports the hypothesis that adiabatic expansion is the main driver of the observed radio delay.
It is interesting to note a possible connection between the expansion and the polarisation measurements. According to  \cite{Yusef2007}, for an expanding blob, the polarisation rotation measure (RM) follows the trend $RM\propto NRB$, which for our setup (constant $N(t)R^{3}(t)$),  will scale as $RM\propto R^{-3}$ for $m_B=1$, and  as $RM\propto R^{-4}$ for $m_B=2$. Therefore, in the case of an expansion we expect the RM to decrease at lower radio frequencies, and would  also expect a correlation between the RM polarisation and the radio light curves.

We note that the change in the SSA frequency with time, and the consequent delay, might also be related to a bending jet. Indeed, the  SSA frequency will change according to the observing angle for a slab   geometry, as discussed by \cite{Ghisellini1985}. In this case, we would expect a change of the beaming factor, and also a shift of the SSA frequency (to higher values if the bending results in a lower   synchrotron optical thickness). Therefore, if the bending orientates the shorter side of the  slab
towards the line of sight of the observer, a further delay is to be expected. On the contrary, if the rotation orientates the larger side of the slab toward the line of sight of the observer, no additional delay should be observed.  We can provide a rough estimate of the distance travelled by the emitting region during the observed delay using Equation \ref{eq:delat_r}, which for $\delta\Gamma\approx 100$ and $\beta_{\Gamma}\approx 1$ would result in $\Delta_{r}$ of the order of $3\times 10^{12}\frac{\Delta^{\rm obs}}{\rm 1 s}$ cm, that is of the order of $10^{18-19}$ cm for observed delays of the order of a few weeks to months. During this timescale, the jets should bend towards the observer in such a way as to move the SSA frequency down to the GHz window. We should also consider that the total amount of requested rotation  will also depend on the slab aspect ratio, and that, for a square aspect ratio, we would expect almost no delay. We notice that  a rotation towards the line of sight of the observer would also result in a larger beaming factor compared to initial one observed during the $\gamma-$ray flaring episode, adding an extra achromatic variability pattern induced by the change in the beaming factor.
All these effects make it difficult to understand the  extent to which bending alone, that is, without expansion, could explain the observed delays. Nevertheless,  it is possible  that both processes happen, and therefore it would be interesting to look for unambiguous signatures, such as an achromatic variability pattern.

\begin{figure}
   \centering
   \begin{tabular}{l}
   \includegraphics[width=0.45\textwidth,angle=-0]{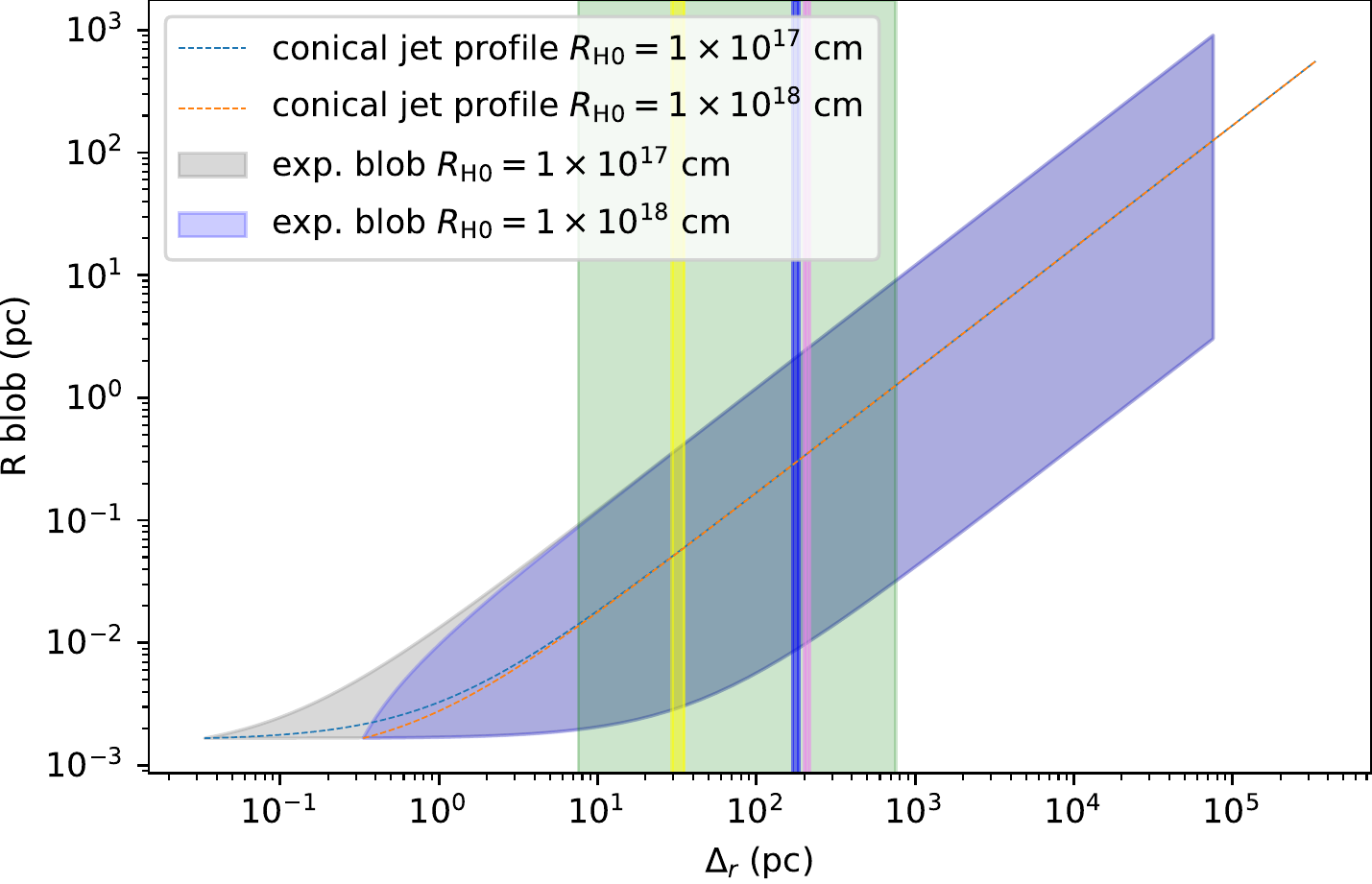}\\
   \includegraphics[width=0.45\textwidth,angle=-0]{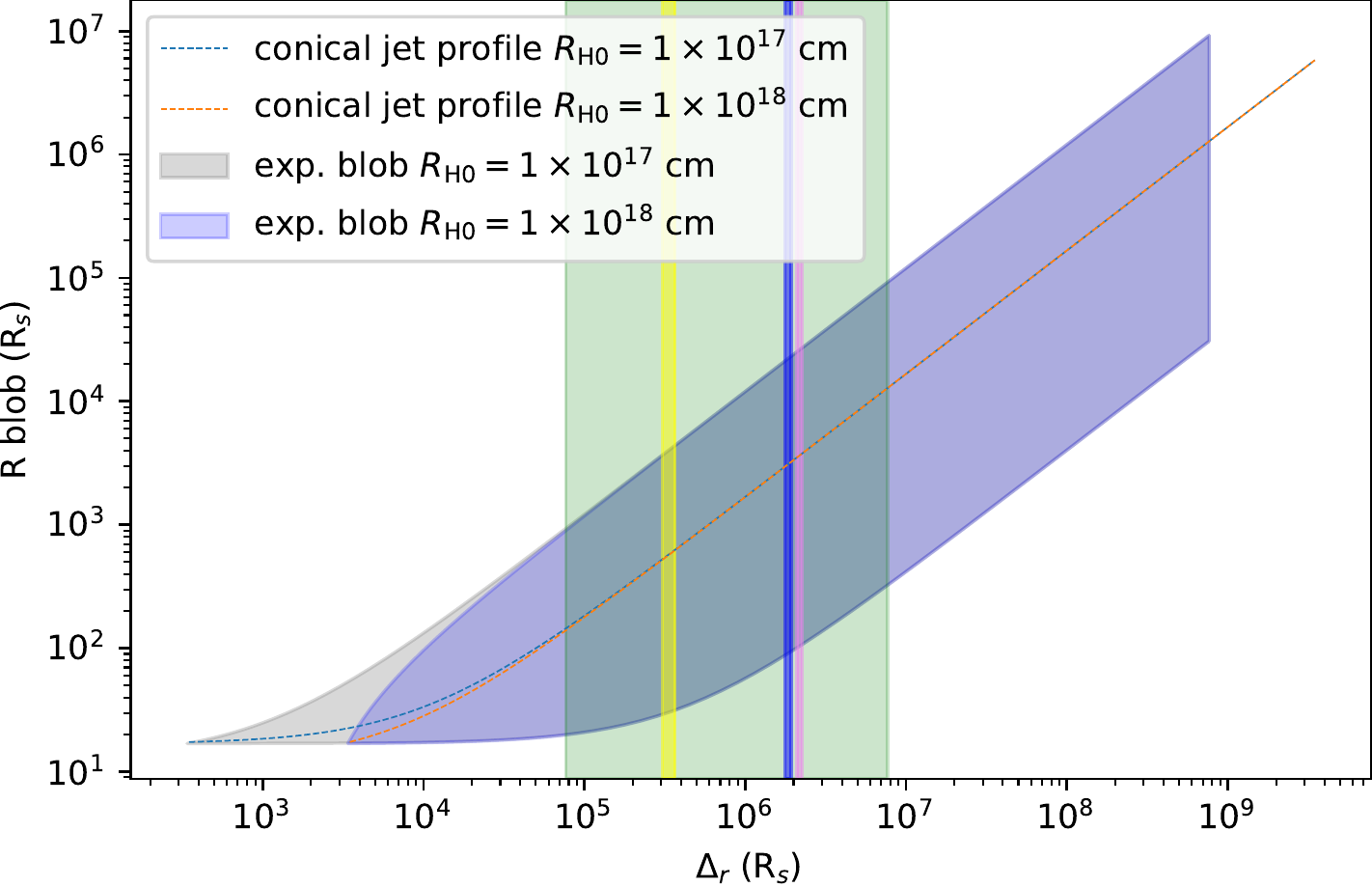}
   \end{tabular}
   \caption{\textit{Top Panel}: 
   { Evolution of blob size ($R$), versus distance travelled by the blob across the jet ($\Delta_r$).}
   The grey shaded area  represents the evolution of $R$ for a blob expanding with constant velocity, with a range of  $\beta_{\rm exp}=[0.001-0.3$, for a Bulk Lorentz factor of 25, an observing angle of 1.5 deg,  $t_{\rm exp}=0$, $R_{0}=5\times 10^{15}$ cm, and  $R_{H0}=1\times10^{17}$ cm. The green shaded area represents the corresponding $\Delta_r$ interval for a range of observed delay of  $\Delta_t^{\rm obs}=[10,1000]$ days.
    The blue shaded area represents the same trend for the case of $R_{H0}=1\times10^{18}$ cm. $R_{H0}$ is the site of the $\gamma$-ray flare. The blue dashed line represents a conical jet profile for $R_{H0}=1\times10^{17}$ cm  and the orange dashed line for $R_{H0}=1\times10^{18}$ cm, both starting from an initial value of $R_{0}=5\times 10^{15}$  cm. The yellow, blue, and purple shaded regions represent the $\Delta_t^{\rm obs}$ for Mrk 421, Mrk 501, and 3C273, respectively. 
   \textit{Bottome Panel}: Same as in the left panel,  replacing the spatial units with the Schwarzschild radius for a BH mass of $10^{9} M_{\sun}$.}
   \label{fig:blob_size_vs_Delta_r}
\end{figure}

We might speculate that, in agreement with the conclusions presented in \cite{Kovalev2020}, the place where the $\gamma$-ray flare occurs is upstream of or close to the parabolic-to-conical transition site, and should provide ideal conditions for shock formation or for magnetic reconnection, which could lead to first- and second-order particle acceleration, both included in our model.  A $\gamma-$ray flaring region at parsec scale is easily compatible with our analysis. In the top panel of figure \ref{fig:blob_size_vs_Delta_r}, we plot (grey shaded area) the evolution of $R,$ i.e. the blob size, versus $\Delta_{r}$, the distance travelled by the blob across the jet, for a constant expansion velocity, with a range of  $\beta_{\rm exp}=[0.001-0.3]$, for a Bulk Lorentz factor of 25, and an observing angle of 1.5 deg. We also set  $t_{\rm exp}=0$, $R_{0}=5\times 10^{15}$ cm, and  $R_{H0}=1\times10^{17}$ cm. 
We notice that for all three sources in our analysis, we find an estimate of $\Delta_r$ above 10 pc. For the case of Mrk 421, $\Delta_r\approx 30$ pc, and for both Mrk 501 and 3C 237,  $\Delta_r$ is above $\approx100$ pc.
The estimate  $\Delta_r$  is particularly relevant for FSRQs. Indeed, \cite{Costamante2018} found that, due to the lack of broad line region (BLR) absorption features in the Fermi-LAT spectra of 106~ FSRQs, the site of the $\gamma$-ray  emission has to be placed beyond the BLR. For the case of Fermi-LAT FSRQs, assuming  an average disk luminosity of $L_{\rm d}\approx 1.5\times10^{46}$ erg s$^{-1}$, and a radius of the BLR given by $R_{\rm BLR}=10^{17} L_{\rm d,45}^{1/2}$ \citep{Ghisellini2010}, the expected average value for $R_{\rm BLR}$ is  $\approx 4\times 10^{17}$ cm. For the case of 3C273, where we find a delay of $\approx 276$ days, and the disk luminosity is $L_{\rm d}\approx 5\times10^{46}$ erg s$^{-1}$, the corresponding value of the BLR size is $R_{\rm BLR}\approx 7\times 10^{17}$ cm ($\approx 0.23$ pc). Hence, our estimate for 3C 273 of $\Delta_r\gtrsim$ 100 pc for  $0.001<\beta_{\rm exp}<0.3]$ is compatible with a $\gamma$-ray site beyond the BLR, in agreement with the analysis in \cite{Costamante2018}. We conclude that the scales inferred from the radio delays analysis are in agreement with those returned by the analysis of the radio jet profiles, with the jet showing a conical profile above $\sim$ [1-10] pc, and with the $\gamma$-ray emission happening beyond the BLR in FSRQs.

Finally, the fact that  we can reproduce the radio light curve as a response to the $\gamma$-ray light curve over long-term timescales and with a single response has several implications. First, we notice that this might suggest a quite stable configuration of the expansion process, and also that the expansion is persistent along with the other physical mechanisms responsible for the flaring activity. Also, we notice that this could disfavour the possibility of lepto-hadronic emission, at least as a persistent mechanism. Indeed, in lepto-hadronic models, the core of the $\gamma$-ray emission is due to the proton synchrotron, whilst the synchrotron bump is mainly of leptonic origin \citep{Zech2017}. This would lead to an almost uncorrelated variability in the synchrotron and in the IC component, making it almost impossible to reproduce the radio light curve with a single response to the $\gamma$-ray emission. 

\section{Conclusions}
\label{sect:conclusions}
In this work, we present a fully self-consistent numerical and phenomenological framework that  incorporates a response function and a physical model of the relevant timescales. We use this framework to explain the radio and $\gamma$-ray responses and delays observed in the long-term light curves of several blazars.
We implemented a self-consistent simulation to reproduce the radio--$\gamma$ delays using the \jetset~ code. In this first approach, we focused our analysis on the derivation of phenomenological trends
and consequent verification with numerical simulations. Our main goals are to identify the main physical drivers of the delay, and to derive and validate a response function able to reproduce the delayed radio flare as a response to the $\gamma$-ray flare. We used a basic simple assumption of a single blob moving with a bulk Lorentz factor $\Gamma$ and constant expansion velocity $\beta_{exp}$. The main findings of our analysis can be summarised as follows:
\begin{itemize}
    \item From our phenomenological relations, we predict that the delay of the radio emission occurs at a lower frequency than the initial SSA frequency, with an off-set given by the time interval  between the flaring episode $t_{\rm exp}$, which without any loss of generality can be set to zero.
    \item  Our phenomenological trends, summarised in Equation \ref{eq:times_obs_t_R0}, also suggest that both rise and decay times are inversely proportional to $\beta_{exp}$ and are directly 
    proportional to the distance in frequency between the initial SSA frequency and the frequency at which the radio-delayed light curve
    is observed. Moreover, as the magnetic field is bound to the size of the emitting region through the magnetic index $m_B\in [1,2]$ by Equation \ref{eq:B(t)}, we also predict an inverse proportionality between the decay time and $m_B$.
    \item We set up our simulations in order to follow the trends with respect to  both $\beta_{\rm epx}$ and the frequency of the radio light curves. We estimated the relevant timescales  using the best-fit results
    of the convolution of  the response function given in Equation \ref{eq:Resp_func},  applied to the simulated $\gamma$-ray and  radio light curves.
    We verified that the delayed light curves are reproduced with excellent accuracy by the convolution of the  $\gamma$-ray light curve with the proposed response function.
    \item  We also verified that the phenomenological trends given by Equation \ref{eq:fit_trends} are in very good agreement with the simulations when the adiabatic cooling is dominant, and we evaluated the deviations when the radiative cooling is competing. By fitting these trends to the estimated parameters  of response functions, we are able to recover the relevant input parameters of the simulations, such as the initial size of the emitting region ($R_0$), the initial SSA frequency, $\beta_{\rm exp}$, and $m_B$. This proves that it is possible to derive parameters connected to the jet and magnetic field topology from MW long-term campaigns.
    \item We also verified that when the expansion is active, a sudden drop in the Compton dominance is observed, and therefore investigating the time profile of the IC emission during the  flaring episode could provide information on whether or not there is an expansion starting during the flare.
    \item We applied our model to observations of the two HBLs, Mrk 421 and Mrk 501, and the FSRQ 3C 273. First, we verified that a single radio--$\gamma$ response accurately reproduces the long-trend radio light curves, even though short-term features are smoothed out. Then, by means of an MCMC approach that combines the phenomenological trends and the output of the convolution analysis, we are able to estimate the source size, the initial SSA, $m_B$, and the electron distribution index $p$  for these sources, finding satisfactory agreement between the proposed scenario and the observed phenomenology in the literature.
   \item We compared our results with other results in the literature \citep{Boula2018,Potter2018} finding a qualitative agreement, even though the setup of our simulation is not fully equivalent to those used elsewhere. We also verified that recent results presented in \cite{MAGIC_Mrk421_2021} regarding the correlation patterns of Mrk 421 in 2017 require a framework based on adiabatic expansion with confined emitters, confirming that the expansion can occur as early as the $\gamma$-ray flaring episode, and providing further validation of expansion as the mechanism responsible for the radio--$\gamma$ delay.
   \item We instigated possible applications and predictions of our analysis. In particular, we find that the limit on the initial size of the emitting region derived from the delay analysis is in agreement with typical values returned from MW SED fitting. We discuss the possible implications in terms of polarisation measurements, predicting a correlation between the RM of the polarisation and the radio light curves, with the RM decreasing at lower frequencies.
   \item We find that the  blob expansion presented in our model and constrained from the observed radio responses $(\beta_{\rm exp}<0.3)$ predicts a conical jet profile at the scale above $\approx$ [1-10] pc in agreement with the results presented in \cite{Kovalev2020}. This is compatible with the hypothesis that the expansion begins when the jets become plasma dominated, a condition that is ideal for the formation of the shock. Hence, the $\gamma$-ray flare can be used as a signature of the shock formation, indicating the onset of the jet expansion, i.e. the transition from parabolic to conical profiles, and the radio delay is an indicator of the expansion velocity, and consequently of the jet collimation profile beyond the shock.
   \item Finally, we find that the strong correlation between the radio and $\gamma-$ray emission 
   predicted by our model casts doubt on a lepto-hadronic mechanism, at least as a persistent process acting in the core of blazar jets.
    
\end{itemize}

We hope to further extend this work, expanding the parameter space of the current analysis,
exploring the impact of variable bulk Lorentz factors, and $m_{B}$, and investigating 
potential observational signatures of their interconnection by linking these parameters to different jet profiles. We also plan to investigate the role of the light-crossing time as a function of the jet observing angle, and the impact of the EC on the adiabatic--radiative interplay. The code and the analysis presented in this paper are publicly available and fully reproducible (see Appendix \ref{sect:code}).

\begin{acknowledgements}
   This research has made use of data from the OVRO 40-m monitoring program \citep{2011ApJS..194...29R}, supported by private funding from the California Insitute of Technology and the Max Planck Institute for Radio Astronomy, and by NASA grants NNX08AW31G, NNX11A043G, and NNX14AQ89G and NSF grants AST-0808050 and AST- 1109911.
   The following open source codes have been used: \texttt{Numpy} \cite{Numpy}, \texttt{Scipy} \cite{Scipy}, \texttt{Matplotlib} \cite{Matplotlib}, \jetset~ \cite{jetset2020,Tramacere2011,Tramacere2009},  \texttt{emcee} \cite{emcee}.
   The authors thank the anonymous referee for the useful and helpful comments on the manuscript.
  
\end{acknowledgements}

\bibliography{42003_final_layout}
\bibliographystyle{aa}

\begin{appendix}
\section{Code availability and reproducibility}
\label{sect:code}
The work presented here is fully reproducible by following the instructions in the git repository\footnote{ \url{https://github.com/andreatramacere/adiabatic_exp_radio_gamma_delay}}. 
In this repository you can find the notebooks to reproduce the analysis and the instructions to install 
the  \jetset~ version  1.2.0rc11. We remind  the reader that this is a  pre-release of the forthcoming 1.2.0 \jetset~ version.
Further details on the \jetset ~ code can be obtained by contacting the author Andrea Tramacere (\email{andrea.tramacere@gmail.com}).

\section{JetSeT temporal evolution}
\label{app:fp_eq}
In the following, we give a detailed description of the cooling terms used 
in the numerical solution of the FP equation:
\begin{eqnarray}
\label{eq:fp_cooling}
|\dot\gamma_{synch}(t)|&=&\frac{ 4\sigma_Tc}{3 m_ec^2}\gamma^2 U_B(t)=C_0\gamma^2U_B(t)
\\
|\dot\gamma_{IC}(t)|&=&\frac{ 4\sigma_Tc}{3 m_ec^2} \gamma^2 \int
f_{KN}(4\gamma\epsilon_0)\epsilon_0
n_{ph}(\epsilon_0,t)d\epsilon_0\\&=&C_0\gamma^2F_{KN}(\gamma,t)\nonumber \\ \nonumber 
|\dot\gamma_{ad}(t)| &=&\frac{1}{3}\frac{\dot{V}}{V}\gamma = \frac{\dot
R(t)}{R(t)}\gamma = \frac{\beta_{\rm exp}c}{R(t)}\gamma  \nonumber \\
C(\gamma,t) &=&|\dot\gamma_{synch}(t)|+|\dot\gamma_{IC}(t)| + |\dot\gamma_{ad}(t)|
\nonumber 
\end{eqnarray}
where $U_B=B^2/8\pi$, is the energy density of the magnetic field, $\epsilon_0=h
\nu_0/m_ec^2$ is the IC seed photon energy in units of $m_ec^2$, and $n_{ph}(\epsilon_  0)$ is
the number density of IC seed photons with the corresponding photon energy density
$U_{ph}=m_ec^2\int \epsilon_0 n_{ph}(\epsilon_0)d\epsilon_0$. The function $f_{KN}$ results from
the analytical integration of the \cite{Jones1968} Compton kernel, fully taking into 
account Klein-Nishina (KN) effects for an isotropic seed photon field \citep[see][appendix C]{Moder2005},
and $F_{KN}(\gamma)$ represents its convolution with the seed photon field.    
We note that $F_{KN}$ plays a crucial role in the cooling process,
depending both on the IC regime (Thomson (TH) limit for $4\gamma\epsilon_0<<1$, KN limit for 
$4\gamma\epsilon_0>>1$), and on $\epsilon_0 n_{ph}(\epsilon_0)\propto B^2/R^2$.   
The acceleration terms in Eq. \ref{eq:fp_eq},  and their related  timescales,
can be expressed as a power-law in terms of the Lorentz factor ($\gamma$):
\begin{equation}
\label{eq:fp_coeff}
\begin{cases} 
D_{p}(\gamma)&=D_{p0}\left(\frac{\gamma}{\gamma_{0}}\right)^{q},~~~~~~~~ t_D =\frac{1}{D_{p0}}\left(\frac{\gamma}{\gamma_{0}}\right)^{2-q} \\
D_A(\gamma)&=2D_{p0}\left(\frac{\gamma}{\gamma_{0}}\right)^{q-1},~~~t_{DA} = \frac{1}{2D_{p0}}\left(\frac{\gamma}{\gamma_{0}}\right)^{2-q}\\
A(\gamma)&=A_{p0}\gamma,~~~~~~~~~~~~~~~~ t_{A} = \frac{1}{A_{0}}.\\
\end{cases} 
\end{equation}

\section{MCMC analysis validation}
\label{sect:mcmc_validation}

\begin{figure*}[!h]
  
   \includegraphics[width=0.9\textwidth,angle=-0]{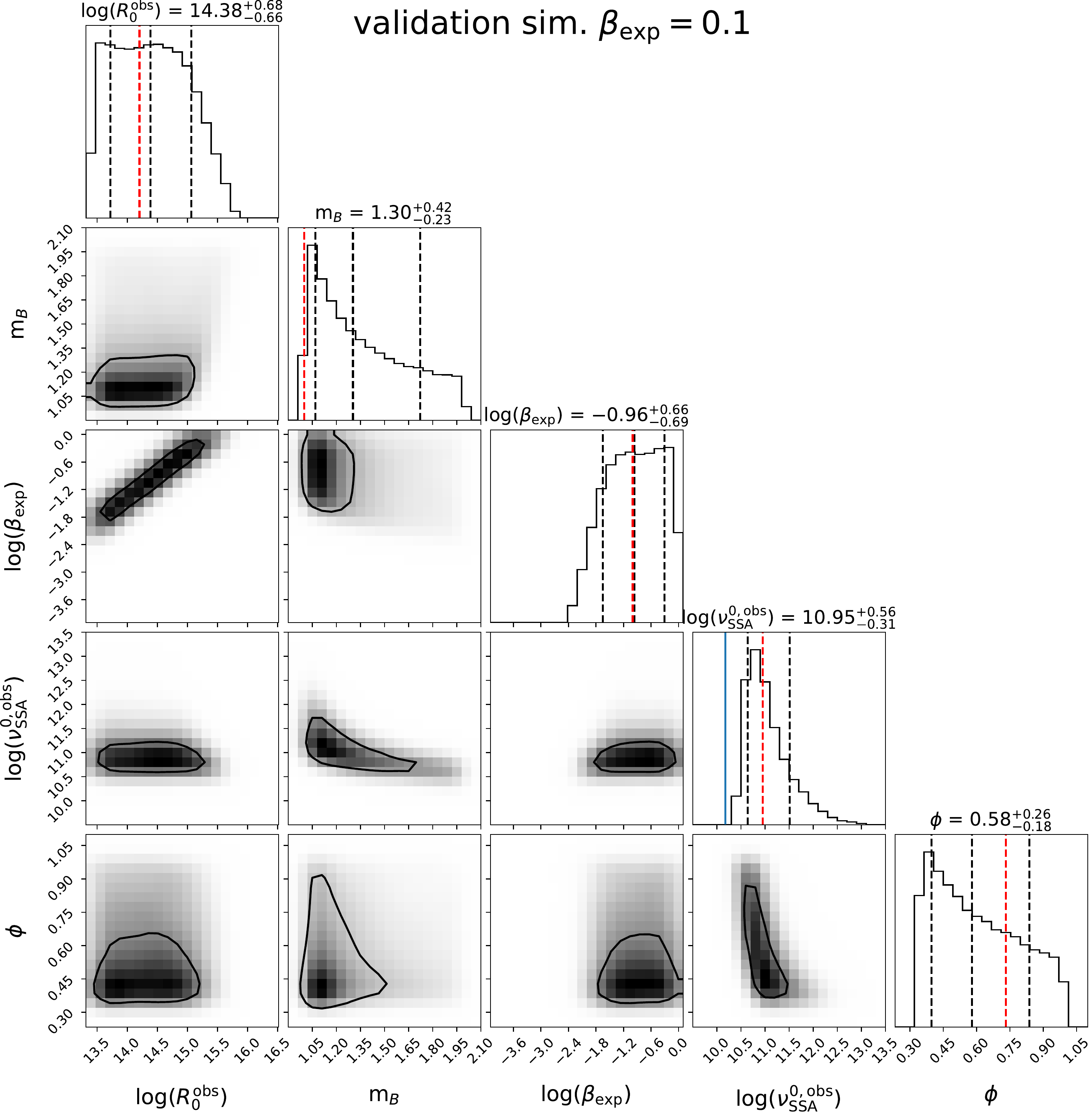}
   \caption{  
   MCMC sampling for the validation of the analysis presented in Section \ref{sect:discussion}. The 
   validation is performed against the simulations for $\beta_{\rm exp}=0.1$. The vertical red dashed lines indicate the input values from the simulation.
   }
   \label{fig:mcmc_validation}
\end{figure*}

\begin{figure}
   \begin{tabular}{c}
   \includegraphics[width=0.4\textwidth,angle=-0]{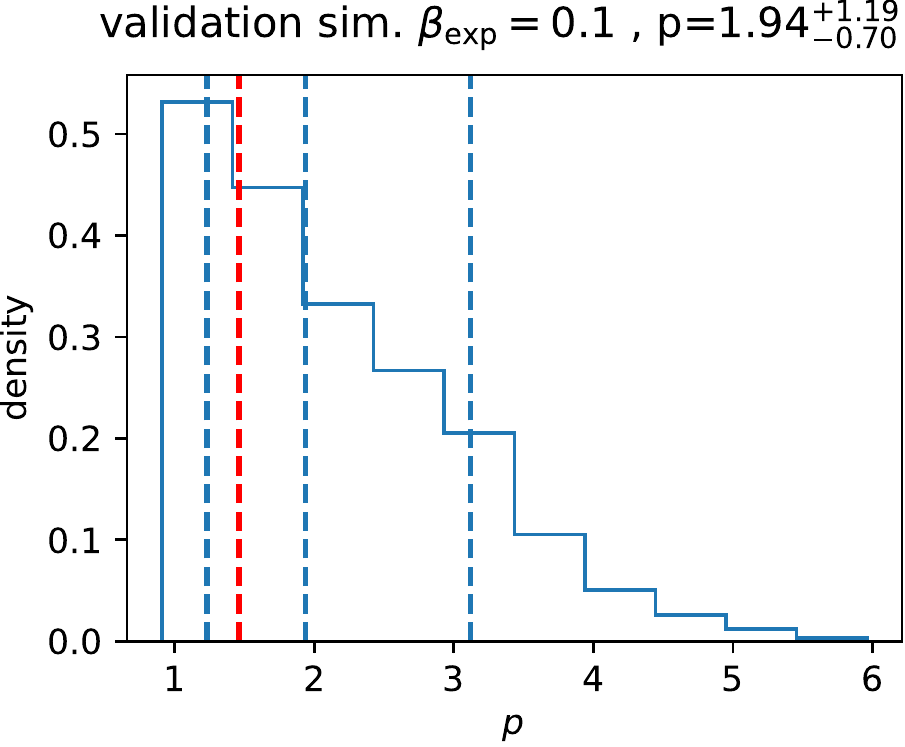}\\
   
   \end{tabular}
   \caption{  
    Same as in Figure \ref{fig:mcmc_p}, but for the $p$ index regarding the MCMC validation reproduced in Figure \ref{fig:mcmc_p_validation}.
    The vertical red dashed line indicates the input value of $p$ for the simulation.}
   
\label{fig:mcmc_p_validation}
\end{figure}

In this section we validate the MCMC analysis presented in Section \ref{sect:conclusions} using as a benchmark the long-term simulations for $\beta_{\rm exp}=0.1$. We use the same configuration as in the case of the analysis presented in Section  \ref{sect:conclusions}. In Figure \ref{fig:mcmc_validation}, we plot the MCMC results, and with a vertical red dashed line we report the `true' simulation values. We notice that, in all cases except $m_B$, the simulation value is contained within the  1-$\sigma$ confidence range returned by the MCMC posterior. For the case of $m_B$, we notice that peak of the posterior PDF matches the simulation value of $m_B=1$. In Figure \ref{fig:mcmc_p_validation}, we show the same check for  the electron distribution index $p$. Also in this case, the agreement is excellent.

\end{appendix}

\end{document}